\documentclass[english, 12pt, a4paper, 
numbers=noenddot 
]{scrartcl}
\usepackage[T1]{fontenc}
\usepackage[utf8]{inputenc}  
\usepackage[english]{babel} 
\usepackage{lmodern} 

\usepackage[onehalfspacing]{setspace} %
\usepackage{scrlayer-scrpage}
\usepackage{amsmath, amssymb, amsthm, amstext} 
\usepackage{thmtools}
\usepackage{bm}

\usepackage[table]{xcolor}
\usepackage{float} 
\usepackage[section]{placeins} 
\usepackage{multirow}
\usepackage{booktabs} 
\usepackage{longtable} 
\usepackage{tabularx} 
\newcolumntype{Y}{>{\centering\arraybackslash}X}

\usepackage{dcolumn}
\newcolumntype{d}[1]{D{.}{.}{#1}}

\usepackage{siunitx}
\usepackage{graphicx}  
\usepackage{epstopdf} 

\usepackage[inline]{enumitem} 

\usepackage[babel, english = british]{csquotes} 

\usepackage[backend=biber,
			style=apa, 
			sorting = nyt,
			maxnames = 2,
            maxbibnames=8,
            minbibnames=4,
			doi=false,
			isbn=false,
			url=false,
            dashed=false,
			eprint=false,
            uniquename=false
            ]{biblatex}
\DeclareDelimFormat{nameyeardelim}{\addcomma\space} 
\usepackage{varioref} 
\definecolor{darkblue}{rgb}{0.0, 0.0, 0.55} 
\usepackage[pdfencoding=auto, psdextra, 
 hyperfootnotes=false, 
 colorlinks=true,
 linkcolor=darkblue,  
 citecolor=darkblue,  
 urlcolor=darkblue    
 ]{hyperref} 
\usepackage{nameref}
\usepackage[capitalize, nameinlink]{cleveref} 

\newtheoremstyle{Definition}
  {0.2cm}                 
  {0.2cm}                 
  {\normalfont}           
  {}                      
  {\bfseries}  						
  {.}                     
  { }              				
  {}
                          
\newtheoremstyle{Theorem}
  {0.2cm}                 
  {0.2cm}                 
  {\itshape}           	  
  {}                      
  {\bfseries}  						
  {.}                     
  { }              				
  {}
\theoremstyle{Theorem}
    	\Crefname{cor}{Corollary}{Corollaries}
		\Crefname{prop}{Proposition}{Propositions}
		\Crefname{lem}{Lemma}{Lemmas}
    	\Crefname{thm}{Theorem}{Theorems}
		\Crefname{assum}{Assumption}{Assumptions}
	
		\Crefname{conjecture}{Conjecture}{Conjectures}
		\Crefname{defn}{Definition}{Definitions}

\theoremstyle{Definition}
    	\Crefname{example}{Example}{Examples}	
    \newtheorem{rem}{Remark}
    	\Crefname{rem}{Remark}{Remarks}      
		\Crefname{algo}{Algorithm}{Algorithms}

\renewcommand*{\theenumi}{\arabic{enumi}}
\renewcommand*{\theenumii}{(\alph{enumii})}
\renewcommand*{\theenumiii}{(\roman{enumiii})}

\makeatletter
\renewcommand*{\p@enumii}{\theenumi\,}
\renewcommand*{\p@enumiii}{\p@enumii.\theenumii} 
\renewcommand*{\p@enumiv}{\p@enumiii.\theenumiii}
\makeatother

\allowdisplaybreaks
\graphicspath{{./figures/}} 
\hypersetup{pdfauthor = {Martin C. Arnold and Thilo Reinschluessel},
            pdftitle = {Information-Enriched Selection of Stationary and Non-Stationary Autoregressions using the Adaptive Lasso}
}

\input{ee.sty} 
\makeatletter
\renewenvironment{abstract}{%
\if@abstrt
    \small
    \begin{center}
      {\normalfont\sectfont\nobreak\abstractname 
        \vspace{-.5em}\vspace{\z@}}%
    \end{center}
\fi
    \singlespacing\quotation 
}{%
\endquotation
} 
\makeatother

\newsavebox\extrainfobox

\newcommand{\DOI}[1]{%
  \gdef\@DOI{DOI: #1}%
}

\addtokomafont{disposition}{\rmfamily} 
\addtokomafont{title}{} 
\addtokomafont{author}{\large}
\addtokomafont{date}{\large}
\addtokomafont{date}{\vspace*{0.75em}}
\title{Information-Enriched Selection of Stationary and Non-Stationary Autoregressions using the Adaptive Lasso\protect\footnotemark\footnotetext{We thank Christoph Hanck for valuable comments that significantly improved the paper. We are indebted to Anders B. Kock for his supportive feedback on several parts of this paper. We also thank Yannick Hoga, Matei Demetrescu, and the Winter 2022/23 Ruhrmetrics Research Seminar participants for their helpful suggestions. The authors acknowledge partial financial support from TRR 391 Spatio-temporal Statistics for the Transition of Energy and Transport (520388526) by the Deutsche Forschungsgemeinschaft (DFG, German Research Foundation). }}
\author{Martin Arnold\thanks{Faculty of Business Administration and Economics, University of Duisburg-Essen, Universit\"atsstra{\ss}e 12, 45117 Essen, Germany}\\{\vspace{-1ex}\small \href{mailto:martin.arnold@vwl.uni-due.de}{martin.arnold@vwl.uni-due.de}}  \and Thilo Reinschlüssel\footnotemark[2]~\thanks{RGS Econ - Ruhr Graduate School in Economics, Research Academy Ruhr, Universitätsstra{\ss}e 150, 44801 Bochum, Germany}\\{\small \href{mailto:thilo.reinschluessel@vwl.uni-due.de}{thilo.reinschluessel@vwl.uni-due.de}}}
\date{October 12, 2026} 

\begin{document}
\maketitle
\vspace*{-2.5em} %

\begin{abstract}
\noindent\textit{\textbf{Abstract}}\\[.5ex]
We propose a novel approach to elicit the weight of a potentially non-stationary regressor in the consistent and oracle-efficient estimation of autoregressive models using the adaptive Lasso. The enhanced weight builds on a statistic that exploits distinct orders in probability of the OLS estimator in time series regressions when the degree of integration differs. We develop a new theoretical framework focusing on activation knots to establish results on the benefit of our approach for detecting stationarity when a tuning criterion selects the $\ell_1$ penalty parameter. Monte Carlo evidence shows that our proposal is superior to using OLS-based weights, as suggested by Kock [Econom. Theory, 32, 2016, 243--259]. We apply the modified estimator to model selection for German inflation rates after the introduction of the Euro. The results indicate that energy commodity price inflation and headline inflation are best described by stationary autoregressions. 
\\[2ex]
\end{abstract}
\renewcommand*{\thefootnote}{\arabic{footnote}}
\setcounter{footnote}{0}
\newpage %
\section{Introduction}
\setcounter{footnote}{0}
Over the last two decades, shrinkage estimators have become increasingly popular in the econometric literature and relevant for economic applications. A substantial body of work considers Lasso-type estimators (\cite{Tibshirani1996}) and establishes conditions for convergence to their unpenalised analogue in the correct model: such \enquote{oracle efficient} procedures select the set of relevant variables with probability converging to 1 and ensure that the coefficient estimators have the identical asymptotic distributions as the unpenalised estimators if the relevant variables were known from the outset (cf. \cite{FanLi2001}). 

Recent contributions apply shrinkage estimators to dependent and non-identic\-ally distributed data, which often arise in macroeconomic modelling. For example, \textcite{CanerHan2014} use group bridge estimation to select the correct factor number in approximate factor models. \textcite{MedeirosMendes2012} establish the oracle property for the adaptive Lasso of \textcite{zou2006adaptive} in sparse high-dimensional models for stationary time series. \textcite{KockCallot2015} present oracle inequalities for high-dimensional stationary vector autoregression (VAR) models.

Lasso-type estimators have also been suggested for models with non-stationary variables. \textcite{Leeetal2022} propose the twin adaptive Lasso for oracle-efficient estimation of predictive regressions with mixed root predictors. \textcite{LiaoPhillips2015} devise a consistent adaptive shrinkage procedure that selects the cointegration rank and the lag order in cointegrated VARs. \textcite{Wilms2016} discuss forecasting high-dimensional time series based on $\ell_1$-penalised maximum likelihood estimators for sparse cointegration. \textcite{CanerKnight2013} show that bridge estimators achieve conservative model selection in stationary and non-stationary autoregressions. \textcite{kock2016consistent} proves that the (BIC-tuned) adaptive Lasso is an oracle-efficient estimator in augmented Dickey-Fuller (ADF) regressions. Similarly to the multivariate procedure in \textcite{LiaoPhillips2015}, adaptive Lasso may thus distinguish between stationary and non-stationary data and simultaneously select the correct sparsity pattern in the lags of the dependent variable.

The key feature of adaptive Lasso is an adaptive penalty applied to the individual model coefficients. To implement the adaptive Lasso for ADF models, \textcite{kock2016consistent} suggests using weights based on ordinary least squares (OLS) estimates from an auxiliary ADF regression. This approach aligns with \textcite{zou2006adaptive} who shows that using weights based on consistent coefficient estimates is sufficient for the oracle property. To our knowledge, no contributions in the literature address tailored weight specification for autoregressive (AR) models beyond using OLS estimates. In a high-dimensional cross-section framework, \textcite{Huangetal2008} prove that using marginal regression to obtain weights that need not be consistent for the true coefficients ensures the oracle property. Their simulation results indicate an improved selection performance under liberal tuning using the AIC. However, their asymptotic analysis relies on critical assumptions like partial orthogonality and a variation of the irrepresentable condition for the Lasso, which is not generally satisfied in time series regressions.\footnote{See \textcite{meinshausen2006high} for the original irrepresentable condition of the Lasso.}

If information indicative of a regressor's relevance other than its OLS coefficient estimate is available, incorporating this information into the adaptive weight may yield a Lasso estimator with improved properties. In ADF models, the potentially non-stationary regressor $y_{t-1}$ is of particular interest since its coefficient determines the order of integration. The unit root test literature suggests a variety of identification principles for the latter that could be informative for the weight. Combining two identification principles, we propose an enhanced weight for $y_{t-1}$ in estimating ADF models using the adaptive Lasso. This information-enriched weight implies several properties of the Lasso solution path that are beneficial for (consistent) tuning, especially under stationarity. The resulting estimator is shown to share the oracle property of the adaptive Lasso estimator in \textcite{kock2016consistent} and can be consistently tuned using BIC. Moreover, we prove that the estimator resulting from the modified penalty weight enjoys favourable properties like a relaxed shrinkage when required. Simulation evidence demonstrates that our modification enhances the correct selection rate of stationary and non-stationary models in finite samples, further motivating the use of adaptive Lasso for unit root testing. Our method, therefore, improves upon the estimator in \textcite{kock2016consistent}, particularly in this respect.

The remainder of this article is organised as follows. In \Cref{sec:dnsapr}, we discuss the statistical framework and model selection of ADF regressions using the adaptive Lasso. \Cref{sec:alieirw} introduces the modified adaptive weight and presents several theoretical results on the estimator's asymptotic properties, emphasising the detection of a stationary stochastic component. \Cref{sec:nmm} elaborates on the merit of information enrichment to mitigate spurious selection of an irrelevant variable using the example of a zero-mean model. Simulation studies highlighting the benefit of the modified weight are presented in \Cref{sec:simev}, where we also discuss handling deterministic trends. In \Cref{sec:empapp}, we apply the procedure to model selection for German energy and headline inflation rates. \Cref{sec:conc} concludes and gives an outlook to potential further applications of the information enrichment principle. \Cref{sec:amcr} presents additional simulation results. \Cref{sec:proofs} contains proofs for the theoretical results of \Cref{sec:alieirw,sec:nmm}.

We will use the following conventions throughout the manuscript. $\xrightarrow{p}$ and $\xrightarrow{d}$ denote convergence in probability and distribution, respectively. Coefficients of the true ADF model have a superscript $\star$. We write $\lambda=\Theta(T^\kappa)$ for $0<\liminf_{T\rightarrow\infty}\lambda/T^\kappa\leq\limsup_{T\rightarrow\infty}\lambda/T^\kappa<\infty$, $\kappa\in\mathbb{R}$. Several (estimated) quantities often denoted by Greek symbols depend on the $\ell_1$ penalty and the sample size, which we express by sub-indexing with $\lambda$ and $T$ when important for the exposition. Coefficients $\beta_i^\star$ of the relevant variables identify the data-generating model $\mathcal{M}:=\left\{i:\beta_i^\star\neq0\right\}$. We refer to an estimator $\widehat{\mathcal{M}}$ as model selection consistent if $\lim_{T\to\infty}\Prob{\widehat{\mathcal{M}} = \mathcal{M}} = 1$ and conservative if $\lim_{T\to\infty}\Prob{\widehat{\mathcal{M}} \supseteq \mathcal{M}} = 1$ with $\lim_{T\to\infty}\Prob{\widehat{\mathcal{M}}\supset\mathcal{M}} > 0$, where $\widehat{\mathcal{M}}$ denotes the 
estimated model.

\section{Model selection of autoregressions using the adaptive Lasso}
\label{sec:dnsapr}
We consider time series $y_t$ obtained from an autoregressive data-generating process (DGP)
    \begin{align}
        \label{eq:thedgp}
        y_t = \vz_t'\vpsi + x_t, \qquad x_t = \varrho x_{t-1} + u_t, \qquad t=1,\dots,T,
    \end{align}
    where $\vz_t = (1,t,\dots,t^q)$ gathers deterministic regressors with coefficients $\vpsi$. Leading cases are intercept-only models where $\vz_t = 1$ and models with a linear time trend, $\vz_t = (1, t)'$. The stochastic component $x_t$ is driven by the error process $u_t$, introducing a unit root into the observable variable $y_t$ if $\varrho = 1$ and is weakly stationary if $\lvert\varrho\rvert<1$. We make the following assumptions about $u_t$.
    \begin{restatable}[Linear process errors]{assum}{lperrors}
        \label{assum:lperrors}
        $u_t=\phi(L)\varepsilon_t= \varepsilon_t + \sum_{j=1}^\infty\phi_j\varepsilon_{t-j}$, where the lag polynomial satisfies $\phi(z)\neq0$ for all $\lvert z\rvert\leq1$ and $\sum_{j=1}^\infty j\lvert\phi_j\rvert<\infty$. The innovations $\varepsilon_t$ satisfy $\varepsilon_t\sim i.i.d.\, (0,\sigma^2)$, $\sigma^2<\infty$, and $\E(\varepsilon_t^4)<\infty$.
    \end{restatable}
    \Cref{assum:lperrors} states that $u_t$ has a linear process representation with \enquote{long-run} variance (LRV) $\omega^2 := \lim_{T\rightarrow\infty} \E[T^{-1}(\sum_{t=1}^T u_t)^2]<\infty$. This is common in the time series literature and covers various serially dependent processes, including stationary ARMA models. The lag polynomial $\phi(L)$ is invertible, meaning that $u_t$ has an AR representation with potentially indefinitely many lags. This ensures the existence of the ADF($\infty$) representation of $y_t$,
\begin{equation}
	\Delta y_t = \, d_t + \rho^\star y_{t-1} + \sum_{j=1}^\infty \delta_j^\star \Delta y_{t-j} + \varepsilon_t,
        \label{eq:adf_dgp}
\end{equation}
with degree-$q$ deterministic time polynomial $d_t := \vz_t'\vpsi$, $\rho^\star\in(-2,0]$ and $\vdelta^\star := (\delta^\star_1,\dots,\delta^\star_p)'$, where $\sum_{j=1}^p\delta_j^\star<1$. The stochastic component of $y_t$ is stationary if and only if \textit{all} solutions to the characteristic polynomial
    	\begin{align}
        	(1-z) - \rho^\star z - \sum_{j=1}^p (1-z)z^j\delta_j^\star = 0 \label{eq:charpolyadf}
	    \end{align}
    	are outside the unit circle, viz. $\rho^\star\in(-2,0)$. If $\rho^\star = 0$, then \eqref{eq:charpolyadf} has a unit root. Selecting a model that includes (omits) $y_{t-1}$ thus is equivalent to classifying $y_t$ as stationary (non-stationary). We thus refer to $y_{t-1}$ as the \textit{inference regressor}.
    	    
In this paper, we deal with the $\ell_1$-penalised estimation of ADF($p$) models
	\begin{align}
        \Delta y_t =& \, d_t + \rho_p y_{t-1} + \sum_{j=1}^p \delta_{p,\,j} \Delta y_{t-j} + \varepsilon_{p,\,t},
        \label{eq:adfreg}
	\end{align}     
which either incorporate or approximate the true model \eqref{eq:adf_dgp}, dependent on the choice of $p$ and the DGP's lag order. If the DGP is not sparse, the coefficients $\rho_p$ and $\vdelta_p$ are flawed with an approximation error, with $\varepsilon_{p,\,t}$ subject to a truncation error inducing autocorrelation. \textcite{ChangPark2002} establish that $p=o(T^{1/3})$ as $T\to\infty$ ensures consistent estimation with the approximation and truncation errors vanishing asymptotically.

 This setting is a generalisation of the sparse processes considered in \textcite{kock2016consistent} and aligns with the concept of \enquote{approximate sparsity} in the high-dimensional regression literature, cf. \textcite{Chernozhukov2013}. Most of the results presented in this paper hold under \Cref{assum:lperrors}. We indicate exceptions by invoking the following assumption, which ensures sparsity of the DGP \eqref{eq:thedgp} and hence an ADF representation with $p<\infty$.
 \begin{restatable}[Sparsity]{assum}{sparsity}
    \label{assum:sparsity}
        $u_t = \phi(L)\varepsilon_t$ such that $\phi(L)^{-1}$ exists and is a finite-order AR lag polynomial. The innovations $\varepsilon_t$ satisfy the conditions of \Cref{assum:lperrors}.
    \end{restatable}
            
The adaptive Lasso optimisation problem to \eqref{eq:adfreg} with $d_t=0$ and given the hyperparameter $\lambda\geq0$ governing the $\ell_1$ penalty obtains as
\begin{align}
		\Psi_T(\dot\rho,\dot\vdelta\vert\lambda) := \sum_{t=1}^T \left(\Delta y_t - \dot\rho y_{t-1} - \sum_{j=1}^p \dot\delta_j \Delta y_{t-j} \right)^2 +&\, 2\lambda\left( w_1^{\gamma_1} \lvert\dot\rho\rvert + \sum_{j=1}^p w_{2,\,j}^{\gamma_2} \left\lvert\dot\delta_j\right\rvert\right),
    \label{eq:adaptive lassooptim}
  \end{align}
with its solution denoted by  
\begin{align}
    \widehat{\vbeta}_\lambda := (\widehat{\rho}_\lambda, \widehat\vdelta_\lambda')' = \argmin_{\dot\rho,\, \dot\vdelta} \Psi_T(\dot\rho,\dot\vdelta\vert\lambda).\label{eq:ALestimator}
\end{align}
All adaptive weights $w_1 := \lvert\widehat{\rho}\rvert^{-1}$, $w_{2,\,j} := \lvert\widehat{\delta}_j\rvert^{-1}$ are determined by the initial (OLS) estimates $\widehat{\rho}$ and $\widehat{\delta}_j$ of \eqref{eq:adfreg} and the parameters $\gamma_1>0,\gamma_2>0$ are chosen a priori. The loss function \eqref{eq:adaptive lassooptim} is as in \textcite{kock2016consistent} except that we add a factor of 2 in front of the $\ell_1$-penalty term for mathematical convenience.

Assuming sparsity, \textcite{kock2016consistent} proves oracle efficiency of $\widehat\vbeta_\lambda$ for a suitable sequence of the tuning parameter $\lambda\in\lambda_{\mathcal{O}}$, with $\lambda_{\mathcal{O}}$ denoting a continuum of all sequences satisfying certain growth rate conditions. He further establishes for $\rho^\star\in(-2,0]$, i.e., irrespective of whether the model is stationary or not, that 
\begin{align*}
    \lim_{T\to\infty}\Prob{\widehat{\mathcal{M}}_{\lambda_\text{BIC}} = \mathcal{M}} = 1, \quad\text{with}\quad
    \lambda_\text{BIC} := \argmin_\lambda\ \log\left(\frac{\widehat{\vvarepsilon}_\lambda'\widehat{\vvarepsilon}_\lambda}{T}\right) + \left\lVert\widehat{\vbeta}_\lambda\right\rVert_0\frac{\log(T)}{T}
\end{align*}
with the non-zero entries of $\vbeta^\star := (\rho^\star,\vdelta^{\star'})'$ defining the true model $\mathcal{M}$ and $\widehat{\vvarepsilon}_\lambda = \Delta \vy_t - \widehat\rho_\lambda \vy_{t-1} - \sum_{j=1}^p \widehat\delta_{j,\,\lambda} \Delta \vy_{t-j}$.\footnote{$\lVert\vx\rVert_0 = \sum_{i=1}^p \mathbb{I}(\lvert x_i\rvert>0)$ denotes the Hamming distance of a $p$-vector $\vx$ and $\mathbb{I}$ the indicator function.} Hence, the BIC-tuned adaptive Lasso estimator $\widehat\vbeta_{\lambda_\textup{BIC}}$ is model selection consistent. In particular, $\widehat\vbeta_{\lambda_\textup{BIC}}$ consistently classifies stationary and non-stationary models. 
\textcite{kock2016consistent} cautions that consistent selection via the BIC does not imply the estimation error $\widehat{\vdelta}_{\lambda_\textup{BIC}} - \vdelta^\star$ to converge to zero, but rather to be bounded.\footnote{We thank Anders B. Kock for drawing our attention to this detail.}  Nevertheless, the oracle property establishes the existence of an estimator that is asymptotically equivalent to the unpenalised OLS estimator of the true model $\mathcal{M}$. It thus appears logical that $\lambda_\mathcal{O}$ asymptotically provides the global minimum of the BIC's objective function.

\section{An information-enriched weight}
\label{sec:alieirw}

Due to its oracle properties and the availability of fast algorithms for the optimisation problem underlying the computation like LARS (\cite{Efronetal2004}), the adaptive Lasso is an attractive tool for model selection in \eqref{eq:adfreg}. However, it requires initial estimators to elicit weights for all coefficients. While OLS weights perform reasonably in finite samples, a targeted weight specification may enhance the small-sample performance in discriminating stationary and non-stationary models.\footnote{We thank Anders B. Kock for kindly providing his supplementary simulation results to \textcite{kock2016consistent} upon which we base our Monte Carlo setup.}  
    
It is an open research question of how weights should be chosen in a potentially non-stationary model where it is often of prime interest to make a well-substantiated choice concerning inclusion or omission of $y_{t-1}$, especially for small $T$. In the following, we suggest an alternative weight for $\rho^\star$ that enhances the performance of the adaptive Lasso in distinguishing between stationary and non-stationary autoregressions.

\subsection{Simulation-based weight estimation}
\label{sec:sbwe}

\begin{figure}[t]
    \centering
    \caption{Simulated distribution of $J_\alpha$ and joint distributions of regressor weights}
    \vspace{.25cm}
    \label{fig:jvsrho}
    \resizebox{\textwidth}{!}{
    \includegraphics{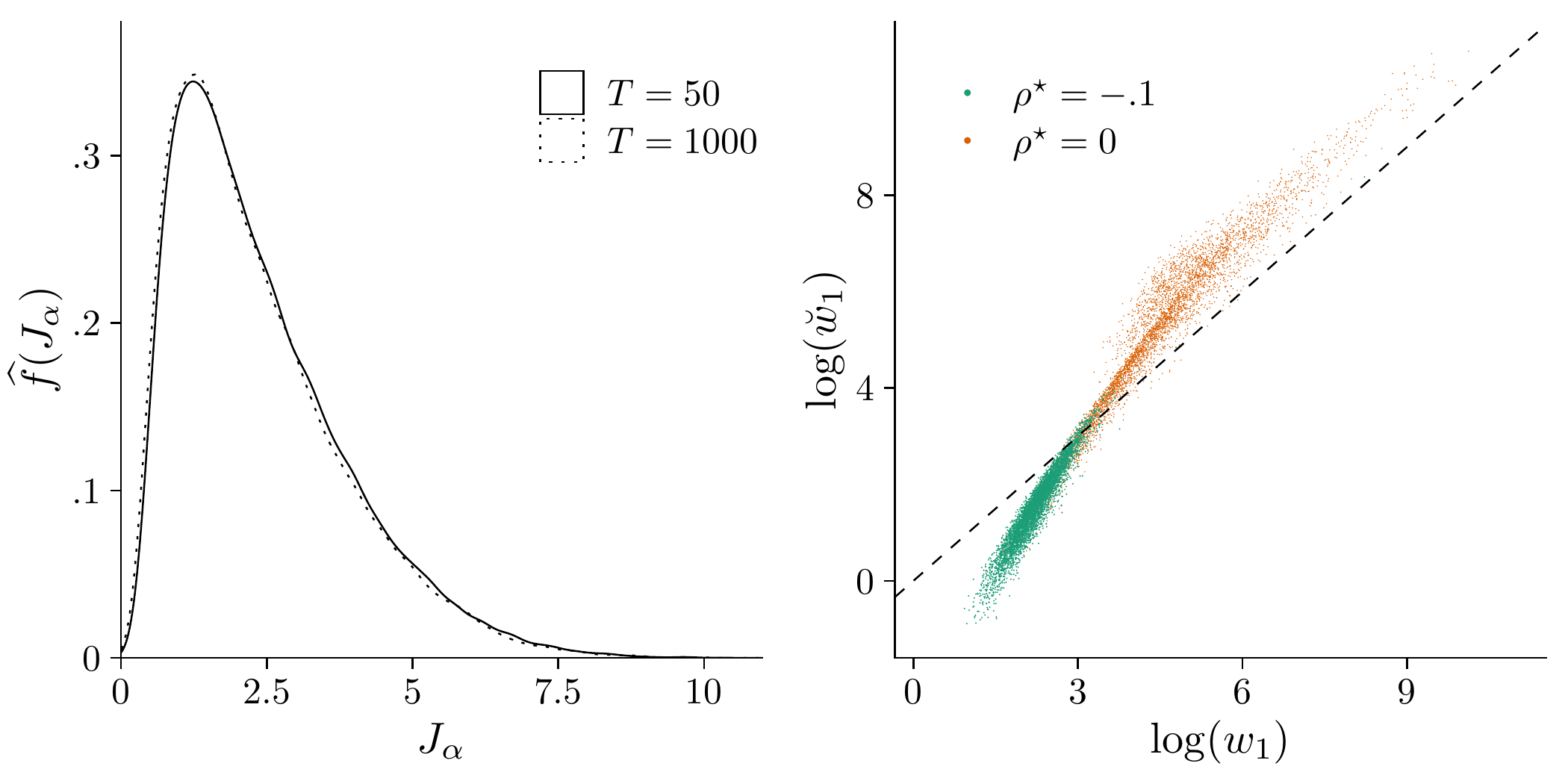}
    }
    \begin{minipage}{\textwidth}
        \vspace{.2cm}
        \scriptsize\textit{Notes:} Left: Gaussian Kernel density estimates of simulated null distributions of $J_\alpha$ with $\alpha = .1$ and $R = 150$, bandwidth selected using the \textcite{Silverman1986} rule. Right: Samples of OLS and modified inference regressor weights for $T=100$. The dashed bisector line marks the set of identical weights. 50000 replications based on Gaussian random walks.
    \end{minipage}
\end{figure}

To improve the performance of the adaptive Lasso in discriminating between stationary and non-stationary models, we modify the adaptive weight to exhibit more attractive properties than for $w_1 = \lvert\widehat\rho\rvert^{-1}$. Instead of using only the OLS coefficient estimate, we suggest enriching $w_1$ using a statistic that exploits distinct stochastic orders of stationary and non-stationary $y_t$ to obtain a more favourable adaptive penalty. To this end, we suggest simulation-based computation of a scaling factor to $\widehat{\rho}$ that behaves antithetically to the OLS estimator regarding the limits attained in stationary and non-stationary models. We use the quantile range statistic $J_\alpha$ by \textcite{HerwartzSiedenburg2010}, which is based on distinct convergence rates of OLS estimators in balanced and unbalanced time series regressions. 
\begin{restatable}[]{assum}{simrw}
    \label{assum:simrw}    
    $y_t$ has DGP \eqref{eq:thedgp} with $d_t=0$ and $u_t$ satisfies \Cref{assum:lperrors}. $q_t$ is generated by $q_t = q_{t-1} + \nu_t$, $\nu_t\sim i.i.d.\,N(0,\sigma_\nu^2)$, and $q_0=0$. $\{u_t\}_{t=1}^T$ and $\{\nu_t\}_{t=1}^T$ are independent.
\end{restatable}
\textcite{Phillips1986} shows that for $y_t$ and $q_t$ satisfying Assumption \ref{assum:simrw},
\begin{align}
    \widehat{\zeta} := \frac{\sum_{t=1}^T q_t y_t}{\sum_{t=1}^T q_t^2} \, \xrightarrow{d}\,\frac{\omega}{\sigma_\nu}\int_0^1 W_w(a)W_y(a)\textup{d}a \biggr/ \int_0^1 W_w(a)^2\textup{d}a,
    \label{eq:phillips1986}
\end{align}
where $\widehat\zeta$ is the OLS estimator in the regression $y_t = \zeta q_t + \epsilon_t$.\footnote{Implications of this result are known as \emph{spurious regression}.} $W_w(a)$ and $W_w(a)$ are independent standard Brownian motion. Hence, $\widehat\zeta$ is inconsistent for the true parameter $\zeta = 0$ but converges to a r.v. with a non-standard distribution. If $y_t$ is weakly-stationary, then $\widehat{\zeta}\xrightarrow[]{p}0$ where $\widehat\zeta= O_p(T^{-1})$. Given these characteristics, we suggest the modified adaptive weight $\Breve{w}_1$.
\begin{samepage}
\begin{restatable}[$\Breve{w}_1$]{algo}{wbrewe}
    \noindent
    \begin{enumerate}
        \item Scale $y_t$ using a consistent estimate of the long-run standard deviation $\widehat{\omega}$.
        \item Obtain Monte Carlo OLS estimates $\widehat\zeta^{(r)}$, $r=1,\dots,R$ in the regression 
            \begin{align}
                y_t = \zeta^{(r)} q_t^{(r)} + \epsilon_t^{(r)}, \ t=1,\dots,T, \label{eq:simres}    
            \end{align} with $q_t^{(r)}$ simulated as stated in \Cref{assum:simrw}. 
        \item Compute the inter-quantile range $J_\alpha := \left\lvert\widehat{\zeta}_{1-\alpha/2}-\widehat{\zeta}_{\alpha/2}\right\rvert$ for some $\alpha\in(0,1)$, where $\widehat{\zeta}_{\alpha}$ denotes the empirical $\alpha$ quantile of $\{\widehat\zeta_r\}_{r=1}^R$.\footnote{One could use some discrete variant of $\bar{J}:= \int_0^1 J_\alpha\mathrm{d}\alpha$. However, this is less favourable for our purpose than choosing a small value for $\alpha$, as we further discuss in Section \ref{sec:alphatuning}.}  
        \item Compute the modified adaptive weight
        \begin{align} 
            \Breve w_1:=w_1\cdot J_\alpha = \biggr\lvert\frac{\widehat{\rho}}{J_\alpha}\biggr\rvert^{-1}.
            \label{eq:mintweights}
        \end{align}
    \end{enumerate}\qed
\end{restatable}
\end{samepage}
We refer to $\Breve w_1$ as an information-enriched weight and the resulting Lasso estimator as \emph{adaptive Lasso with information enrichment} (ALIE). For intuition as to why \eqref{eq:mintweights} is useful, note that $J_\alpha= O_p(1)$ for $\rho^\star = 0$. While the asymptotic distribution of $J_\alpha$ has no closed form, simulation results indicate that it is well approximated for small $T$, cf. the left panel in Figure \ref{fig:jvsrho}. The distribution has much probability mass on the interval $(1, \infty)$ and so \eqref{eq:mintweights} has the tendency to upscale the penalty $w_1$ when $y_{t-1}$ is non-stationary. Under stationarity, $J_\alpha$ approaches a degenerate limit at zero at rate $T$, and so \eqref{eq:mintweights} may downscale $w_1$. \Cref{lem:crates} states the asymptotic behaviour of $\Breve{w}_1$.

\begin{restatable}[Stochastic order of $\Breve{w}_1$]{lem}{crates}
\label{lem:crates}
	Let $\Breve w_1 := \lvert\widehat{\rho}/J_\alpha\rvert^{-1}$.
	\begin{enumerate*}
		\item\label{item:lem:crates:H0} If $\rho^\star=0$, the modified weight $\Breve{w}_1$ diverges as $T\rightarrow\infty$ and $\Breve{w}_1^{-1} = O_p(T^{-1})$.
		\item\label{item:lem:crates:H1} If $\rho^\star\in(-2,0)$, we have $\Breve{w}_1= O_p(T^{-1})$.
	\end{enumerate*}
\end{restatable}

\begin{proof}
    See \Cref{sec:proofs}.
\end{proof}
Note that $\Breve w_1$ grows unlimited as $T\to\infty$ in a non-stationary model by part \ref{item:lem:crates:H0} of \cref{lem:crates}, just as $w_1$. However, small-sample properties of the weights differ as is illustrated in the right panel of \Cref{fig:jvsrho}, which shows samples of $\Breve w_1$ and $w_1$ in $\log$-$\log$ space for DGP \eqref{eq:thedgp} with $u_t\sim i.i.d.\,N(0,1)$ and $T=100$. We deduce that the joint distribution for $\rho^\star=0$ (orange dots) has the most probability mass above the dashed bisector line, i.e., $\Breve w_1 > w_1$ on average, implying a higher penalty from setting $\dot\rho\neq0$. 

Under stationarity, $w_1-\lvert\rho^\star\rvert^{-1}=O_p(T^{-1/2})$ whereas $\Breve{w}_1\xrightarrow[]{p}0$ at rate $O_p(T^{-1})$ by part \ref{item:lem:crates:H1} of \cref{lem:crates}. The effect is observed in the right panel of \Cref{fig:jvsrho} where samples (green dots) indicate that $\Breve{w}_1<w_1$ for $\rho^\star=-.1$, on average. The faster convergence rate and the zero limit of $\Breve w_1$ are expected to benefit the finite sample performance since a small penalty from setting $\dot\rho\neq0$ renders ALIE less likely to exclude $y_{t-1}$ from the model, misclassifying $y_t$ as non-stationary. We have several remarks.

\begin{rem}
     \label{rem:whyJ}
      Other enriching statistics could be used for $\Breve{w}_1$. The class of statistics discussed by \textcite{Stock1999} seems promising. Similarly to $J_\alpha$, these exploit different orders of probability of stationary and non-stationary $y_t$ and obtain by transforming $v_T(\nu) := (\widehat{\omega}^2T)^{-1/2}y_{\lfloor T\nu\rfloor}$, $0\leq\nu\leq1$ by a continuous functional $g(\cdot)$, with $\widehat{\omega}^2$ a LRV estimate.\footnote{Generally, $g: \mathcal{D}[0,1]\rightarrow\mathbb{R}$ with $\mathcal{D}[0,1]$ the Skorokhod space such that the continuous mapping theorem (CMT) can be applied.} For $\rho^\star = 0$ and under suitable conditions, $v_T \xrightarrow[]{d} W_y$ and so $g(v_T)\xrightarrow[]{d} g(W_y)$, while for $\rho^\star\in(-2,0)$, $v_T\xrightarrow[]{p}0$ and thus $g(v_T)\xrightarrow[]{p}g(0)$. Generally, one could work backwards from some asymptotic null distribution suitable for weighting $y_{t-1}$ to a corresponding sample analogue of $g(\cdot)$. However, this is outside the scope of this paper.  
\end{rem}

\begin{rem}
     \label{rem:Jtuning}
    $J_\alpha$ can be modified: $\alpha$ governs the estimated quantile range, thereby determining the distribution's scale and shape. Also, the scale and the location can be controlled via $\sigma_v$. Both parameters impact $\Breve w_1$ and hence the performance of ALIE. This flexibility is beneficial when accommodating a deterministic trend: for $\rho^\star=0$ the limit of $J_\alpha$ involves projections that affect $\Breve{w}_1$. We elaborate on this in \Cref{sec:hdt,sec:alphatuning}.
\end{rem}
We next present results on consistency and consistent tuning of ALIE.
\begin{restatable}[Oracle properties and consistent tuning]{prop}{plugandplim}
    \label{prop:plugandplim}
	Consider DGP \eqref{eq:thedgp} under \Cref{assum:lperrors}. The oracle properties of the adaptive Lasso in model \eqref{eq:adfreg} as presented in \textcite{kock2016consistent} are retained by ALIE for a sequence of tuning parameters $\lambda_\mathcal{O}$ satisfying conditions (i) $\lambda_\mathcal{O} \big/ \sqrt{T} \to 0$, (ii) $\lambda_\mathcal{O} / T^{1-\gamma_1} \to \infty$ and (iii) $\lambda_\mathcal{O} \big/ T^{(1 - \gamma_2)/2} \to \infty$, where $\gamma_1>1/2$, $\gamma_2>0$. Selecting $\lambda$ by BIC renders ALIE model selection consistent under \Cref{assum:sparsity}.
\end{restatable}
\begin{proof}
	See \Cref{sec:proofs}.
\end{proof}

\Cref{prop:plugandplim} establishes the asymptotic (oracle-)equivalence of ALIE and the adaptive Lasso with OLS-based weights of \textcite{kock2016consistent}, henceforth AL, for sparse processes.\footnote{Different lower bounds on $\gamma_1$ and $\gamma_2$ are due to distinct convergence rates of the initial estimates in a non-stationary model where $\widehat{\rho}= O_p(T^{-1})$ and $\widehat{\delta}_j-\delta_j^\star = O_p(T^{-1/2})$.} This implies consistency of ALIE for sequences $\lambda \in \lambda_\mathcal{O}$ satisfying the \textit{same} asymptotic conditions as required for AL: Given $\lambda \in \lambda_\mathcal{O}$, ALIE discriminates consistently between the sets of relevant and irrelevant regressors and the estimated coefficients are asymptotically equivalent to OLS estimates in the true model. Furthermore, ALIE can be tuned using BIC to achieve consistent model selection, giving practical guidance on choosing $\lambda$ in applications.\footnote{The Lasso literature also considers plug-in estimators for $\lambda_\mathcal{O}$, but to our knowledge, no method performs as well as the BIC in our application.}

\subsection{Asymptotics of the adaptive Lasso estimators}
\label{sec:aotale}

In the following, we present several asymptotic results for ALIE regarding consistent classification of $y_{t-1}$, conditional on the regularisation parameter $\lambda$. For ease of exposition and comparability with the results for AL in \textcite{kock2016consistent}, we consider an ADF(0) model and set $\gamma_1 = 1$.

\begin{restatable}[]{thm}{aronelpnull}
    \label{thm:aronelpnull}
	Let $\Delta y_t = \rho^\star y_{t-1} + u_t$ with $\rho^\star=0$ and $u_t$ satisfies \Cref{assum:lperrors}. Denote $\widehat{\rho}_\lambda$ the minimiser of $\Psi(\dot\rho)=\sum_{t=1}^T(\Delta y_t - \dot\rho y_{t-1})^2 + 2\lambda\lvert\dot\rho\rvert\breve{w}$ with $\breve{w}=\lvert \widehat{\rho}/J_\alpha\rvert^{-1}$, for some $\alpha\in(0,1)$ and denote $\widehat{\rho}$ the OLS estimator of $\rho^\star$.
\begin{enumerate}
	\item If $\lambda \rightarrow 0$, then $\lim_{T\rightarrow\infty}\Prob{\widehat\rho_\lambda=0} = 0$.\label{item:thm:aronelpnull:P1}
	\item If $\lambda \rightarrow c\in(0,\infty)$, then $\lim_{T\rightarrow\infty}\Prob{\widehat\rho_\lambda=0} = F_{\Breve{\mathcal{H}}}(c)>0$ with $F_{\Breve{\mathcal{H}}}$ the CDF of the r.v. $\Breve{\mathcal{H}}$ as defined in the proof of this theorem.\label{item:thm:aronelpnull:P2}
	\item If $\lambda \rightarrow \infty$, then $\lim_{T\rightarrow\infty}\Prob{\widehat\rho_\lambda=0} = 1$.\label{item:thm:aronelpnull:P3}
\end{enumerate}
\end{restatable}
\begin{proof}
	See \Cref{sec:proofs}.
\end{proof}
\Cref{thm:aronelpnull} is along the lines of Theorem 3 in \textcite{kock2016consistent} and considers properties in non-stationary models for $T\to\infty$. Part \ref{item:thm:aronelpnull:P1} states that $\lambda\to0$ yields an inconsistent classifier. By part \ref{item:thm:aronelpnull:P3}, $\lambda\to\infty$ is required for ALIE to correctly classify a non-stationary model in the limit. Both results also apply to AL. Part \ref{item:thm:aronelpnull:P2} states that if $\lambda$ converges to some constant $c>0$, then $\widehat{\rho}_\lambda = 0$ with non-zero probability $F_{\Breve{\mathcal{H}}}(c)$.

We next discuss the asymptotic behaviour of ALIE in a stationary model conditional on $\lambda$.

\begin{restatable}[]{thm}{aronelpalt} 
\label{thm:aronelpalt}
Let $\Delta y_t = \rho^\star y_{t-1} + u_t$ with $\rho^\star\in(-2,0)$ under \Cref{assum:lperrors}. Denote $\widehat{\rho}_\lambda$ the minimiser of $\Psi(\dot\rho)=\sum_{t=1}^T(\Delta y_t - \dot\rho y_{t-1})^2 + 2\lambda\lvert\dot\rho\rvert\breve{w}_1$ with $\lambda=\Theta(T^{1+\kappa})$, $\kappa\in\mathbb{R}$, $\breve{w}_1=\lvert \widehat{\rho}/J_\alpha\rvert^{-1}$ with $\alpha\in(0,1)$ and $\widehat{\rho}$ the OLS estimator of $\rho^\star$ with $\widehat{\rho}\xrightarrow[]{p}\rho^{\star\star}$.
	\begin{enumerate}
		\item\label{item:thm:aronelpalt:kappa<1} If $\lambda=\Theta(T^{1+\kappa})$ with $\kappa<1$, then $\lim_{T\rightarrow\infty}\Prob{\widehat{\rho}_\lambda=0} = 0$.
		\item\label{item:thm:aronelpalt:kappa=1} For $\lambda=\Theta(T^{1+\kappa})$ with $\kappa=1$ so that $J\lambda/T\xrightarrow{d}\mathcal{C}$, denote by $c$ a realisation of the r.v. $\mathcal{C}$.
		\begin{enumerate}
			\item\label{item:thm:aronelpalt:kappa=1lim0} If $c\in\left[0,\,{\rho^{\star\star}}^2\E\left(y_{t-1}^2\right)\right)$, then $\lim_{T\rightarrow\infty}\Prob{\widehat{\rho}_\lambda = 0\vert\mathcal{C}=c} = 0$. 
			\item\label{item:thm:aronelpalt:kappa=1limp} If $c = {\rho^{\star\star}}^2\E\left(y_{t-1}^2\right)$, then $\lim_{T\rightarrow\infty}\Prob{\widehat{\rho}_\lambda=0\vert\mathcal{C}=c} = F(0)$ with $F$ the CDF of the sum of two zero-mean Gaussians as defined in the proof of this theorem.
			\item\label{item:thm:aronelpalt:kappa=1lim1} If $c\in\left({\rho^{\star\star}}^2\E\left(y_{t-1}^2\right),\,\infty\right)$, then $\lim_{T\rightarrow\infty}\Prob{\widehat{\rho}_\lambda = 0\vert\mathcal{C}=c} = 1$. 
		\end{enumerate}
		\item\label{item:thm:aronelpalt:kappa>1} If $\lambda=\Theta(T^{1+\kappa})$ with $\kappa>1$, then $\lim_{T\rightarrow\infty}\Prob{\widehat{\rho}_\lambda=0}= 1$.
	\end{enumerate}
\end{restatable}
\begin{proof}
	See \Cref{sec:proofs}.
\end{proof}

\Cref{thm:aronelpalt} is our main result and describes the asymptotic behaviour of $\Prob{\widehat{\rho}_\lambda=0}$ for ALIE if $\rho^\star\in(-2,0)$, conditional on $\lambda$. Part \ref{item:thm:aronelpalt:kappa<1} states that $\lambda=\Theta(T^{1+\kappa})$ with $\kappa<1$ is required for $\Prob{\widehat{\rho}_\lambda = 0} = 0$ asymptotically, implying 100\% power. By part \ref{item:thm:aronelpalt:kappa>1}, ALIE will incorrectly classify the model as non-stationary with probability one as $T\to\infty$ if $\kappa>1$. Part \ref{item:thm:aronelpalt:kappa=1} considers $\lambda=\Theta(T^2)$, the knife edge between parts \ref{item:thm:aronelpalt:kappa<1} and \ref{item:thm:aronelpalt:kappa>1}. The classification result then is random. This finding generalises an analogous result in \textcite{kock2016consistent}, who imposes an additional assumption on the mode of convergence of a specific quantity. We prove in \cref{lem:rhostarstar,lem:knifeCLT} in \Cref{sec:proofs} that this assumption is not necessary, resulting in a different distribution of the activation probabilities. 

We conclude that ALIE performs consistent classification for $\rho^\star\in(-2,0]$ if $\lambda=\Theta(T^{1+\kappa})$, $\kappa<1$, satisfying part \ref{item:thm:aronelpnull:P3} of \Cref{thm:aronelpnull} ($\lambda\to\infty$ required if $\rho^\star = 0$) and part \ref{item:thm:aronelpalt:kappa<1} of \Cref{thm:aronelpalt} ($\kappa<1$ required if $\rho^\star\in(-2,0))$. 
By Theorem 4 in \textcite{kock2016consistent}, AL only has power for $\lambda = O(T)$. \Cref{thm:aronelpalt} relaxes this constraint to $\lambda = O(T^2)$ for ALIE, which is a direct consequence of the properties of $\Breve{w}_1$ stated in part \ref{item:lem:crates:H1} of \cref{lem:crates}. 

While the $\lambda$-conditions of \Cref{thm:aronelpalt} are weaker than $\lambda_\mathcal{O} = o(\sqrt{T})$, as required for the oracle property, cf. \Cref{prop:plugandplim}, they suffice for ALIE to consistently select stationary models. If the goal is distinguishing stationary and non-stationary data by classification, the lack of oracle efficiency is negligible. Importantly, \Cref{thm:aronelpalt} foreshadows a power advantage of ALIE over AL in stationary ADF($p$) models under consistent tuning of $\lambda$, and thus also for oracle-efficient tuning using BIC. We will discuss this in the next section.

\subsection{Solution path properties}
\label{sec:ctuning}
The asymptotic analysis for the ADF(0) model presented in the previous section builds on the activation probability of $y_{t-1}$, which is analysed based on the first-order condition (FOC) for $\dot\rho_\lambda = 0$. We next transfer the results of the previous section to general ADF models in terms of the penalty parameter $\lambda$ by examining the Lasso solution path. For this, we use the activation threshold $\lambda_0$.

\begin{restatable}[Activation knots]{defn}{lambda0_definition}
    \label{def:lambda0_definition}
    Be $x_i\in(y_{t-1}, \Delta y_{t-1}, \dots, \Delta y_{t-p})'$, $\dot\vbeta := (\dot\rho,\,\dot\vdelta')'$ and consider the coefficient functions $\dot{\beta}_i(\lambda) \in \mathbb{R}$, $i=1,\dots,p+1$ along the adaptive Lasso solution path to \eqref{eq:adaptive lassooptim}. Define $\lambda_{0,\,\beta_i^\star}$ the earliest activation threshold of $x_i$, i.e.,
	\begin{align*}
        \lambda_{0,\, \beta_i^\star} :=& \max \Lambda_{0,\, \beta_i^\star}, \\
	\Lambda_{0,\, \beta_i^\star} :=& \left\{\lambda : \ \ \partial_{+} \dot\beta_i(\lambda) = 0 \quad \land \quad \partial_{-} \dot\beta_i(\lambda) \neq 0 \quad \land \quad \dot\beta_i(\lambda) = 0\right\},
	\end{align*}
    where $\partial_{+}\dot\beta_i(\lambda)$ and $\partial_{-}\dot\beta_i(\lambda)$ denote derivatives of $\dot\beta_i(\lambda)$ with respect to $\lambda$ from above and from below, respectively. 
        \qed
	\end{restatable}

For every $\lambda \in \Lambda_{0,\, \beta_i^\star}$, the FOC for $\dot{\beta}_i(\lambda) = 0$ is just binding, i.e., it holds with weak inequality whilst the coefficient $\dot{\beta}_i(\lambda)$ remains set to zero. Following the interpretation in \textcite{Efronetal2004} closely, we declare regressor $x_i$ \textit{active} as soon as the FOC is binding, so that $x_i \in \widehat{\mathcal{M}}_{\lambda_{0,\,\beta_i^\star}}$ even though $\dot{\beta}_{i}(\lambda_{0,\,\beta_i^\star}) = 0$. Note that the FOC for $\dot{\beta}_i(\lambda) = 0$ can also be just binding with a zero coefficient if the regressor is being removed from the active set or is on the verge of a sign reversal.\footnote{A projected sign reversal is the sole reason for a variable to be deactivated by the LARS algorithm with the Lasso modification, see Section 3.1 in \textcite{Efronetal2004}.} By imposing the right-hand derivative to be zero while the left-hand derivative is non-zero, \Cref{def:lambda0_definition} only considers knots on the solution path where variable $i$ is \textit{added} to the active set. Note that the solution path for the ADF(0) model in the previous section features a single \emph{activation} knot $\Lambda_{0,\,\rho^\star} = \lambda_{0,\,\rho^\star}$ since $y_{t-1}$ is the only regressor.

In least angle regression (LAR), the $\Lambda_{0,\, \beta_i^\star}$ are singletons because active variables cannot be deactivated. However, Lasso and its variants allow for repeated activation and deactivation of variables along the solution path.\footnote{See \textcite{Efronetal2004} for details on the LARS algorithm in Lasso mode.} Still, the earliest activation knot of variable $i$, $\lambda_{0,\, \beta_i^\star}$, is the most informative because it is mainly driven by the shrinkage applied to the associated coefficient. \Cref{prop:lambdanaught} describes the asymptotic behaviour of the activation thresholds for the relevant and the irrelevant variables in stationary and non-stationary ADF models. 
\begin{restatable}[]{prop}{lambdanaught}
    \label{prop:lambdanaught}
	Consider model \eqref{eq:adfreg} for $y_t$ generated by \eqref{eq:thedgp} under \Cref{assum:lperrors}. Denote $\lambda_{0,\,\rho^\star}$, $\Breve\lambda_{0,\,\rho^\star}$ and $\lambda_{0,\,\delta_j^\star}$ activation thresholds as defined in \Cref{def:lambda0_definition} for $w_1$, $\Breve{w}_1$, and $w_{2,\,j}$, respectively.
    \begin{enumerate*}
        \item\label{item:prop:lambdanaught:rho=0} If $\rho^\star=0$, then $\lambda_{0,\,\rho^\star}= O_p(T^{1-\gamma_1})$ and $\Breve\lambda_{0,\, \rho^\star}= O_p(T^{1-\gamma_1})$.
        \item\label{item:prop:lambdanaught:rho<0} If $\rho^\star\in(-2,0)$, then $\lambda_{0,\,\rho^\star}=\Theta(T)$ and $\Breve\lambda_{0,\,\rho^\star}=\Theta(T^{1+\gamma_1})$.
        \item\label{item:prop:lambdanaught:delta=0} If $\delta_j^\star=0$, then $\lambda_{0,\,\delta_j^\star} = O_p\left(T^{\frac{1-\gamma_2}{2}}\right)$. If $\delta_j^\star\neq0$, then $\lambda_{0,\,\delta_j^\star} =\Theta(T)$.
    \end{enumerate*}
\end{restatable}
\begin{proof}
    See \Cref{sec:proofs}.
\end{proof}
Part \ref{item:prop:lambdanaught:rho=0} of \Cref{prop:lambdanaught} corroborates an asymptotically equivalent effect of $w_1$ and $\Breve w_1$ when $\rho^\star=0$, cf. \Cref{thm:aronelpnull}. Denote by $\lambda_{0,\,\rho^\star\in(-2,0)}$ the activation threshold for a relevant stationary $y_{t-1}$ for brevity. Part \ref{item:prop:lambdanaught:rho<0} implies that $$\lambda_{0,\,\rho^\star\in(-2,0)} = o\left(\Breve\lambda_{0,\,\rho^\star\in(-2,0)}\right)\quad\forall\quad\gamma_1>0.$$ This is one main feature of information enrichment: $\Breve w_1$ implies $\Breve\lambda_{0,\,\rho^\star\in(-2,0)} =\Theta(T^{1+\gamma_1})$, whereas AL's $w_1$ yields $\lambda_{0,\,\rho^\star\in(-2,0)} =\Theta(T)$ -- just as for the $\lambda_{0,\,\delta_j^\star\neq0}$ of the \emph{relevant} $\Delta y_{t-j}$ with $\sqrt{T}$-consistent weights $w_{2,\,j}$. Part \ref{item:prop:lambdanaught:delta=0} asserts the stochastic orders of the $\lambda_{0,\,\delta_j^\star}$ associated with the relevant and irrelevant $\Delta y_{t-j}$. These orders are the same for ALIE and AL, irrespective of $\rho^\star$.

We conclude from \Cref{prop:lambdanaught} that distinct orders of $w_1$ and $\Breve{w}_1$ influence the selection ability of the adaptive Lasso for $\rho^\star \in (-2,0)$. However, asymptotic performance differences cannot be explained by growth rate considerations for $\rho^\star=0$, where benefits of information enrichment are different, as discussed in the next section. We focus on $\rho^\star \in (-2,0)$ in the remainder.

As for the ADF(0) model considered in \Cref{thm:aronelpalt}, the asymptotic shrinkage applied to $\widehat{\rho}_\lambda$ in an ADF($p$) model depends on the growth rate of the $\lambda$-sequence. ALIE has the virtue that below $\Breve{\lambda}_{0,\, \rho^\star \in (-2,0)}$ the shrinkage on $\widehat{\rho}_\lambda$ vanishes asymptotically, as we assert in \Cref{lem:ZeroShrinkage}.

\begin{restatable}[Zero asymptotic shrinkage]{lem}{ZeroShrinkage}
\label{lem:ZeroShrinkage} 
Denote $\widehat{\varepsilon}_{t,\,p,\,\lambda}$ the period-$t$ residual associated with the ALIE estimators $\widehat{\rho}_{\lambda}$ and $\widehat{\delta}_{j,\,\lambda}$ under the conditions of \Cref{prop:lambdanaught} and consider $\rho^\star \in (-2,0)$. For any sequence $\lambda = o\left(\Breve{\lambda}_{0,\, \rho^\star \in (-2,0)}\right)$, the FOC for $\widehat{\rho}_\lambda = 0$ is asymptotically always binding with 
\begin{equation*}
    \lim_{T \to \infty} \frac{1}{T} \sum_{t=1}^{T} y_{t-1} \widehat{\varepsilon}_{t,\,p,\,\lambda} = 0, 
\end{equation*}
amounting to a vanishing shrinkage applied to $\widehat{\rho}_{\lambda}$ as $T\to\infty$.
\end{restatable}
\begin{proof}
    See \Cref{sec:proofs}.
\end{proof}
Importantly, zero asymptotic shrinkage is not synonymous with unbiasedness or consistency because $\widehat{\rho}_\lambda$ may be subject to omitted variable bias or an approximation error. Unbiasedness is only guaranteed at $\lambda_\mathcal{O}$ where the shrinkage on the coefficients of the relevant variables vanishes for $T\to\infty$, as required for the oracle property. \Cref{lem:ZeroShrinkage} is useful nonetheless: it transfers the findings for the stationary ADF(0) model in \Cref{thm:aronelpalt} to ADF($p$) models. \Cref{lem:ZeroShrinkage} further implies an asymptotic ordering in every possible Lasso solution path across all permissible stationary linear models, as the subsequent corollary details.
\begin{restatable}[Knot asymptotics]{cor}{KnotOneAsymptotics}
\label{cor:knot1asymptotics}
Be $\lambda^{(m)}$ the $m^{th}$ knot on ALIE's solution path characterised by the knots $\lambda^{(1)} > \lambda^{(2)} > ... > 0$ under the conditions of \Cref{lem:ZeroShrinkage} and $\gamma_1,\gamma_2>0$. If $\rho^\star \in (-2,0)$, then
\begin{enumerate*}
    \item\label{item:cor:knot1asymptotics:P1} $\lambda^{(1)} \xrightarrow[]{p} \Breve\lambda_{0,\, \rho^\star}$ and
    \item\label{item:cor:knot1asymptotics:P2} $\widehat{\rho}_{\lambda^{(2)}} \xrightarrow[]{p}  \rho^{\star\star} \in (-2,0)$.
\end{enumerate*}
\end{restatable}
\begin{proof}
    See \Cref{sec:proofs}.
\end{proof}
Part \ref{item:cor:knot1asymptotics:P1} of \Cref{cor:knot1asymptotics} states that ALIE activates $y_{t-1}$ first w.p.~1 in any stationary linear model as $T\to\infty$. This feature is unique to ALIE and has not arisen in any other Lasso variant, to our knowledge. Part \ref{item:cor:knot1asymptotics:P2} asserts that $\widehat\rho_\lambda$ escapes penalisation as some $\Delta y_{t-j}$ enters the active set (with zero coefficient) at $\lambda^{(2)}$, rendering $\widehat\rho_{\lambda^{(2)}}$ equivalent to the unpenalised OLS estimator of $\rho$ in the (possibly underspecified) model $\Delta y_t = \rho y_{t-1} + u_t$, with the non-zero probability limit $\rho^{\star\star}$. Remarkably, although $\rho^{\star \star}$ does not necessarily equal $\rho^\star$, detection of stationarity is asymptotically guaranteed at $\lambda^{(2)}$.\footnote{\Cref{cor:knot1asymptotics} hence motivates another principle to test for a unit root: checking whether $y_{t-1}$ is activated first. This test is consistent but has low power, as simulation evidence suggests.} The results of \cref{cor:knot1asymptotics} are not granted to hold for AL since $\lambda_{0,\,\rho^\star}$ has the same stochastic order as the $\lambda_{0,\,\delta^\star_j}$ of the relevant $\Delta y_{t-j}$. Therefore, the asymptotic activation knot of $y_{t-1}$ is subject to the DGP and thus unknown.

Combining \Cref{prop:plugandplim}, \Cref{lem:ZeroShrinkage} and \Cref{cor:knot1asymptotics}, we obtain a further result on ALIE's solution path.
\begin{restatable}[Perpetual activation]{thm}{PermanentActivation}
\label{thm:permactive}
Under the conditions of \Cref{lem:ZeroShrinkage} and for any sequence $\lambda \in \left[0, \lambda^{(1)}\right]$,
\begin{equation}
    \lim_{T \to \infty} \Prob{y_{t-1} \in \widehat{\mathcal{M}}_\lambda} = 1.
\end{equation}
\end{restatable}
\begin{proof}
    See \Cref{sec:proofs}.
\end{proof}
\Cref{thm:permactive} states that once $y_{t-1}$ has been activated moving down the solution path, the probability to encounter a deactivation knot converges asymptotically to zero. This renders ALIE robust to restrictive tuning, a common finite-sample feature of consistent tuning criteria that is also observed for the BIC, cf. \textcite{Efronetal2004}.

\begin{rem} 
    \label{rem:LARsimilarity}
    The property that $y_{t-1}$ is asymptotically never removed from the active set once activated is reminiscent of the LAR solution path to \eqref{eq:adaptive lassooptim}. Still, the solution paths of ALIE and LAR are not necessarily equal since \Cref{thm:permactive} only applies to $y_{t-1}$ in the case of ALIE but to \emph{every} regressor in LAR, since there are no deactivations. Moreover, coefficient sign changes do not generate knots on the LAR solution path, as is the case with the Lasso solution path. For ALIE, \Cref{thm:permactive} requires any possible deactivation knot of $y_{t-1}$ to be simultaneously an activation knot of $y_{t-1}$, turning it into the functional equivalent of a sign change.
    \end{rem}
\begin{rem}
    \label{rem:countzerorho}
    Due to the piece-wise linearity of the solution path $\dot{\rho}_\lambda$ in $\lambda$, sign changes can occur only once between two neighbouring knots that do not affect the activation status of $y_{t-1}$. Consequently, for an asymptotic solution path with $m$ knots $\lambda^{(m)} < \lambda^{(m-1)} < \cdots < \lambda^{(2)}$ unrelated to $y_{t-1}$, there can only be a maximum of $m-1 \geq p$ instances where $\dot{\rho}_\lambda = 0$.\footnote{Referring to an asymptotic solution path, we do not suggest a certain ordering of knots, but rather asymptotics to have kicked in.} However, these spots constitute a countable set of singletons and thus are negligible with respect to the Lebesgue measure.
\end{rem}

\subsection{Properties of ALIE under tuning}
In applications, $\lambda$ is often selected by data-driven methods, e.g., cross-validation or information criteria, usually targeting some $\lambda = O(\lambda_\mathcal{O})$. Even though the asymptotic properties of these tuning methods have been established under various conditions, their stochastic behaviour in small samples needs to be better understood. \textcite{Efronetal2004} remark that consistent tuning criteria like the BIC tend to choose too large values for $\lambda$ in small samples, thereby neglecting relevant variables. To emulate this behaviour, we assume the growth rate of a tuned parameter sequence $\lambda_\textup{IC}$ to vary stochastically in finite samples.
For comparison of $\lambda$-sequences with varying growth rates, we apply the $\log_T$-transformation. Occasions of restrictive tuning as addressed above translate to $\log_T(\lambda_\textup{IC}) > \log_T(\lambda_\mathcal{O})$.  Note that in $\log_T$-space, all $\lambda$-sequences are asymptotically bounded and can be considered to be drawn from a continuous distribution function $F_{T,\,\lambda_\textup{IC}}(\log_T \lambda \vert \vvarepsilon_T, \vbeta^\star)$, shorthand $F_{T,\,\lambda_\textup{IC}}$. We account for the dependence of $F_{T,\,\lambda_\textup{IC}}$ on the DGP by conditioning on the innovations $\vvarepsilon_T$ and the vector of true coefficients $\vbeta^\star$. The distribution function $F_{T,\,\lambda_\textup{IC}}$ has support on $\mathbb{R}^+$ and converges to its limit $F_{\lambda_\textup{IC}}$, inheriting the asymptotic properties of the tuning criterion applied.\footnote{E.g., $F_{\lambda_\textup{BIC}}$ has all probability mass on $\lambda\in\lambda_\mathcal{O}$ asymptotically while $F_{\lambda_\textup{AIC}}$ distributes some probability mass below $\lambda_\mathcal{O}$, see \textcite{kock2016consistent}.} This convergence also acknowledges the fact that $\lambda_\mathcal{O}$ is only guaranteed to exist as $T \to \infty$. Under consistent tuning using the BIC, deviations of $\lambda_\textup{BIC}$ from $\lambda_\mathcal{O}$ can happen as $\log_T \lambda_\textup{BIC} \to \log_T \lambda_\mathcal{O}$. This is reminiscent of the local alternatives regarding coefficients widely used in the unit root literature. 

The activation probability of $y_{t-1}$ for Lasso-type estimators is bounded by $F_{T,\,\lambda_\textup{IC}}$,
\begin{align} 
    \Prob{\widehat{\rho}_{\lambda_\text{IC}} \neq 0} 
    =&\, \Prob{\widehat{\rho}_{\lambda_\text{IC}} \neq 0 \vert \lambda_\text{IC} < \widetilde{\lambda}_{0,\,\rho^\star}} \cdot \Prob{\lambda_\text{IC} < \widetilde{\lambda}_{0,\,\rho^\star}}\notag\\
    =&\, \Prob{\widehat{\rho}_{\lambda_\text{IC}} \neq 0 \vert \lambda_\text{IC} < \widetilde{\lambda}_{0,\,\rho^\star}} \cdot F_{T,\,\lambda_\text{IC}}\left(\log_T \widetilde{\lambda}_{0,\,\rho^\star}\right), \label{eq:rho_activ_prob} 
\end{align}
with generic first activation threshold $\widetilde{\lambda}_{0,\, \rho^\star}$. \Cref{lem:ZeroShrinkage}, \Cref{cor:knot1asymptotics} and  \Cref{thm:permactive} point to an advantage for ALIE over AL concerning the conditional activation probability $\Prob{\widehat{\rho}_{\lambda_\text{IC}} \neq 0 \vert \lambda_\text{IC} < \widetilde{\lambda}_{0,\,\rho^\star}}$. The same applies for $F_{T,\,\lambda_\text{IC}}\left(\log_T \widetilde{\lambda}_{0,\,\rho^\star}\right)$, which is ultimately driven by the different growth rates of $\lambda_{0,\,\rho^\star}$ and $\Breve\lambda_{0,\,\rho^\star}$ (cf. \cref{prop:lambdanaught}). 
Maximising  $\Prob{\widehat{\rho}_{\lambda_\text{IC}} \neq 0}$ requires maximising $F_{T, \lambda_\text{IC}}\left(\log_T \widetilde{\lambda}_{0,\,\rho^\star} \right)$ which can either be achieved using another tuning method or by lifting $\widetilde{\lambda}_{0,\,\rho^\star}$. However, a different tuning method may deteriorate classification performance for $\rho^\star=0$. Information enrichment via $\Breve{w}_1$ does not share this disadvantage as it manipulates $\widetilde{\lambda}_{0,\,\rho^\star}$, \textit{conditional on} $\rho^\star$. Since the CDF $F_{T,\,\lambda_\text{IC}}$ is monotonically increasing in $\lambda$, raising the growth rate of $\Breve\lambda_{0,\,\rho^\star \in (-2,0)}$ adds robustness against the finite sample characteristics of $\lambda_\textup{IC}$, a prominent feature in our simulation studies.

\begin{figure}[htbp]
    \centering
    \caption{Comparison of $\lambda_0$-distributions in ADF models}
    \label{fig:lambda0dist}
    \vspace{.25cm}
        
    \begin{minipage}{\textwidth}
    \resizebox{\textwidth}{!}{
    \includegraphics{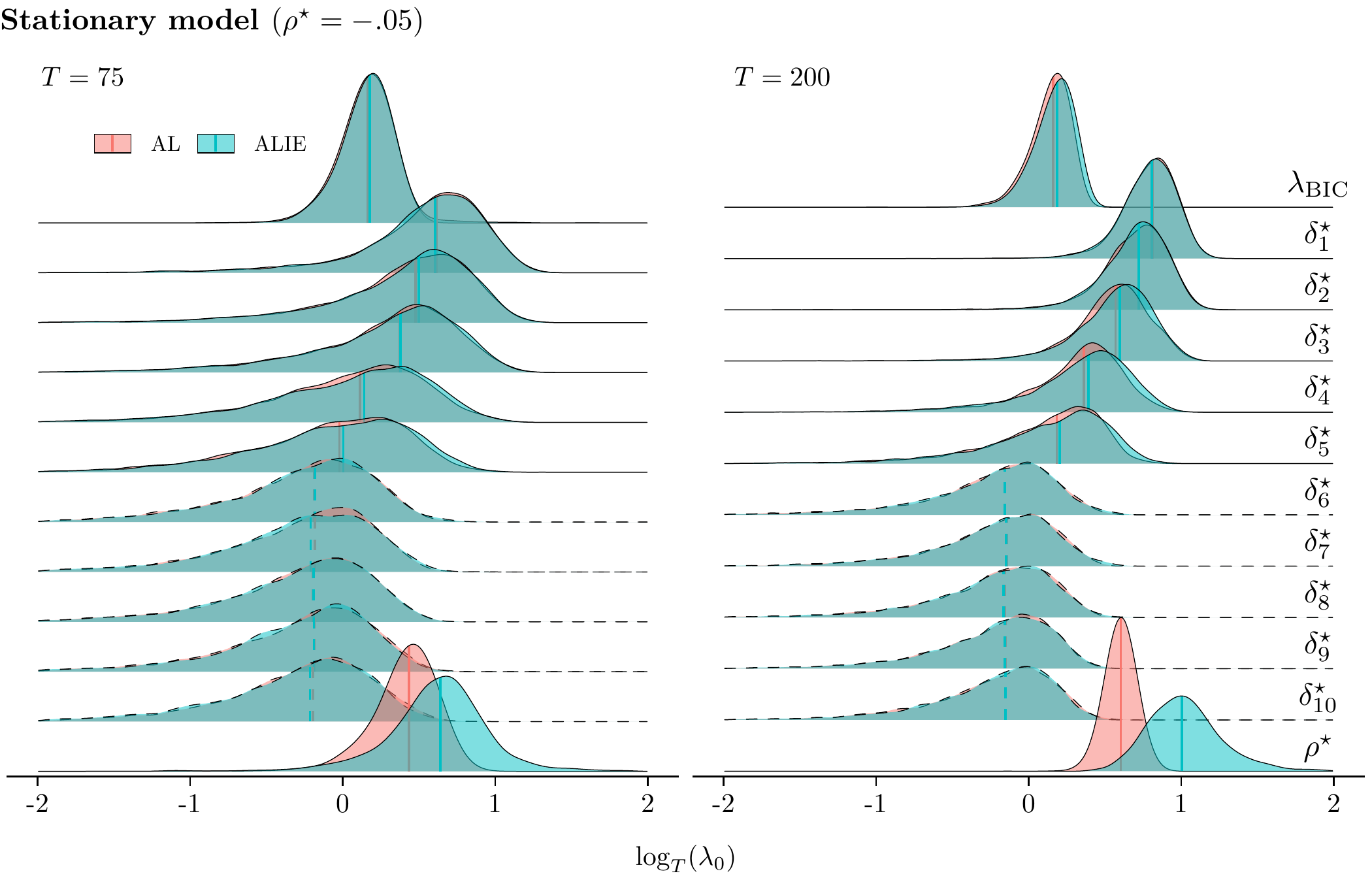}
    }
    \end{minipage}
    \begin{minipage}{\textwidth}
    \resizebox{\textwidth}{!}{
    \includegraphics{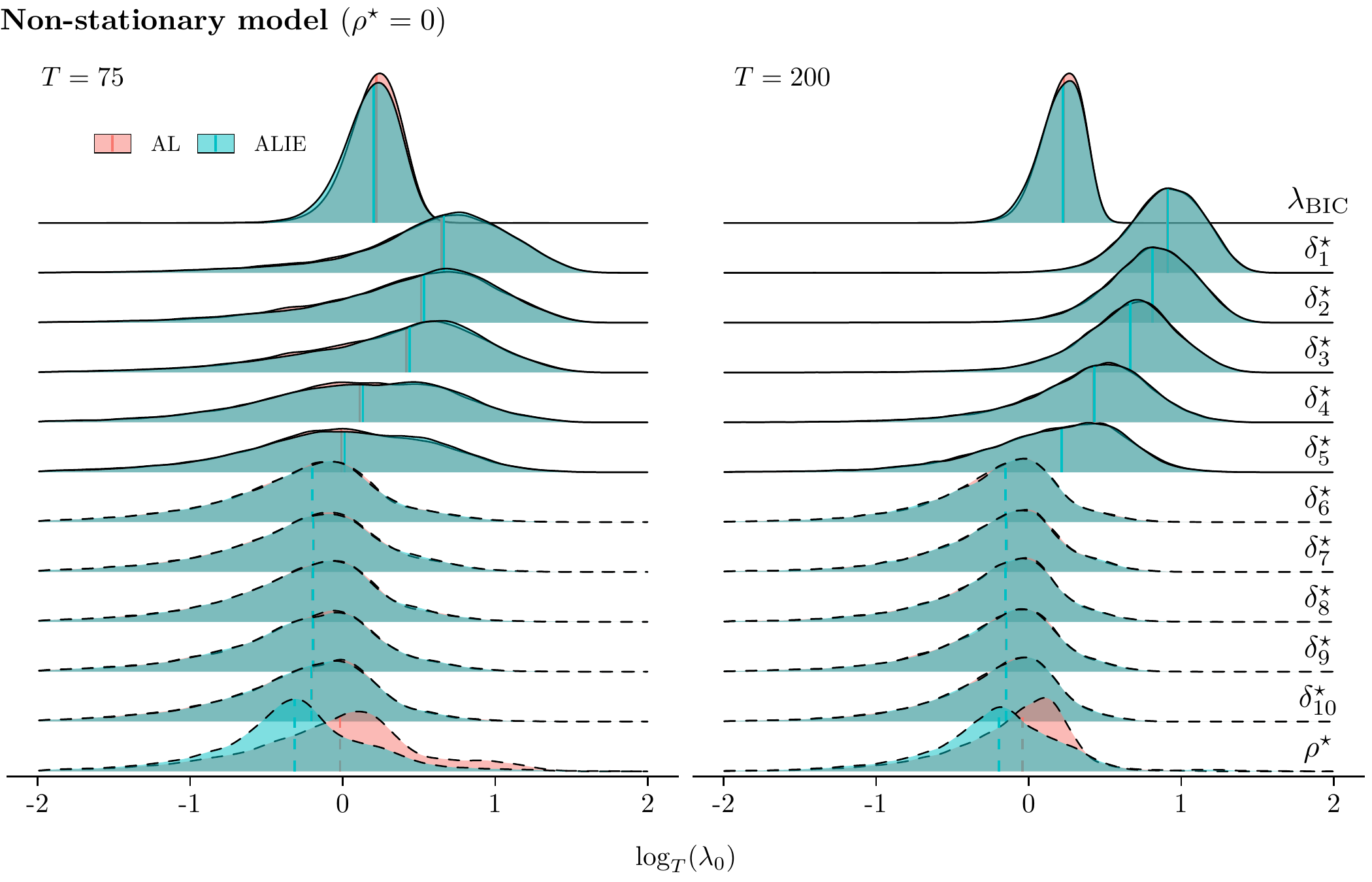}
    }
    \begin{minipage}{\textwidth}
        \scriptsize\textit{Notes:} DGP \eqref{eq:adfdgp}. $\Breve{w}_1$ is computed using the AR spectral density estimator of \textcite{PerronNg1998} with $k=10$. Ridges show Gaussian kernel density estimates of activation thresholds as in \Cref{def:lambda0_definition} with joint bandwidth estimated according to the \textcite{Silverman1986} rule. Vertical lines show medians. Dashed lines indicate \emph{irrelevant} variables. 5000 replications.
    \end{minipage}
    \end{minipage}
\end{figure}
\Cref{fig:lambda0dist} illustrates the finite sample impact of the results in \Cref{prop:lambdanaught} and \Cref{cor:knot1asymptotics}. Here, we compare simulated $\lambda_0$ distributions for the regressors in an over-specified ADF model to the BIC-tuned penalty parameter $\lambda_\text{BIC}$ for ALIE and AL. We set $\rho^\star=0$ or $\rho^\star=-.05$, let $\vdelta^\star=(.275,\,.25,\,.2,\,.15,\,.1)'$ and estimate model \eqref{eq:adfreg} with $p=10$. For ease of interpretation, we adopt the popular choice $\gamma_1 = \gamma_2 = 1$ which translates to $\lambda_0 = O_p(1)$ for \textit{all} irrelevant regressors (dashed lines). Activation thresholds of the relevant regressors (solid lines), i.e., $\lambda_{0,\,\rho^\star\in(-2,0)}$ and $\lambda_{0,\,\delta_j^\star\neq0}$, are $O_p(T)$ and $\Breve\lambda_{0,\,\rho^\star\in(-2,0)} = O_p(T^2)$, by \Cref{prop:lambdanaught}. Note that $F_{T,\,\lambda_\textup{BIC}}$ converges to a stable asymptotic distribution $F_{\lambda_\textup{BIC}}$ with support on $\mathbb{R}^+$. 

For $\rho^\star=-.05$ (top panel), the asymptotic implications of \cref{prop:lambdanaught}, part \ref{item:prop:lambdanaught:rho<0} already emerge for $T=75$. The empirical distribution of $\lambda_{0,\,\rho^\star \in (-2,0)}$ is shifted towards larger values, reducing the overlap with the density of $\lambda_\textup{BIC}$. Following \cref{eq:rho_activ_prob}, this feature suggests an improved activation probability from information enrichment. Consistent with the theory, these effects become more pronounced for $T=200$. We also find high congruence between distributions of the remaining variables for ALIE and AL for $T=75$ and $T=200$, corroborating part \ref{item:prop:lambdanaught:delta=0} of \Cref{prop:lambdanaught}.

The bottom panel of \Cref{fig:lambda0dist} shows results for $\rho^\star = 0$. Unlike for $\rho^\star=-.05$, the $\Delta y_{t-j}$ are weakly correlated with $y_{t-1}$, meaning that estimates are expected to be less sensitive to the inclusion or omission of $y_{t-1}$. We hence find the distributions of the $\lambda_{0,\,\delta_j^\star}$ to be virtually identical for ALIE and AL.\footnote{The distributions of $\lambda_{0,\,\delta_j^\star}$ are identical for the irrelevant lags since $\Delta y_{t-j}$ is always stationary.} However, $\Breve\lambda_{0,\,\rho^\star}$ displays a shift of probability mass towards small values of $\Breve\lambda_{0,\,\rho^\star}$, reducing overlap with the density of $ \lambda_\textup{BIC}$. This is a desirable property that, however, does not stem from the asymptotic results detailed in this section. The shape of this distribution matters most in applications because the oracle properties impose zero overlaps between $F_{T,\, \lambda_\textup{BIC}}$ and $\Breve\lambda_{0,\,\rho^\star = 0}$ only as $T \to \infty$. The following section provides further insights by discussing information enrichment in a zero-mean model.

Comparing the medians (vertical lines) of $\lambda_{0,\, \rho^\star}$ for $\rho^\star = -.05$ and $\rho^\star = 0$ reveals how combining two imperfectly correlated statistics translates into an improved performance via projection on the activation thresholds: information enrichment drives an additional wedge between $\Breve\lambda_{0,\,\rho^\star = 0}$ and $\Breve\lambda_{0,\,\rho^\star \in (-2,0)}$, which is captured in the following remark.
\begin{rem}
    \label{rem:lambda0_score}
   By parts \ref{item:prop:lambdanaught:rho=0} and \ref{item:prop:lambdanaught:rho<0} of \Cref{prop:lambdanaught},
    \begin{equation*}
        \frac{\Breve\lambda_{0,\,\rho^\star\in(-2,0)}}{\Breve\lambda_{0,\,\rho^\star=0}} = O_p(T^{2\gamma_1})
    \quad \textup{and} \quad
        \frac{\lambda_{0,\,\rho^\star\in(-2,0)}}{\lambda_{0,\,\rho^\star=0}} = O_p(T^{\gamma_1}),
    \end{equation*}
    i.e., information enrichment accelerates the separation between the activation thresholds of relevant and irrelevant $y_{t-1}$. Of course, AL could be set up to yield the same discriminatory power as ALIE by doubling $\gamma_1$, thereby putting twice the weight on the OLS estimate $\widehat\rho$ used in $w_1$. However, this is not sensible because the rationale to favour $\Breve{w}_1$ over $w_1$ is that $\widehat\rho$ is deemed insufficient in the first place.\qed
\end{rem}

\subsection{Information enrichment in a zero-mean model}
\label{sec:nmm}
As $\lambda_{0, \rho^\star = 0}$ and $\Breve\lambda_{0, \rho^\star = 0}$ have the same stochastic order, ALIE's size advantage shown in \Cref{fig:lambda0dist} and our simulation results must root from differing shapes of the respective distributions $F_{\lambda_0}$, $F_{\Breve{\lambda}_0}$ for AL and ALIE. Therefore, the relationship between information enrichment and the shape of $F_{\Breve{\lambda}_0}$ will be critical in our considerations below.

Suppose the setup
	\begin{align}
		\begin{split}
				x =&\,\mu + \epsilon, \qquad \eps \sim (0, \sigma_\epsilon), \label{eq:nmm} \\
				z_{j} =&\, \mu + \nu_{j}, \quad \nu_{j}\sim (0,\sigma_{\nu,\,j}),		
		\end{split}
	\end{align}
	with $\sigma_\epsilon,\sigma_{\nu,\,j}<\infty$, $\mu=0$, and statistics $z_j$, $ j=1,\dots,q$, providing additional information on $\mu$. Also, let $\{\epsilon,\nu_1,\cdots,\nu_q\}$ be a set of independent noise terms. The measurements $x$ and $z_j$ can represent single observations or sample averages. By defining $\mu := \rho^\star y_{t-1}$, inference on $\rho^\star$ is tantamount to inference on $\mu$ with the null hypothesis being $\mu = 0$.\\
	An adaptive Lasso estimator for $\mu$ in \eqref{eq:nmm} is the minimiser
	\begin{equation}
		\widehat{\mu}_w := \argmin_{\dot{\mu}} \left( x - \dot{\mu} \right)^2 + 2 \lambda w \lvert \dot{\mu} \rvert, \label{eq:nmmadalasso} 
	\end{equation}
    with the FOC for $\widehat\mu_w = 0$ resulting in
	\begin{align}
    	\lambda w^\gamma \geq \lvert 1 \cdot \left( x - \dot{\mu} \right)\rvert
     \Leftrightarrow\lambda \geq w^{-\gamma} \lvert \mu - \dot{\mu} + \epsilon \rvert. \label{eq:nmmadlfoc}
	\end{align}
	The adaptive Lasso estimator suggested by \textcite{zou2006adaptive} uses exclusively $x$ to determine the penalty weight, $w:=\lvert1/x\rvert^\gamma$. Setting $\dot{\mu} = 0$ and $\mu = 0$ in \eqref{eq:nmmadlfoc}, one obtains $w=\lvert1/\epsilon\rvert^\gamma$ and $\lambda_0:=\lvert\epsilon\rvert^\gamma\lvert\epsilon\rvert$. We suppose the penalty parameter $\lambda_\textup{IC}$ to be drawn from a tuning distribution $F_{\lambda_\textup{IC}}$ independent of $x$ and $z_j$. Because the model \eqref{eq:nmm} has a single regressor, $\Prob{\widehat\mu_w\neq0\bigr\vert\lambda_0>\lambda_\textup{IC}} = 1$ is ensured. The probability to misclassify $\mu$ simplifies to
    \begin{equation*}
        \Prob{\widehat\mu_{w}\neq0} = \Prob{\lambda_0>\lambda_\textup{IC}}.
    \end{equation*} 
An information-enriched adaptive Lasso estimator $\widehat\mu_{\Breve w}$ adds the $z_{j}$, yielding
\begin{align*}
    \Breve{w} := \left\lvert \frac{1}{\epsilon\cdot\prod_{j=1}^q z_{j}} \right\rvert^\gamma, \quad \Breve{\lambda}_0 := \left\lvert\epsilon\cdot\prod_{j=1}^q\nu_{j}\right\rvert^\gamma\lvert\epsilon\rvert, 
\end{align*}
as $z_j = \nu_j$ since $\mu = 0$. The enriched estimator has the activation probability 
\begin{equation}
    \Prob{\widehat\mu_{\Breve w}\neq0} = \Prob{\Breve\lambda_0>\lambda_\textup{IC}} = 1 - \Prob{\Breve\lambda_0 \leq \lambda_\textup{IC}}.
\end{equation}
 Accordingly, reducing the classification error of $\mu$ requires shrinking $\Breve\lambda_0$, i.e., shifting $F_{\Breve\lambda_0}$ closer to zero. In \Cref{thm:nmmlmtwo}, we examine how the (erroneous) activation probabilities compare for $\widehat\mu_w$ and $\widehat\mu_{\Breve w}$. 	
\begin{restatable}[Activation probabilities]{thm}{nmmlmtwo}
    \label{thm:nmmlmtwo}
	Let $\mu=0$ in \eqref{eq:nmm} and consider the adaptive Lasso estimators $\widehat\mu_w$ and $\widehat\mu_{\Breve w}$ with penalty weights $w$ and $\Breve w$, respectively. Denote $\lambda_0\sim F_{\lambda_0}$ and $\Breve{\lambda}_0\sim F_{\Breve{\lambda}_0}$ the thresholds for which the FOC \eqref{eq:nmmadlfoc} is just binding and be $\lambda_\textup{IC}\sim F_{\lambda_\textup{IC}}$ the tuned penalty parameter in \eqref{eq:nmmadalasso}. Let $F_{\lambda_\textup{IC}}$, $F_{\Breve{\lambda}_0}$, and $F_{\lambda_0}$ be continuous distributions that are tight on $\mathbb{R}^+$ and be $F_{\lambda_\textup{IC}}$ exogenously given.
    \begin{enumerate}
        \item\label{item:thm:nmmlmtwo:P1}
        \begin{enumerate}
            \item\label{item:thm:nmmlmtwo:P1:a} $\Prob{\widehat\mu_w\neq0}=1-\int_{0}^\infty F_{\lambda_0}(\lambda_\textup{IC})\,\mathrm{d}F_{\lambda_\textup{IC}}\in[0,1]$.
            \item\label{item:thm:nmmlmtwo:P1:b} If $\lim_{q\to\infty}\E(\Breve{\lambda}_0)=0$, then $\lim_{q\to\infty}\Prob{\widehat\mu_{\Breve{w}}\neq0}=0$.
        \end{enumerate} 
        \item\label{item:thm:nmmlmtwo:P2} Let $A_q := \left\{a\in(0,1): F^{-1}_{\Breve{\lambda}_0}(1-a) < F^{-1}_{\lambda_0}(1-a)\right\}$ for $q\in(0,\infty)$. If $\Prob{\widehat\mu_w\neq0}=a\in A_q$, then $\Prob{\widehat\mu_{\Breve{w}}\neq0}\leq \Prob{\widehat\mu_w\neq0}$.
    \end{enumerate}
\end{restatable}
\begin{proof}
	See \Cref{sec:proofs}.
\end{proof}

Part \ref{item:thm:nmmlmtwo:P1:a} of \Cref{thm:nmmlmtwo} states how the probability for a \textit{general} adaptive Lasso estimator $\widehat\mu_w$ to activate $\mu$ in model \eqref{eq:nmm} depends on the data $x$ through $\lambda_0$ and the employed tuning procedure, viz. $\Prob{\widehat\mu_w\neq0}=\Prob{\lambda_0>\lambda_\textup{IC}}$. Part \ref{item:thm:nmmlmtwo:P1:b} asserts that $\Prob{\widehat\mu_{\Breve{w}}\neq0}\to0$ as $q\to\infty$ for an information-enriched adaptive Lasso estimator $\widehat\mu_{\Breve w}$, i.e., enriching the adaptive weight $\Breve w$ with additional information on $\mu$ obtains an estimator that dominates any $\widehat\mu_w$ in terms of activation probability.

Part \ref{item:thm:nmmlmtwo:P2} of \Cref{thm:nmmlmtwo} gives a condition for $\Prob{\widehat\mu_{\Breve w}\neq0}$ to be bounded from above by the activation probability of any adaptive Lasso estimator for some $q\in(0,\infty)$: we require that $\Prob{\widehat\mu_w\neq0}=a$ corresponds to an upper tail event of $\lambda_0$ that is less likely to occur for $\Breve\lambda_0$. While this statement is independent of $q$, note that $\lim_{q \to \infty} A_q = (0,1)$ is a consequence of part \ref{item:thm:nmmlmtwo:P1:b}. Hence, enriching the information on $\mu$ enlarges the potential of $\widehat\mu_{\Breve w}$ to improve upon $\widehat\mu_w$ as the set of dominating activation probabilities $a$ converges to the entire unit interval.
	
	We next examine the information-enriched adaptive Lasso properties in a \emph{Gaussian} zero mean model.

	\begin{restatable}[Gaussian zero means]{cor}{qlimnmm}\label{cor:qlimnmm}
		Let the assumptions of \Cref{thm:nmmlmtwo} hold and be $\epsilon\sim N(0,\sigma_\epsilon)$ and $\nu_j\sim N(0,\sigma_{\nu,\,j})$. Consider the information-enriched adaptive Lasso estimator $\widehat\mu_{\breve w}$ with penalty weight $\Breve w$.			
		\begin{enumerate}
          	\item\label{item:cor:qlimnmm:P1} If $\prod_{j=1}^q \E(\lvert \nu_j\rvert^\gamma) = o(1)$, then $\lim_{q\rightarrow\infty} \Prob{\widehat\mu_{\Breve w}\neq0} = 0.$
            \item\label{item:cor:qlimnmm:P2} Let $\gamma=1$. If $\prod_{j=1}^q \sigma_{\nu,\,j} = o\left( (\pi/2)^{q/2}\right)$, then $\lim_{q\rightarrow\infty} \Prob{\widehat\mu_{\Breve w}\neq0} = 0.$
        \end{enumerate}
	\end{restatable}
	\begin{proof}
	See \Cref{sec:proofs}.
	\end{proof}

Result \ref{item:cor:qlimnmm:P1} of \Cref{cor:qlimnmm} states that in a zero-mean model, $\prod_{j=1}^q \E(\lvert \nu_j\rvert^\gamma) = o(1)$ is sufficient for the probability of $\widehat\mu\neq 0$  to converge to zero as the number of informative statistics $q$ diverges. Since the $\lvert\nu_j\rvert$ are half-Gaussians, this condition is equivalent to restricting the signal-to-noise ratio of the enriching statistics $z_j$ in $\Breve{w}=(1/\prod_{j=1}^q z_j)^\gamma$. Result \ref{item:cor:qlimnmm:P2} of \Cref{cor:qlimnmm} provides a tangible condition for $\lim_{q\rightarrow\infty}\Prob{\widehat\mu_{\Breve w}\neq0} = 0$, requiring an upper bound on the noise added to $\Breve w$ by the enriching statistics. Note that $\prod_{j=1}^q \sigma_{\nu,\,j} = o\left( (\pi/2)^{q/2}\right)$ is trivially satisfied if $\sigma_{\nu,\,j} = 1 \, \forall j$, i.e., it is permitted to employ a (possibly infinite) set of independent Gaussian statistics to enrich $\Breve{w}$.

\begin{figure}[t]
    \centering
    \caption{Activation probabilities in univariate Gaussian zero-mean models}
    \vspace{.25cm}
    \label{fig:actratenmm}
	\resizebox{\textwidth}{!}{
	\includegraphics{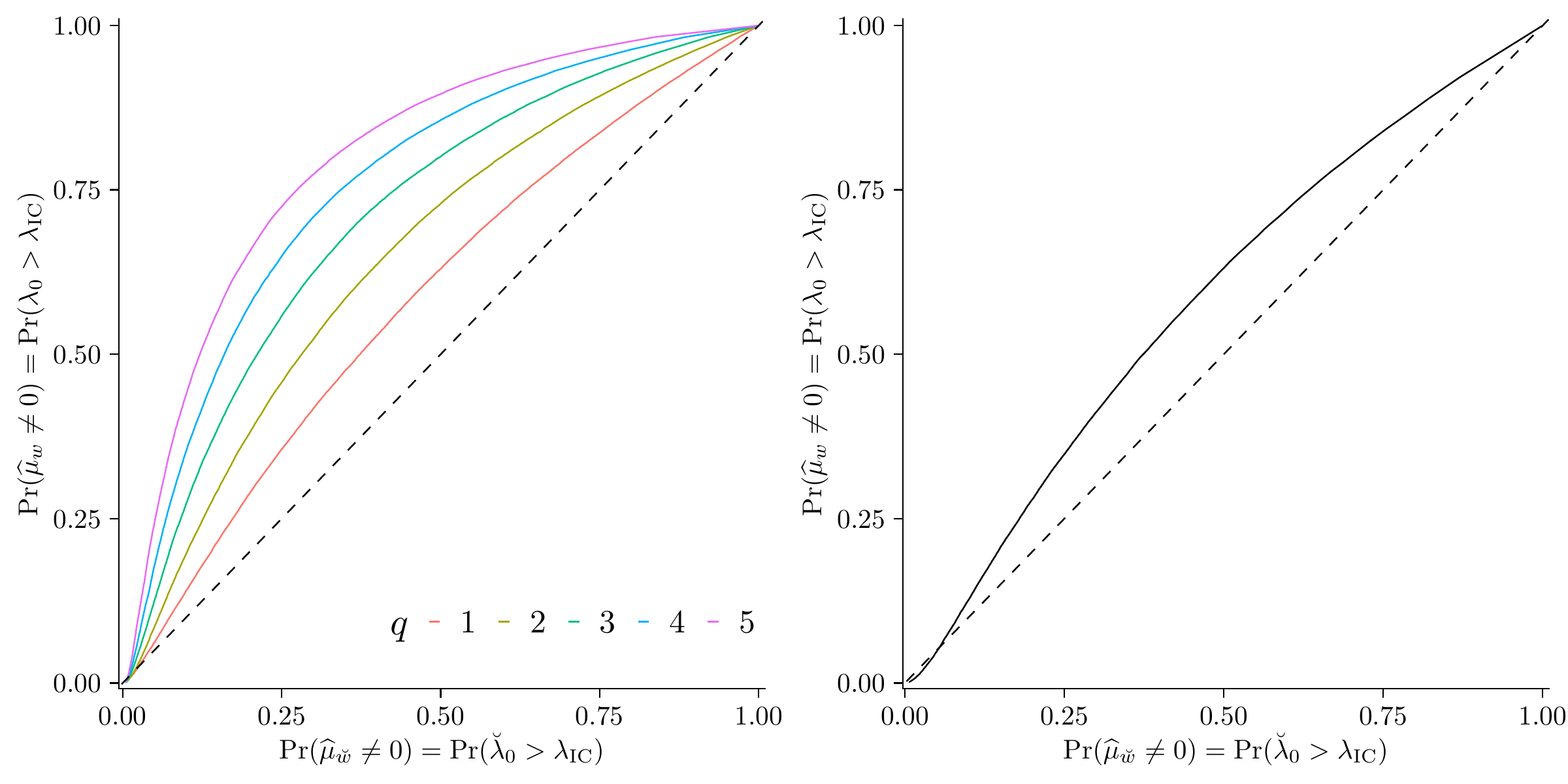}
	}
    \begin{minipage}{\textwidth}
        \scriptsize\textit{Notes}: $\gamma=1$. Left: Simulated activation probabilities for $n=1$. Standard Gaussian enriching statistics used in $\widehat\mu_{\Breve{w}}$, $\sigma_\epsilon=\sigma_{\nu,j}=1$. Right: Simulated activation probabilities for $n=50$ with $w=\lvert\overline{\boldsymbol{x}}\rvert^{-1}$ and $\Breve{w}=\lvert t_{z_1}\cdot\overline{\boldsymbol{x}}\rvert^{-1}$ where $t_{z_1}$ is the $t$-statistic for $H_0:\mu=0$ with $\sigma_\epsilon=1,\ \sigma_{\nu}=2,\ \corr(\epsilon, \nu) = .2$. $5\cdot10^4$ replications.
    \end{minipage}
\end{figure}

The advantage of $\widehat\mu_{\Breve w}$ over $\widehat\mu_w$ can be seen from the left panel of \Cref{fig:actratenmm}, which compares activation probabilities for $\gamma = 1$ and Gaussian $z_j$. Information enrichment may yield significant improvements, even for $q=1$ (red curve). Including additional information results in even superior estimators. A comparison of the activation probabilities for $n = 50$ and a single correlated enriching statistic is shown in the right panel of \Cref{fig:actratenmm}. Here, $w=\lvert\overline{\boldsymbol{x}}\rvert^{-1}$ and we set $\Breve{w}=\lvert t_{z_1}\cdot\overline{\boldsymbol{x}}\rvert^{-1}$ where $t_{z_1}$ is the $t$-statistic for $H_0:\mu=0$. Again, we find the significant potential of $\Breve{w}_1$ to mitigate the risk of spurious activations.

\section{Implementation and Monte Carlo evidence}
\label{sec:simev}

In the following, we investigate the empirical properties of ALIE in more detail and compare its performance to AL using simulation studies. For implementing $\breve{w}_1$, it is crucial to scale $y_t$ by a consistent estimator of the long-run standard deviation $\omega$ such that the limit distribution of $J_\alpha$ given $\rho^\star = 0$ is nuisance-parameter-free. We follow \textcite{HerwartzSiedenburg2010} and estimate $\omega^2$ using the AR spectral density estimator at frequency zero,
\begin{align}
    \widehat{\omega}^2_{\textup{AR}}(k) := \frac{\widehat{\sigma}_k^2}{(1 - \sum_{j=1}^k \widehat{\delta}_j)^2}, \qquad \widehat{\sigma}^2_k := (T-k)^{-1} \sum_{t=k+1}^T \widehat{\varepsilon}_{k,\,t}^2, \label{eq:s2ar}    
\end{align}
suggested by \textcite{PerronNg1998}. The $\widehat{\varepsilon}_{k,\,t}$ are residuals from estimating \eqref{eq:adfreg} by OLS with lag order $p=k$. To compute \eqref{eq:s2ar}, we estimate $k$ as $\widehat{k}:= \argmin_{0\leq k\leq k_{\max}} \textup{IC}(k)$ where $\textup{IC}(k)$ is an information criterion of the form
\begin{align}
    \textup{IC}(k) := \log(\tilde{\sigma}^2_k) - C_T \frac{\tau_T(k) + k}{T-k_{\max}}.\label{eq:icadf}
\end{align}
$k_{\max}$ is a maximum lag order and $\tilde\sigma^2_k := (T-k_{\max})^{-1}\sum_{t=k_{\max}+1}^T \widehat{\varepsilon}_{k,\,t}^2$ a deviance estimate. $C_T$ is a penalty function and $\tau_T(k)$ a stochastic adjustment term. Popular choices are $C_T = \log(T)$ and  $\tau_T(k) = 0$ (BIC), $C_T = 2$ and $\tau_T(k) = 0$ (AIC). \textcite{NgPerron2001} have suggested modified versions for truncating long autoregressions. These are MBIC and MAIC where $\tau_T(k) = \left(\tilde{\sigma}_k^2\right)^{-1}\widehat{\rho}^2\sum_{k_{\max}+1}^T y_{t-1}^2$, respectively.

To estimate the LRV by $\widehat{\omega}^2_{\textup{AR}}(k)$, we use BIC to select $k$ if not indicated otherwise for sparse processes covered by \Cref{assum:sparsity} and MAIC for processes that do not allow consistent model selection when $T<\infty$. The adaptive Lasso estimator \eqref{eq:ALestimator} is computed in a model with $p=k_{\max} = \lfloor12 (T/100)^{1/4}\rfloor$, cf. \textcite{Schwert1989}.

All simulations were performed using the statistical programming language R \parencite{R}. We used the LARS algorithm implemented in the \texttt{lars} package \parencite{pkg-lars} for computing the Lasso solution paths.

\subsection{Rationale for treating \texorpdfstring{$\alpha$}{\textalpha} and \texorpdfstring{$\sigma_v$}{\textsigma} as tuning parameters}
\label{sec:alphatuning}

As outlined in \Cref{rem:whyJ}, $\Breve{w}_1$ can be calibrated via $\alpha$ and $\sigma_v$. The rationale is that larger values of $\alpha$ narrow the distribution of $J_\alpha$ and thus impact the behaviour of $\Breve{w}_1$. By definition of $\Breve{w}_1$, we expect a positive association between $\alpha$ and $\Prob{\widehat{\rho}_\lambda\neq0}$ for $\rho^\star\in(-2,0]$. A similar argument can be made for $\sigma_v$, which governs the variance of $J_\alpha$. 

\begin{figure}[tbp]
    \centering
    \caption{Information enriched weights --- $\alpha$ and $\sigma_v$ as tuning parameters}
    \label{fig:alphatuning}
    \vspace{.25cm}
    \resizebox{\textwidth}{!}{
    \includegraphics{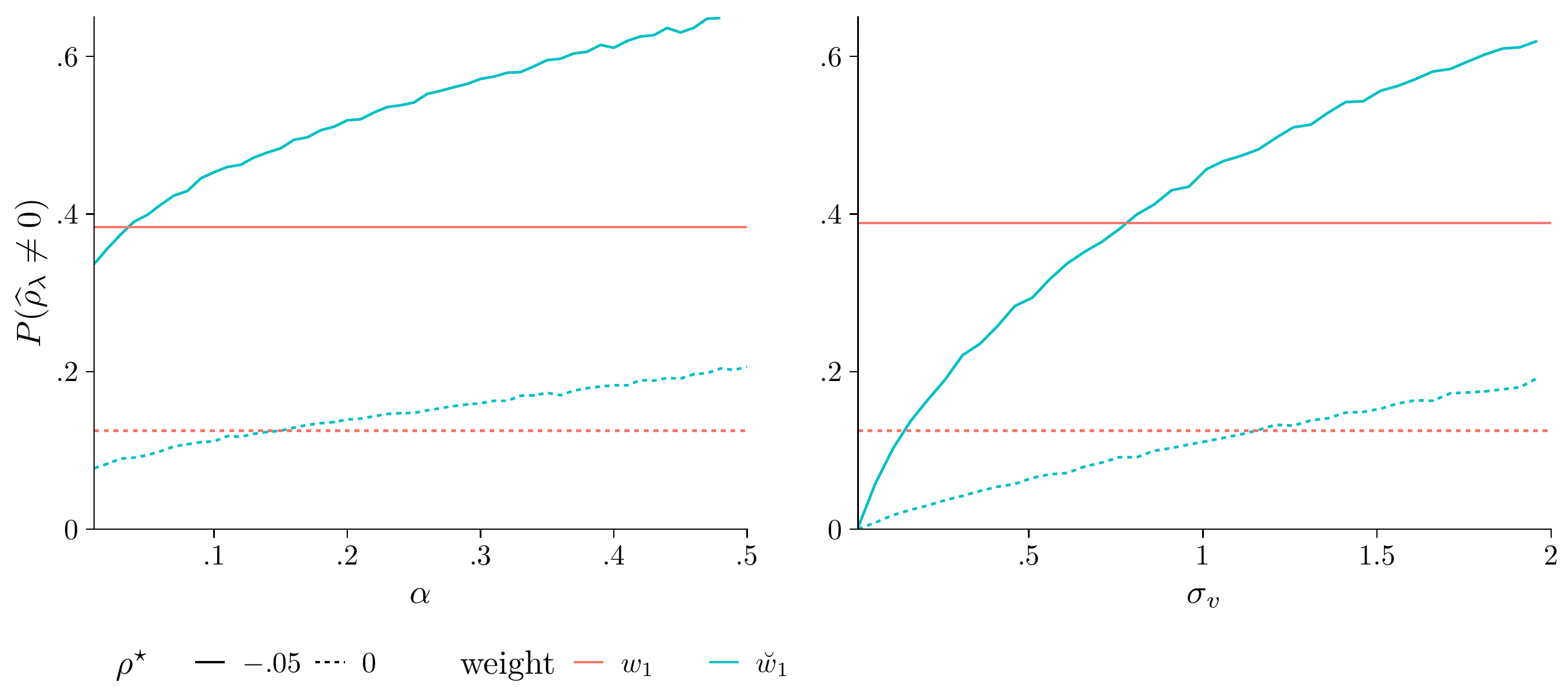}
    }
    \begin{minipage}{\textwidth}
        \vspace{.2cm}
        \scriptsize\textit{Notes:} Model \eqref{eq:adfreg} with $d_t = 0$ and $p=12$. DGP \eqref{eq:thedgp} with $u_t \sim i.i.d.\,N(0,1)$ and $T=100$. $\Breve{w}_{1}$ computed with $J_\alpha$ with LRV estimate $\widehat{\omega}^2_{\textup{AR}}(k=12)$. Activation rates for $\sigma_v = 1$ (left) and $\alpha = .1$ (right). 25000 replications.
    \end{minipage}
\end{figure}

To illustrate the effect of both parameters on the activation rate of $y_{t-1}$, we simulate $y_t$ using DGP \eqref{eq:thedgp} for $T=100$ and consider $\alpha\in(0, .5)$ for $\sigma_v=1$ and $\alpha=.1$ for $\sigma_v\in(0,2]$. The results are shown in Figure \ref{fig:alphatuning}. Activation rates increase with $\alpha$ so that power gains are accompanied by higher misclassification rates for non-stationary $y_t$ (left panel). Similarly, activation rates increase with $\sigma_v$ (right panel). 

Notice that the comparison with AL indicates room for tuning. For example, ALIE could be calibrated to have an activation rate close to that of AL for non-stationary $y_t$ but to exhibit more attractive selection rates under stationarity. Hereafter, we follow \textcite{HerwartzSiedenburg2010} and set $\sigma_v=1$ and $\alpha=.1$ which seem reasonable. However, we note that different choices may be more suitable when the expected loss from a misclassification is not identical for $\rho^\star=0$ or $\rho^\star\in(-2,0)$. This may be relevant in forecasting where the loss from falsely classifying a model as stationary rather than excluding the inference regressor can be quite different than vice-versa.

\subsection{Properties for simple autoregressions}
\label{sec:simpleproc}

To illustrate the effect of $\Breve{w}_1$, we first consider BIC-tuned adaptive Lasso estimates of \eqref{eq:adfreg} with lag order $p=\lfloor12(T/100)^{.25}\rfloor$ for simple processes. For this we simulate $y_t$ according to DGP \eqref{eq:thedgp} without a deterministic component with coefficient $\varrho=\rho^\star+1\in\{.95, 1\}$, and errors $u_t \sim i.i.d.\,N(0,1)$. We consider $T\in\{25,50,100,150,250,500,1000\}$. $\breve{w}_1$ is computed using $J_{.1}$ with the LRV estimated by $\widehat{\omega}_{\text{AR}}^2(0) = \widehat{\sigma}^2_0$.

\begin{table}[tbp]
\setlength{\tabcolsep}{10pt}
\renewcommand{\arraystretch}{.9}
\centering
\caption{Properties of adaptive Lasso estimators for AR(1) processes}
\label{tab:simplear}
\vspace{.3cm}
\resizebox{\textwidth}{!}{
\begin{tabular}{lccccccccc}
\toprule
\multicolumn{3}{c}{ } & \multicolumn{3}{c}{AL} & \multicolumn{1}{c}{ } & \multicolumn{3}{c}{ALIE} \\
\cmidrule(l{3pt}r{3pt}){4-6} \cmidrule(l{3pt}r{3pt}){8-10}
$T$ & $p$ & $\rho^\star$ & $\mathrm{P}(\widehat\rho_\lambda\neq0)$ & $\log(w_1)$ & $\log(\lambda_{0,\,\rho^\star})$ &   & $\mathrm{P}(\widehat\rho_\lambda\neq0)$ & $\log(\breve{w}_1)$ & $\log(\breve{\lambda}_{0,\,\rho^\star})$\\
\midrule
\addlinespace[-.2cm]
\multicolumn{10}{l}{\textbf{}}\\
25 & $8$ &  & $.17$ & $2.14$ & $-.09$ &  & $.1$ & $3.78$ & $-1.7$\\

50 & $10$ &  & $.06$ & $3.29$ & $-.38$ &  & $.05$ & $4.43$ & $-1.41$\\

100 & $12$ &  & $.02$ & $4.18$ & $-.46$ &  & $.03$ & $5.13$ & $-1.39$\\

150 & $13$ &  & $.02$ & $4.6$ & $-.38$ &  & $.02$ & $5.56$ & $-1.31$\\

250 & $15$ &  & $.01$ & $5.19$ & $-.45$ &  & $.02$ & $6.06$ & $-1.29$\\

500 & $22$ &  & $.01$ & $5.94$ & $-.48$ &  & $.01$ & $6.82$ & $-1.36$\\

1000 & $31$ & \multirow{-7}{*}{ $0$} & $0$ & $6.65$ & $-.51$ &  & $.01$ & $7.49$ & $-1.32$\\

\addlinespace[-.2cm]
\multicolumn{10}{l}{\textbf{}}\\
25 & $8$ &  & $.18$ & $1.89$ & $.04$ &  & $.15$ & $3.01$ & $-1.11$\\

50 & $10$ &  & $.1$ & $2.59$ & $.29$ &  & $.12$ & $2.76$ & $.09$\\

100 & $12$ &  & $.12$ & $2.85$ & $.91$ &  & $.17$ & $2.61$ & $1.16$\\

150 & $13$ &  & $.21$ & $2.87$ & $1.35$ &  & $.28$ & $2.34$ & $1.87$\\

250 & $15$ &  & $.41$ & $2.91$ & $1.87$ &  & $.56$ & $2.01$ & $2.76$\\

500 & $22$ &  & $.89$ & $2.95$ & $2.54$ &  & $.96$ & $1.48$ & $4.01$\\

1000 & $31$ & \multirow{-7}{*}{\centering\arraybackslash $-.05$} & $1$ & $2.98$ & $3.23$ &  & $1$ & $.86$ & $5.34$\\
\bottomrule
\end{tabular}
}
    \begin{minipage}{\textwidth}
        \vspace{.25cm}
        \scriptsize\textit{Notes:}
        Model \eqref{eq:adfreg} with $p=\lfloor12\cdot(T/100)^{.25}\rfloor$ and $d_t=0$. DGP \eqref{eq:adfdgp} with $\varepsilon_t \sim i.i.d.\,N(0,1)$. $\Breve{w}_1$ computed with $J_{.1}$. LRV estimated with $\widehat{\omega}^2_\textup{AR}(0)$. $\Prob{\widehat{\rho}_\lambda\neq0}$ is the average activation rate of $y_{t-1}$. $\log(\breve w_1)$ and $\log(w_1)$ are log scale median penalty weights. $\log(\lambda_{0,\,\rho^\star})$ and $\log(\Breve{\lambda}_{0,\,\rho^\star})$ are log scale median activation thresholds for $\lambda$. 5000 replications.
    \end{minipage}
\end{table}

The results are presented in Table \ref{tab:simplear}. We find that in the case of a random walk $(\rho^\star=0)$, there is improvement in the rate of correct classification of ALIE over AL for $T<150$, where ALIE never exceeds the probability of activating $y_{t-1}$ of AL. This benefit diminishes for large $T$. A comparison of the inference regressor weights shows that $\Breve{w}_1$ implies a higher average penalty from including $y_{t-1}$ in the model. We also find that the average activation threshold on the Lasso solution path decreases faster for ALIE than for AL as $T$ grows.

For stationary autoregressions ($\rho^\star = -.05$), we also find improvement in the discriminatory power of the procedure. For larger $T$, $\Breve{w}_1$ yields a pronounced advantage. E.g., for $T=250$, ALIE has approximately 16\% higher probability of activating $y_{t-1}$ than AL. Finite sample implications of our theoretical results in \Cref{sec:sbwe,sec:dnsapr} are mirrored by the average weights and activation thresholds: while $w_1$ converges at rate $\sqrt{T}$ towards its probability limit $\lvert\rho^\star\rvert^{-1} = 20$, the modified $\Breve{w}_1$ yields a lower adaptive penalty for $T>50$. We also find that $\Breve{\lambda}_{0,\,\rho^\star}$ grows at a faster rate than $\lambda_{0,\,\rho^\star}$.

\subsection{Handling deterministic terms}
\label{sec:hdt}

To our knowledge, no literature has examined the properties of adaptive Lasso in the ADF regression \eqref{eq:adfreg} with $d_t\neq0$ to date. For trend filtering, \textcite{tibshirani2011solution} give a comprehensive overview of past contributions to fitting polynomial trends or structural breaks components using the Lasso and its variants. While these approaches to removing deterministic components appear promising, a theoretical analysis is beyond the scope of this article. We instead investigate the performance of the estimators on detrended data by simulation.\footnote{\textcite{kock2016consistent} does not discuss properties of AL when $d_t\neq0$ and notes that oracle properties carry over to models with detrended data.}  

\begin{figure}[t]
    \centering
    \caption{Distribution of $J_\alpha$ and CDFs of weights for $y_{t-1}$ under trend adjustment}
    \vspace{.25cm}
    \label{fig:jdetr}
    \resizebox{.95\textwidth}{!}{
    \includegraphics{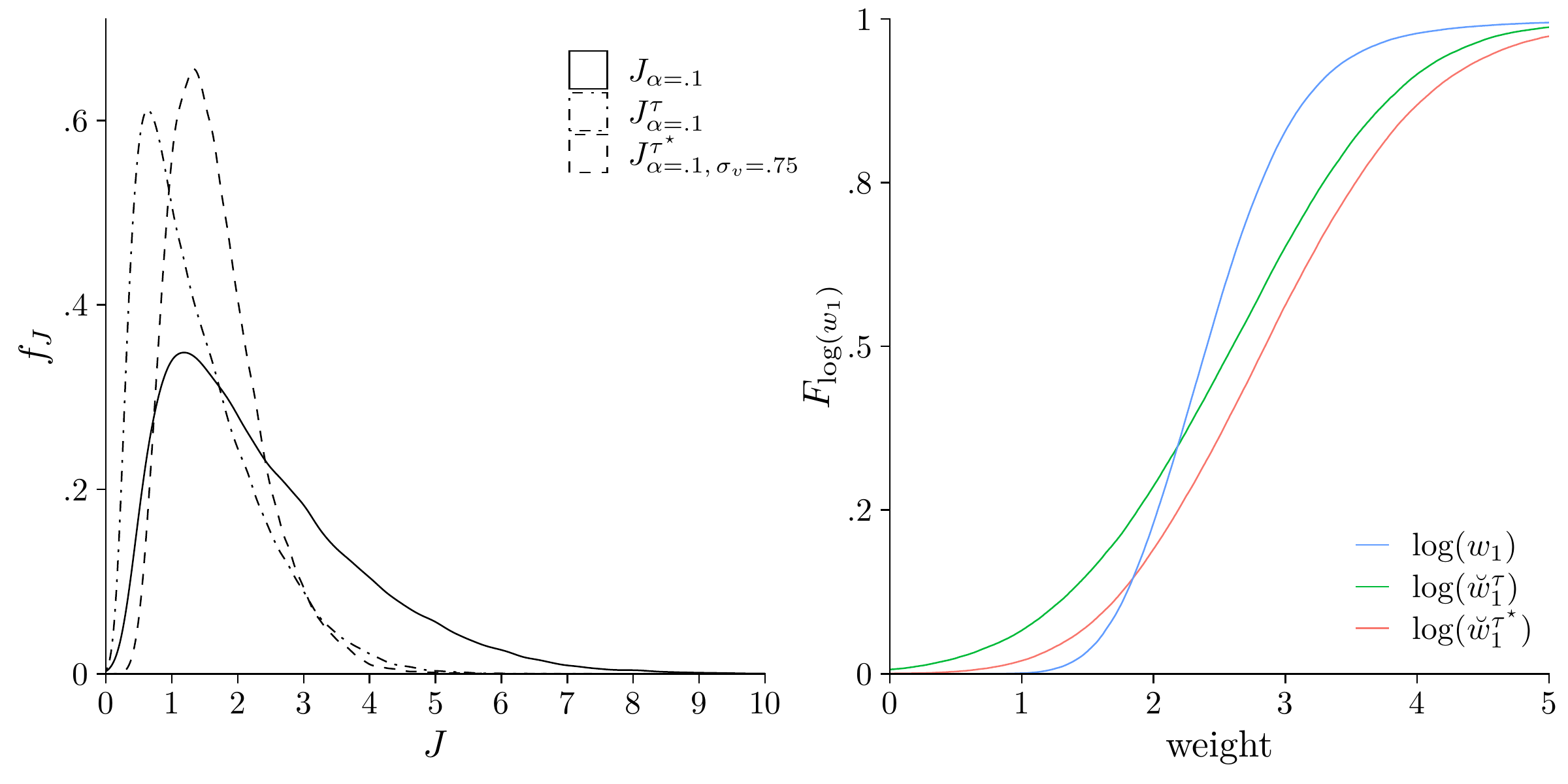}
    }
    \begin{minipage}{\textwidth}
        \vspace{.2cm}
        \scriptsize\textit{Notes:} DGP \eqref{eq:thedgp} with $u_t\sim i.i.d.\,N(0,1)$. $T = 100$. Left: Gaussian kernel density estimates. $J$ computed without deterministic components ($J_\alpha$), for OLS-detrended data ($J_\alpha^\tau$), and with simulation-based trend adjustment ($J_{\alpha,\,\sigma_v}^{\tau^\star}$) with $\sigma_v = .75$. $\alpha = .1$ and LRV estimated by $\widehat{\omega}^2_{\textup{AR}}(k = 0) = \widehat{\sigma}_0^2$ throughout. Bandwidth selected using the \textcite{Silverman1986} rule. Right: empirical CDFs of inference regressor weights. $10^5$ replications based on Gaussian random walks.
    \end{minipage}
\end{figure}

Computation of $\Breve{w}_1$ requires special care in the presence of deterministic terms. For the asymptotic distribution of $J_\alpha$ to be invariant to $\vpsi\neq0$ one needs to adapt \eqref{eq:simres} and the LRV estimator \eqref{eq:s2ar}, for which OLS or quasi-difference detrending has been suggested, cf. \textcite{PerronQu2007}. Detrending affects the distribution of the OLS estimator $\widehat\zeta$ in \eqref{eq:phillips1986} and hence alters the null distribution of $J_\alpha$. To see this for OLS detrending, note that $W_y$ in \eqref{eq:phillips1986} is replaced by
\begin{align}
    W_y^D(a) := W_y(a) - \left(\int_0^1 \mD(s)\mD(s)'\text{d}s\right)^{-1}\left(\int_0^1 \mD(s)W_y(s)\text{d}s\right) \mD(a),
\end{align}
with $\mD(a) := \lim_{n\rightarrow\infty}\mS_n^{-1}\vz_{[na]}$, $a\in[0,1]$ for a suitable scaling matrix $\mS_n$. $\mD(a)=1$ under demeaning where $\vz_t = 1$ and $\mD(a)=(1,a)'$ if $\vz_t = (1,t)'$. Simulation results for $\vz_t = (1,t)'$ indeed reveal higher kurtosis, accompanied by a shift of the distribution of $J_\alpha$ towards zero. The simulated null distribution of $J_{\alpha}^{\tau}$ under OLS detrending has more probability mass below the threshold of $1$ than the distribution of $J_\alpha$, implying an undesired deflation of $\Breve{w}_1^\tau$ relative to $\Breve{w}_1$ in the (no-)intercept model, see the left panel of Figure \ref{fig:jdetr}.\footnote{This shift is even stronger pronounced for QD detrending where $W_y^d$ is replaced by the corresponding projection of the QD-detrended data, see \textcite{Phillips1998}.} This tendency to yield smaller weights for non-stationary $y_t$ is also seen from a comparison of the cumulative distribution functions (CDFs) of $\Breve{w}_1^\tau$ and $w_{1}$ in the right panel of \ref{fig:jdetr}.
We thus expect OLS detrending to inflate activations for $\rho^\star\in(-2,0]$. 

We propose the trend-adjusted simulated weight $\Breve{w}_{1}^{\tau^\star}$ based on a modified trend-adjusted statistic $J_{\alpha,\,\sigma_v}^{\tau^\star}$.
\begin{restatable}[$\Breve{w}_1^{\tau^\star}$]{algo}{wbrewedet}
    \label{algo:wbrewedet}
    \noindent
    \begin{enumerate}
    \item Estimate the LRV using model \eqref{eq:adfreg} with $\vz_t = (1,\,t)'$ by
    \begin{align*}
        \widehat{\omega}^2_{\textup{AR}}(k) =&\, \frac{\widehat{\sigma}_k^2}{(1 - \sum_{j=1}^k \widehat{\delta}_j)^2}, \quad \widehat{\sigma}^2_k = (T-k-2)^{-1} \sum_{t=k+1}^T \widehat{\varepsilon}_{k,\,t}^2
    \end{align*}
    for suitable $k$.
    \item Scale $y_t$ by $\widehat{\omega}_{\textup{AR}}(k)^{-1}$.
    \item Obtain the range statistic $J_{\alpha,\,\sigma_v}^{\tau^\star} := \left\lvert\widehat{\zeta}^{(r)}_{1-\alpha/2}-\widehat{\zeta}^{(r)}_{\alpha/2}\right\rvert$ for simulated OLS estimates $\widehat\zeta^{(r)}$, in the model $y_t = \mu^{(r)} + \varsigma^{(r)} t + \zeta^{(r)} x_t^{(r)} + \nu_t^{(r)}$ with $x_t^{(r)}$ satisfying \Cref{assum:simrw} and $r=1,\dots,R$.
    \item Compute $\Breve{w}_{1}^{\tau^\star} := \left\lvert\widehat{\rho} / J_{\alpha,\,\sigma_v}^{\tau^\star}\right\rvert$ with $\widehat{\rho}$ the OLS estimator of $\rho^\star$ in model \eqref{eq:adfreg} with $d_t$ as in step 1.\qed
\end{enumerate}
\end{restatable}
The procedure is similar to \ref{sec:sbwe} with the difference that we incorporate a linear time trend in the regressions underlying the LRV estimation and the simulation. Following the results in section \ref{sec:alphatuning}, we suggest simulating random walks with $\sigma_v\leq1$ in step 3 to attenuate spurious activations of $y_t$. Simulation results for $J_{\alpha,\,\sigma_v}^{\tau^\star}$ and the associated $\Breve{w}_{1}^{\tau^\star}$ in Figure \ref{fig:jdetr} indeed reveal null distributions with more favourable properties than for the OLS-detrended variants $J^\tau_\alpha$ and $w^\tau_1$.\footnote{We use the ad-hoc choice $\sigma_v = .75$ for better size control of $J_{\alpha,\,\sigma_v}^{\tau^\star}$ for small $T$.}

\subsection{Sparse processes}
\label{sec:adfdgp}

We next investigate finite sample properties of ALIE and AL for sparse processes and also evaluate the ability of the (inconsistent) plain Lasso (PL) to distinguish between stationary and non-stationary models. Data are simulated with the ADF-DGP
\begin{equation}
    y_t(1-B(L)) = v_t, \qquad v_t\overset{i.i.d.}{\sim}N(0,1), \qquad t=1,\dots,T, \label{eq:adfdgp}
\end{equation}
with
\begin{align*} 
    B(L)=\,(\rho^\star + \delta_1^\star) L+ (\delta_2^\star - \delta_1^\star) L^2 + \dots + (\delta^\star_{k^\star-1}-\delta^\star_{k^\star}) L^{k^\star}- \delta^\star_{k^\star} L^{k^\star+1},
\end{align*}
and $k^\star+1$ zero initial values where the lag order is $k^\star := \dim(\vdelta^\star)$. To investigate the model selection performance when there is sparsity in the coefficients of the $\Delta y_{t-j}$, we set $\vdelta_A^\star := (.4, .3, .2)'$ and $\vdelta_B^\star := (.4, .3, .2, 0, 0, 0, -.2, 0, 0, .2)'$.  $\vdelta_C^\star := .7$ yields a short AR process. $\vdelta_D^\star:= (-.4, 0, .7)'$ is taken from \textcite{CanerKnight2013} and is a low power setting for unit root tests. We compute Lasso estimates of model \eqref{eq:adfreg} with lag order $p=\lfloor12\cdot(T/100)^{.25}\rfloor$ and tune $\lambda$ using BIC. We consider $T\in\{25,50,100,150,200,250\}$, let $\rho^\star=0$ for non-stationarity and $\rho^\star = -.05$ to investigate the stationary case. For $\Breve w_1$ we estimate the LRV by $\widehat\omega^2_\text{AR}(k)$ and compute $J_\alpha$ as outlined in Sections \ref{sec:sbwe} and \ref{sec:hdt}. 

Accommodating linear trends is non-trivial. While model selection consistency should be possible for detrended data, unreported simulations indicate that OLS detrending leads to prohibitive activation rates of $y_{t-1}$ for the DGPs under consideration when $\rho^\star=0$. To provide some guidance for empirical application, we present results for first-difference (FD) detrending (\cite{SchmidtPhillips1992}), which seems more reliable for sparse processes.\FloatBarrier
  
\begin{figure}
    \centering
    \caption{Activation rates of Lasso estimators under sparsity}
    \label{fig:sp_rates}
    \vspace{.25cm}
    \resizebox{.97\textwidth}{!}{
        \includegraphics{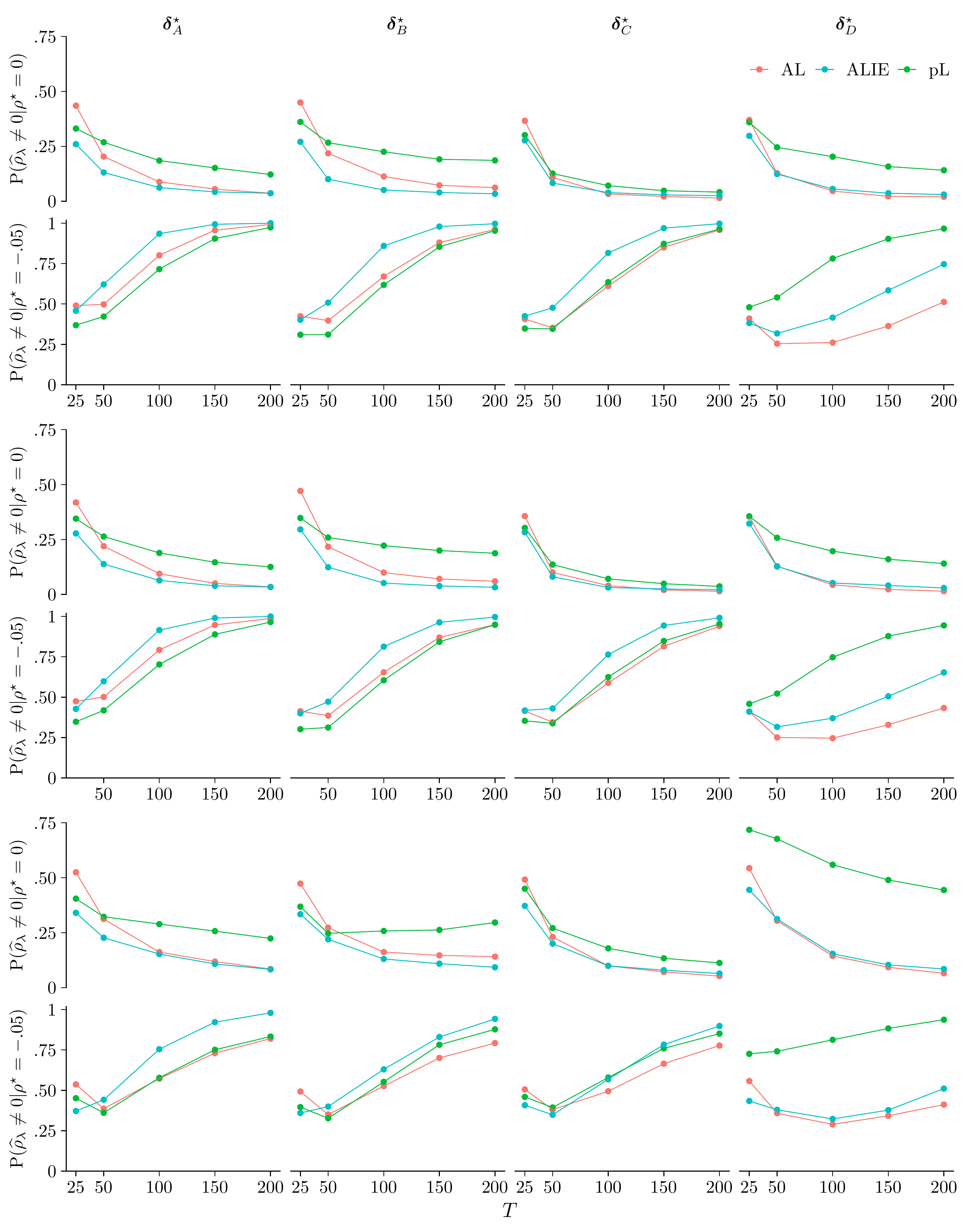}
    }
    \begin{minipage}{\textwidth}
        \vspace{.2cm}
        \scriptsize\textit{Notes:} DGP \eqref{eq:adfdgp}. $\vdelta_A^\star := (.4, .3, .2)'$, $\vdelta_B^\star := (.4, .3, .2, 0, 0, 0, -.2, 0, 0, .2)'$, $\vdelta_C^\star := .7$, $\vdelta_D^\star:= (-.4, 0, .7)'$. Model \eqref{eq:adfreg} with $p=\lfloor 12 \cdot (T/100)^{.25}\rfloor$. AL and ALIE are adaptive Lasso estimators based on $w_1$ and $\Breve{w}_1$, respectively. PL is the plain Lasso. Top panel: no adjustment. Middle panel: FD-demeaned data. Bottom panel: FD-detrended data. $\Breve{w_1}$ computed as discussed in \Cref{sec:sbwe,sec:hdt} with LRV estimate $\widehat\omega^2_{\textup{AR}}(k)$. $k$ selected by BIC with $k_{\max} = p$. 5000 Monte Carlo replications.
    \end{minipage}
\end{figure}
\begin{table}[t]
\centering
\caption{Classification metrics of Lasso estimators for sparse processes}
\label{tab:cmetrics}
\vspace{.25cm}
\setlength{\tabcolsep}{12pt}
\renewcommand{\arraystretch}{.6}
\resizebox{\textwidth}{!}{
\begin{tabular}{lcd{3.2}d{3.2}cd{3.2}d{3.2}cd{3.2}d{3.2}}
\toprule
\multicolumn{1}{c}{ } & \multicolumn{1}{c}{ } & \multicolumn{2}{c}{PL} & \multicolumn{1}{c}{ } & \multicolumn{2}{c}{AL} & \multicolumn{1}{c}{ } & \multicolumn{2}{c}{ALIE} \\
\cmidrule(l{3pt}r{3pt}){3-4} \cmidrule(l{3pt}r{3pt}){6-7} \cmidrule(l{3pt}r{3pt}){9-10}
  & $\vdelta^\star$ & \multicolumn{1}{c}{PPV} & \multicolumn{1}{c}{NPV} &   & \multicolumn{1}{c}{PPV} & \multicolumn{1}{c}{NPV} &   & \multicolumn{1}{c}{PPV} & \multicolumn{1}{c}{NPV}\\
\midrule
\addlinespace[.2cm]
\multicolumn{10}{l}{no adjustment}\\
\addlinespace[.2cm] & $\vdelta_A^\star$ & .79 & .74 & \ & .90 & .82 & \ & .94 & .93\\
 & $\vdelta_B^\star$ & .73 & .67 & \ & .86 & .73 & \ & .94 & .87\\
 & $\vdelta_C^\star$ & .90 & .72 & \ & .95 & .71 & \ & .95 & .84\\
 & $\vdelta_D^\star$ & .79 & .78 & \ & .85 & .56 & \ & .88 & .62\\
\addlinespace[.2cm]
\multicolumn{10}{l}{constant}\\
\addlinespace[.2cm] & $\vdelta_A^\star$ & .79 & .73 & \ & .89 & .81 & \ & .93 & .92\\
 & $\vdelta_B^\star$ & .73 & .66 & \ & .87 & .72 & \ & .94 & .83\\
 & $\vdelta_C^\star$ & .90 & .71 & \ & .94 & .70 & \ & .96 & .80\\
 & $\vdelta_D^\star$ & .79 & .76 & \ & .85 & .56 & \ & .88 & .60\\
\addlinespace[.2cm]
\multicolumn{10}{l}{linear trend}\\
\addlinespace[.2cm] & $\vdelta_A^\star$ & .67 & .63 & \ & .78 & .66 & \ & .83 & .78\\
 & $\vdelta_B^\star$ & .68 & .62 & \ & .76 & .64 & \ & .83 & .70\\
 & $\vdelta_C^\star$ & .76 & .66 & \ & .83 & .64 & \ & .85 & .68\\
 & $\vdelta_D^\star$ & .59 & .70 & \ & .67 & .55 & \ & .68 & .56\\
\bottomrule
\end{tabular}
}
\begin{minipage}{\textwidth}
    \vspace{.2cm}
        \scriptsize\textit{Notes:} See \Cref{fig:sp_rates} for the DGP and the model. $T=100$. PPV and NPV denote the positive and negative predictive values. Adjustments are done using FD-detrending.
    \end{minipage}
\end{table}

\Cref{fig:sp_rates} shows activation rates of $y_{t-1}$ without deterministic components (top panel), for FD demeaning (middle panel) and for FD detrending (bottom panel). Spurious activation rates for $\rho^\star = 0$ (first sub-panel) indicate deficiencies for AL in small samples and across most settings for $\vdelta^\star$. For instance, at $T=25$ observations, the activation probability of $y_{t-1}$ in settings $\vdelta_A^\star$ and $\vdelta_B^\star$ exceeds 40\%, whereas ALIE achieves considerably lower rates. Similar outcomes are found for demeaned and detrended data. Although the discrepancy between ALIE and AL diminishes in $T$, these results indicate an advantage of $\Breve{w}_1$ over $w_1$ in finite samples.

Results for $\rho^\star=-.05$ (second sub-panel) corroborate favourable finite sample implications of our theoretical results in \Cref{sec:ctuning} under stationary. ALIE's activation rates for $y_{t-1}$ are above AL's across most settings. For example, ALIE has power advantages of up to 20\% for sample sizes between $T=100$ and $T=150$ in setting $\vdelta_B^\star$. Outcomes are qualitatively similar when adjusting for deterministic components. However, all methods suffer from reduced power under detrending, where gains from information enrichment are smaller. This is most pronounced in scenario $\vdelta_D^\star$ where none of the methods dominates.

\Cref{fig:sp_rates} also indicates the mediocre performance of the BIC-tuned plain Lasso in most settings: PL often cannot compete with AL, let alone ALIE in power, despite comparable activation rates under non-stationarity. At worst, the inconsistency of the plain Lasso entails erratic behaviour, cf. setting $\vdelta_B^\star$ for $\rho^\star = 0$ where rejection rates display upward surges or plateau as $T$ increases. 

Further insight into the methods' suitability as classifiers for stationary and non-stationary models is given in \Cref{tab:cmetrics}. Here, we report the positive predictive value (PPV) and the negative predicted value (NPV),
\begin{align*}
    \textup{PPV} = \frac{\sum_{s\in\mathcal{S}} \mathbb{I}\left(\widehat{\rho}^{(s)}_\lambda\neq0 \big\vert \rho^\star\in(-2,0)\right)}{\sum_{s\in\mathcal{S}} \mathbb{I}\left(\widehat{\rho}^{(s)}_\lambda\neq0 \big\vert \rho^\star\in(-2,0]\right)}, \quad \textup{NPV} = \frac{\sum_{s\in\mathcal{S}} \mathbb{I}\left(\widehat{\rho}^{(s)}_\lambda=0 \big\vert \rho^\star=0\right)}{\sum_{s\in\mathcal{S}} \mathbb{I}\left(\widehat{\rho}^{(s)}_\lambda=0 \big\vert \rho^\star\in(-2,0]\right)},
\end{align*}
based on pooled simulation results $\mathcal{S}$ for $\rho^\star\in(-2,0]$ for $T=100$.\footnote{We consider $\widehat{\rho}_\lambda<0$ a \enquote{positive}.} The comparison indicates that ALIE is the superior classifier. Especially for NPV, ALIE dominates the other methods. Regarding PPV, ALIE scores at least as high as AL and is always ahead of PL.

Extended results on model selection capabilities of AL and ALIE are presented in Tables \ref{tab:sp_c_extended_size}--\ref{tab:sp_ct_extended_power}. These report correct ($\widehat{\mathcal{J}} = \mathcal{J}$) and conservative ($\mathcal{J}\in\widehat{\mathcal{J}}$) selection of the lag pattern, correct model selection ($\widehat{\mathcal{M}} = \mathcal{M}$), as well as log-scale medians of $\lambda_{0,\,\rho^\star}$ and $\Breve{\lambda}_{0,\,\rho^\star}$. The outcomes largely align with our theoretical results in \Cref{sec:ctuning}. Discrepancies in medians of the $\lambda_{0\,\rho^\star}$ indicate potential for improved selection of $y_{t-1}$ by ALIE. The finite sample impact of $\Breve{w}_1$ on the selection of the $\Delta y_{t-j}$ is minor under non-stationarity, since $\mathrm{P}(\widehat{\mathcal{J}} = \mathcal{J})$ and $\mathrm{P}(\mathcal{J}\in\widehat{\mathcal{J}})$ are often similar for both procedures. However, ALIE is more likely to select the correct lag structure in stationary regressions in larger samples. Moreover, we find evidence that BIC tuning of ALIE achieves correct model selection and that ALIE often has higher $\mathrm{P}(\widehat{\mathcal{M}} = \mathcal{M})$ than AL.\footnote{Notice that $\lvert\vdelta_B^\star\rvert>p = k_{\max}$ if $T\in(0, 50)$ due to Schwert's rule so that selection of the true model is infeasible in this setting.} 

\subsection{ARMA processes}
\label{sec:asparsity}

\begin{figure}[htbp]
    \centering
    \caption{Activation rates of Lasso estimators for MA errors}
    \label{fig:arimarates}
    \vspace{.25cm}
    \resizebox{\textwidth}{!}{
    \includegraphics{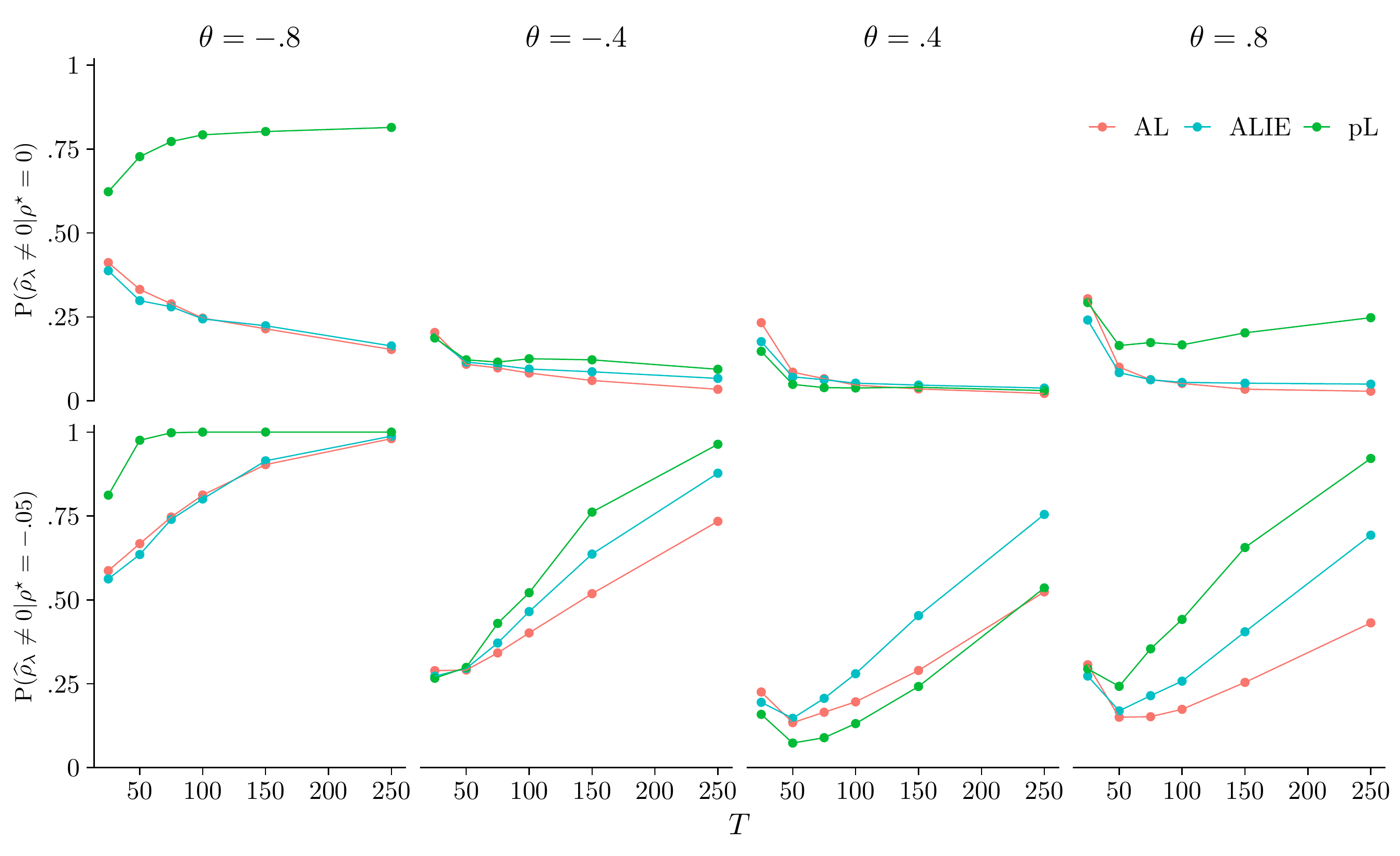}
    }
    \begin{minipage}{\textwidth}
        \vspace{.2cm}
        \scriptsize\textit{Notes:} DGP \eqref{eq:thedgp} with $\vpsi=0$ and $u_t$ as in \eqref{eq:maerrors}. Model \eqref{eq:adfreg} with $d_t=0$ and $p=\lfloor 12 \cdot (T/100)^{.25}\rfloor$. AL and ALIE are adaptive Lasso estimators based on $w_1$ and $\Breve{w}_1$, respectively. PL is the plain Lasso. $J_{\alpha}$ with $\alpha=.1$ computed with LRV estimate $\widehat\omega^2_{\textup{AR}}(k)$, $k$ selected by MAIC with $k_{\max} = p$. 5000 Monte Carlo replications.
    \end{minipage}
\end{figure}

To investigate the effect of information enrichment when $u_t$ satisfies \Cref{assum:lperrors} but not the stronger condition of \Cref{assum:sparsity}, we consider a small simulation study for moving average errors. Although selecting the correct model is infeasible for $T<\infty$ in an ADF($p$) framework, it is interesting to study the implications of using $\Breve{w}_1$ given the empirical relevance of such processes. 

We generate $y_t$ as in \eqref{eq:thedgp} with $\varrho=1$ or $\varrho=.95$ and the errors $u_t$ follow a first-order moving average,
\begin{align}
    u_t = \varepsilon_t + \theta \varepsilon_{t-1}, \qquad \varepsilon_t\overset{i.i.d.}{\sim}N(0,1).
    \label{eq:maerrors}
\end{align}
We consider $\theta\in\{-.8, -.4, .4, .8\}$ and sample sizes $T\in\{25,50,100,150,250, 500\}$.

The results are shown in \Cref{fig:arimarates}. Error processes with large negative MA coefficients are particularly challenging for all methods in non-stationary models, which is reminiscent of the large size distortions reported for standard unit root tests under these conditions, cf. \textcite{NgPerron2001}. Overall, we find the discrepancy in AL and ALIE activation rates for $\rho^\star=0$ to be small. ALIE has some advantages in small samples occasionally, but activation rates of AL improve quickly with $T$ and are slightly superior to ALIE in large samples. Depending on the DGP, the plain Lasso may exhibit a strong tendency to activate $y_{t-1}$ spuriously. Especially for MA coefficients with large magnitude, this issue does not ease and may even exacerbate with $T$.

For $\rho^\star = -.05$, we find that ALIE can be considerably more powerful than AL. An exception is $\theta=-.8$, where both methods perform equally well. Again, we find PL to be inferior due to its inconsistency. High detection rates of stationary models are associated with a tendency to misclassify non-stationary data. PL thus is significantly less reliable than its adaptive competitors.

\section{Empirical application}
\label{sec:empapp}

The recent turmoil in the energy sector due to the Russian invasion of Ukraine and prevailing market frictions from the COVID pandemic have thrust price inflation into the forefront of public and scientific discussions in a way not witnessed in decades. It is broadly recognised that accurate projections of key variables are essential for the success of monetary policy-making. For monetary targeting, inflation forecasts rank among the most critical measures. Workhorse tools for medium-term inflation forecasting include structural macroeconomic models built on Phillips curves. While these models typically entail a forward-looking component summarising survey-based inflation expectations, they also incorporate a stochastic part for sufficiently capturing the dynamics over time. Model selection for this time series component usually includes testing for a unit root and lag truncation.
\begin{figure}[htbp]
    \centering
    \caption{Quarterly German consumer price indices and inflation rates}
    \vspace{.25cm}
    \label{fig:cpiinf}
    \resizebox{\textwidth}{!}{
    \includegraphics{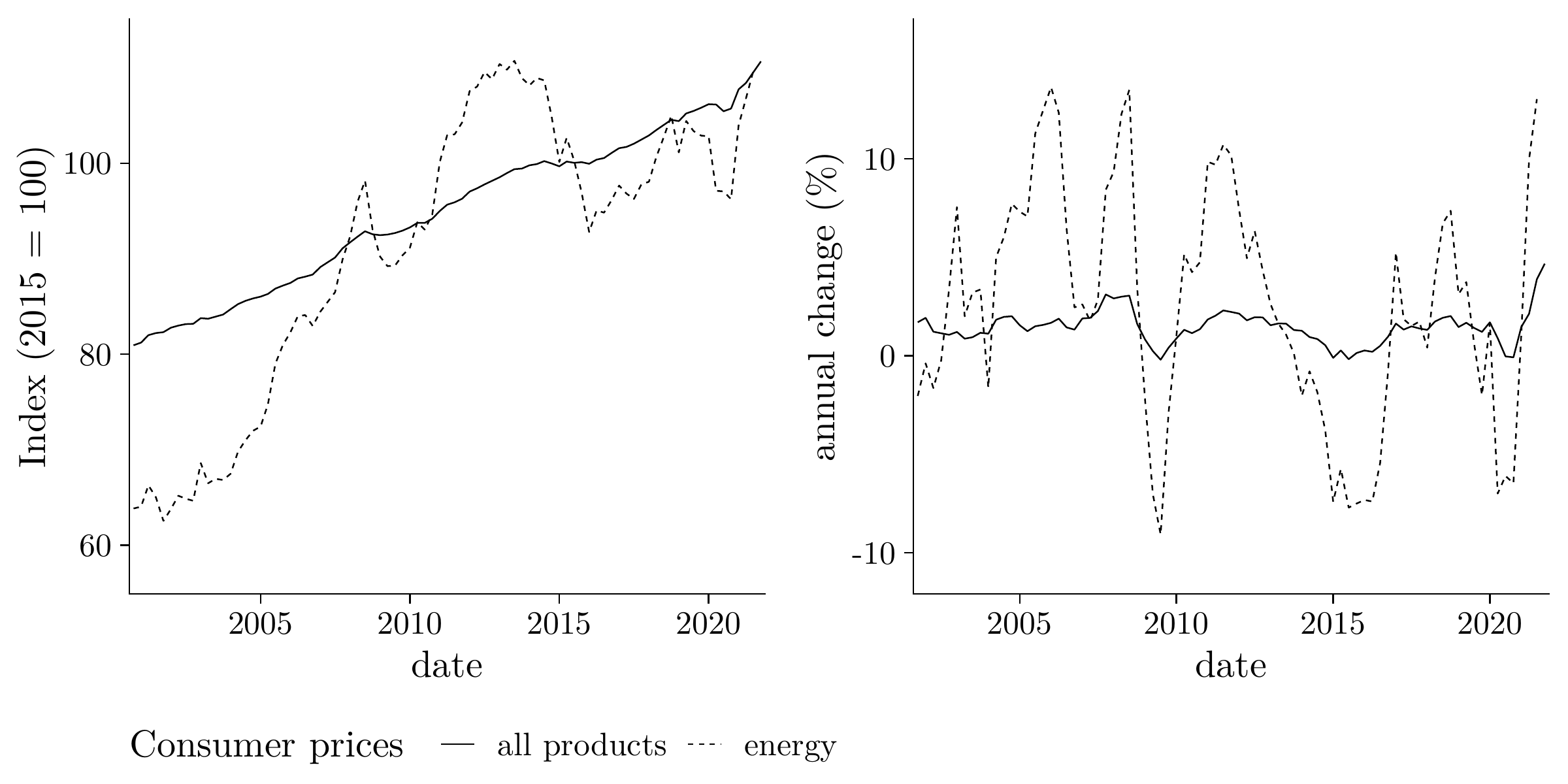}
    }
    \begin{minipage}{\textwidth}
        \vspace{.25cm}
        \scriptsize\textit{Notes:} Left: seasonally adjusted quarterly German consumer price indices for all products (\cite{CPIall2022}) and energy commodities (\cite{CPIenrg2022}) including fuel, electricity, and gasoline from 2001-Q4 to 2021-Q4. The base year is 2015. Right: quarterly year-on-year inflation rates.
    \end{minipage}
\end{figure}
\begin{table}[htbp]
\centering
\caption{Model selection outcomes for German price inflation rates}
\label{tab:infoutcomes}
\vspace{.25cm}
\setlength{\tabcolsep}{6pt}
\renewcommand{\arraystretch}{1.1}
\resizebox{\textwidth}{!}{
\begin{tabular}{lcc d{3.2} cccc  d{3.2} cc}
\toprule
& \multicolumn{4}{c}{all products} & & \multicolumn{4}{c}{energy commodities} &\\
\cmidrule(l{0pt}r{0pt}){2-5} \cmidrule(l{0pt}r{0pt}){7-10}
procedure & IC & lags & $t$ & $\pi_{t-1}$ & & IC & lags & $t$ & $\pi_{t-1}$ &\\
\hline
\multirow{2}{*}{ADF} & $\text{BIC}^\star$ & 1, 4, 8 & -.47 & exclude & & $\text{BIC}^\star$ & 1, 4, 8 & -.82 & exclude &\\
                     & AIC                & 1--8 & -1.78 & exclude & & AIC & 1--11 & -2.61 & exclude &\\
\\
\multirow{2}{*}{DFQD} & $\text{BIC}^\star$ & 1, 4, 8 & -.66 & exclude & & $\text{BIC}^\star$ & 1, 4, 8 & -.66 & exclude &\\
                      & MAIC               & -- & -1.41 & exclude & & MAIC & 1--8 & -1.40 & exclude &\\
\\
PL & & 1, 4, 8 & & exclude & & & 1, 4, 6, 8 & & exclude &\\
AL & & 1, 4, 8 & & exclude & & & 1, 4, 8 & & exclude &\\
ALIE  & BIC ($J_\alpha$) & 1, 4, 8 & & \textit{\textbf{include}} & & BIC ($J_\alpha$) & 1, 4, 8 & & \textit{\textbf{include}} &\\
ALIE  & $p_\textup{max}$ ($J_\alpha$) & 1, 4, 8 & & \textit{\textbf{include}} & & $p_\textup{max}$ ($J_\alpha$) & 1, 4, 8 & & \textit{\textbf{include}} &\\
ALIE  & MAIC ($J_\alpha$) & 1, 4, 8 & & \textit{\textbf{include}} & &  MAIC ($J_\alpha$) & 1, 4, 8 & & exclude &\\
\bottomrule
\addlinespace[.3cm]
\end{tabular}
}
\begin{minipage}{\textwidth}
        \vspace{.25cm}
        \scriptsize\textit{Notes:} Model selection results for \eqref{eq:infreg}. Lasso estimators are computed using FD-demeaned data and for $p=p_\textup{max}=\lfloor12\cdot(100/T)^{.25}\rfloor=11$ with $\lambda$ tuned by BIC. IC reports the criterion for selecting the lag pattern/truncation lag of the unpenalised procedures. $\text{BIC}^\star$ denotes exhaustive lag pattern selection with BIC. For ALIE, we report the criterion used in the LRV estimation for $J_\alpha$. $5\%$ critical values used for ADF and DFQD are $-2.89$ and $-1.95$, respectively.
    \end{minipage}
\end{table}

We apply the penalised estimators considered in this article to German seasonally adjusted consumer price index (CPI) data, performing model selection for energy commodities and all-product inflation rates (\cite{CPIenrg2022, CPIall2022}). The dataset encompasses 81 quarterly CPI observations from 2001-Q4 to 2021-Q4, beginning with the introduction of the Euro and ending before the markets were shocked by the war in Ukraine.\footnote{We restrict ourselves to this period to mitigate the effects of stylised features of long macroeconomic time series, such as structural change in the first and second moments.} We consider variable selection in the ADF regression
\begin{align}
    \Delta\pi_t = \mu + \rho\pi_{t-1} + \sum_{j=1}^p \delta_j \Delta\pi_{t-j} + e_t. \label{eq:infreg}
\end{align}
\Cref{eq:infreg} includes a constant since the ECB monetary policy strives to maintain the inflation rate $\pi_t$ in a corridor around a positive target, implying a stationary process with a non-zero mean.

Besides model selection using the shrinkage estimators, we apply information criteria for lag selection in conjunction with the classical ADF and quasi-difference demeaned ADF tests (DFQD) of \textcite{Elliottetal1996}. For the OLS-estimated models, we consider an exhaustive search for a sparsity pattern using BIC (denoted $\text{BIC}^\star$) as well as an estimation of the truncation lag (by AIC/MAIC). As before, $\lambda$ for PL, AL, and ALIE is tuned using BIC. For ALIE, we use $\Breve{w}_1$ with $J_\alpha$ computed based on $\widehat\omega^2_\textup{AR}(k)$ for $k$ selected by BIC, MAIC, or $k=11$ according to the \textcite{Schwert1989} rule, which is the maximum lag order ($p_{\max}$) considered across all procedures.

The outcomes are shown in Table~\ref{tab:infoutcomes}. For both inflation rates, the consistent $\text{BIC}^\star$ procedure estimates the lag pattern $\widehat{\mathcal{J}}=\{1, 4, 8\}$ which indicates short-run dependence on the seasonal lags $j=4$ and $j=8$. Reassuringly, the Lasso estimators select the same lag pattern as $\text{BIC}^\star$, except for PL for energy commodities. None of the classical tests reject a non-stationary model at the 5\% level for either series. ALIE classifies the all-products inflation rate as stationary for all three specifications of $\widehat\omega_\textup{AR}^2(k)$. It also includes $\pi_{t-1}$ in the energy inflation rate model for LRV estimates based on BIC and $p_\textup{max}$. PL and AL classify both series as non-stationary.

\section{Conclusion}
\label{sec:conc}
Previous research on penalised estimation has considered single identification principles to elicit data-dependent penalty weights. A prominent example is the adaptive Lasso, for which \textcite{zou2006adaptive} recommends using consistent coefficient estimates. While oracle properties grant asymptotic equivalence between such adaptive procedures and their non-penalised counterparts under appropriate conditions, simulation studies often point to mediocre finite sample performance of ad-hoc implementations. 

In this article, we proposed combining two identification principles in computing penalty weights for consistent and oracle-efficient estimation using the adaptive Lasso. With ADF regressions being our use case, we propose to enrich a consistent coefficient estimate with additional information from a (simulation-based) statistic that exploits different stochastic orders of stationary and non-stationary data. We established that the enhanced weight promotes a \emph{zero shrinkage} property and ensures \emph{perpetual activation} of the corresponding regressor in stationary models as $T\to\infty$. In particular, we highlighted that beneficial consequences of these features for inference by consistent selection arise from an improved adaptive sorting of the activation knots on the Lasso solution path. We have further extended the theoretical analysis of the asymptotic properties of the adaptive Lasso for the AR(1) models examined in \textcite{kock2016consistent} to general ADF($p$) and ARMA models, even allowing $p\to\infty$ beyond the (approximate) sparsity assumption.

Simulation evidence supports our theoretical results, showing that the information-enriched ALIE  estimator dominates the adaptive Lasso and is significantly more reliable than the (possibly inconsistent) plain Lasso regarding both the selection of $y_{t-1}$ and recovering the entire model. Additionally, ALIE improves the classification of stationary and non-stationary models in finite samples under detrending, which appears to be a weakness common to all Lasso variants compared to unpenalised regression-based tests. Abstracting from ADF regressions and as another novelty, we showcase the ability of information enrichment to combine hypothesis tests without type I error accumulation or, under consistent tuning, even with shrinking size. In contrast to most popular combination methods, ALIE does not require standardised evidence, e.g., p-values. Hence, ALIE can be considered a testbed for combining different estimators without establishing asymptotic distributions.

We apply the information-enriched estimator in a model selection exercise for German consumer price data from 2001-Q4 to 2021-Q4. The results indicate that a stationary ADF model is better suited to model the headline inflation rate than a non-stationary autoregression. We also find evidence supporting a stationary model for the German energy commodity inflation rate.

There are various avenues for further research: we consider it relevant to investigate whether the performance can be further improved by alternative specifications of the adaptive weight $\Breve{w}$. A first idea would be to swap the simulation-based statistic $J_\alpha$ with a different statistic. Our theory suggests benefits from adding further weakly correlated estimators to the computation of $\Breve{w}$. Second, since our method focuses exclusively on the weight for $y_{t-1}$, another path would be to elicit improved weights for the lags of $\Delta y_t$ to enhance further the classification rates concerning model selection. In this regard, it would also be worthwhile to investigate adaptive Lasso estimators that include deterministic components, perhaps in a flexible fashion as suggested in \textcite{tibshirani2011solution} and compare them with detrending approaches.

Another aspect yet to be investigated is the benefit of using ALIE for forecast averaging. Due to its superior selection power under consistent tuning, ALIE could help sort out bad predictors better than adaptive Lasso or comparable procedures can by drawing on other forecasts of the same predictor. In more general terms, it would also be instructive to analyse ALIE's performance under conservative model selection and to benchmark against pre-test or averaging estimators as suggested by \textcite{Hansen2007, Hansen2010}. We consider it also promising to extend our approach to multivariate models, e.g., for penalised estimation of the cointegration rank in VAR models as in the framework of \textcite{LiaoPhillips2015}.

Certain violations of our assumptions, e.g., smooth transitions in the (unconditional) variance, have received much attention in the unit root test literature during the last two decades as features of economic time series. Studying the behaviour of our approach in such a framework and making it adaptive would be of interest to a broad range of applications in empirical macroeconomic research. One opportunity is to devise a procedure robust to non-stationary volatility by employing scaled estimators based on non-parametric estimation of the series' variance profile as in \textcite{Beare2018}.

Finally, tuning the adaptive Lasso to detect local alternatives while exhibiting reliable model selection properties is an unresolved problem that is particularly challenging in the context of correlated regressors. We are currently investigating this issue.\\

%
\newpage 
\appendix
\addcontentsline{toc}{section}{Appendix}
\renewcommand{\thesubsection}{\Alph{subsection}} 
\renewcommand{\theequation}{\thesubsection\arabic{equation}} 
\setcounter{table}{0}
\setcounter{figure}{0}
\renewcommand{\thetable}{\thesubsection\arabic{table}}
\renewcommand{\thefigure}{\thesubsection\arabic{figure}}
%
\subsection{Additional Monte Carlo results}
\label[appendix]{sec:amcr}

\begin{table}[h!]
\centering
\caption{Model selection outcomes for non-stationary autoregressions \newline (no adjustment)}
\label{tab:sp_nc_extended_size}
\vspace{.25cm}
\setlength{\tabcolsep}{2pt}
\renewcommand{\arraystretch}{1.2}
\resizebox{\textwidth}{!}{
    
\centering
\begin{tabular}{l *{5}{S} c *{5}{S}}
\toprule
\multicolumn{1}{c}{ } & \multicolumn{5}{c}{AL} & \multicolumn{1}{c}{ } & \multicolumn{5}{c}{ALIE} \\
\cmidrule(l{3pt}r{3pt}){2-6} \cmidrule(l{3pt}r{3pt}){8-12}
$T$ & {$\Prob{\widehat{\rho}_\lambda\neq0}$} & {$\Prob{\widehat{\mathcal{J}} = \mathcal{J}}$} & {$\Prob{\mathcal{J}\in\widehat{\mathcal{J}}}$} & {$\Prob{\widehat{\mathcal{M}} = \mathcal{M}}$} & {$\lambda_{0,\,\rho^\star}$} &   & {$\Prob{\widehat{\rho}_\lambda\neq0}$} & {$\Prob{\widehat{\mathcal{J}} = \mathcal{J}}$} & {$\Prob{\mathcal{J}\in\widehat{\mathcal{J}}}$} & {$\Prob{\widehat{\mathcal{M}} = \mathcal{M}}$} & {$\Breve{\lambda}_{0,\,\rho^\star}$}\\
\midrule
\addlinespace[.3cm]
\multicolumn{12}{l}{$\vdelta_A^\star$}\\
25 & 0.4352 & 0.0066 & 0.0540 & 0.0038 & 0.7490838 & \ & 0.2600 & 0.0108 & 0.0716 & 0.0070 & -1.303423\\
50 & 0.2034 & 0.0462 & 0.0968 & 0.0380 & -0.0442521 & \ & 0.1310 & 0.0502 & 0.1098 & 0.0426 & -1.167635\\
100 & 0.0882 & 0.1944 & 0.3192 & 0.1794 & -0.2981228 & \ & 0.0622 & 0.1910 & 0.3178 & 0.1774 & -1.274264\\
150 & 0.0554 & 0.3406 & 0.4860 & 0.3242 & -0.3253123 & \ & 0.0428 & 0.3530 & 0.4996 & 0.3352 & -1.269272\\
200 & 0.0366 & 0.4522 & 0.6180 & 0.4392 & -0.3725005 & \ & 0.0366 & 0.4528 & 0.6112 & 0.4376 & -1.204364\\
500 & 0.0112 & 0.7864 & 0.9364 & 0.7784 & -0.4104954 & \ & 0.0200 & 0.7894 & 0.9430 & 0.7742 & -1.267216\\
\addlinespace[.3cm]
\multicolumn{12}{l}{$\vdelta_B^\star$}\\
25 & 0.4494 &  &  &  & 0.7076118 & \ & 0.2706 &  &  &  & -1.286534\\
50 & 0.2180 & 0.0082 & 0.0176 & 0.0070 & -0.0991647 & \ & 0.1006 & 0.0092 & 0.0230 & 0.0070 & -1.445803\\
100 & 0.1128 & 0.0348 & 0.0778 & 0.0284 & -0.1977521 & \ & 0.0514 & 0.0456 & 0.0878 & 0.0412 & -1.486709\\
150 & 0.0728 & 0.0954 & 0.1730 & 0.0844 & -0.2632591 & \ & 0.0400 & 0.0892 & 0.1816 & 0.0822 & -1.517475\\
200 & 0.0620 & 0.1578 & 0.2994 & 0.1442 & -0.3406312 & \ & 0.0340 & 0.1656 & 0.3098 & 0.1566 & -1.466582\\
500 & 0.0344 & 0.4654 & 0.8514 & 0.4516 & -0.4142805 & \ & 0.0432 & 0.4788 & 0.8702 & 0.4588 & -1.459265\\
\addlinespace[.3cm]
\multicolumn{12}{l}{$\vdelta_C^\star$}\\
25 & 0.3662 & 0.2886 & 0.7132 & 0.2498 & 0.3385810 & \ & 0.2770 & 0.3152 & 0.7558 & 0.2780 & -0.952846\\
50 & 0.1084 & 0.6660 & 0.9772 & 0.6254 & -0.3169525 & \ & 0.0828 & 0.6732 & 0.9820 & 0.6408 & -1.079834\\
100 & 0.0334 & 0.8194 & 0.9998 & 0.8020 & -0.4250211 & \ & 0.0396 & 0.8190 & 1.0000 & 0.7924 & -1.241122\\
150 & 0.0218 & 0.8716 & 1.0000 & 0.8606 & -0.3965683 & \ & 0.0292 & 0.8730 & 1.0000 & 0.8530 & -1.314045\\
200 & 0.0148 & 0.8866 & 1.0000 & 0.8786 & -0.4117321 & \ & 0.0254 & 0.8854 & 1.0000 & 0.8684 & -1.299121\\
500 & 0.0070 & 0.9242 & 1.0000 & 0.9192 & -0.5040257 & \ & 0.0106 & 0.9282 & 1.0000 & 0.9194 & -1.396408\\
\addlinespace[.3cm]
\multicolumn{12}{l}{$\vdelta_D^\star$}\\
25 & 0.3700 & 0.1316 & 0.4182 & 0.1100 & 0.1757619 & \ & 0.2976 & 0.1374 & 0.4250 & 0.1206 & -1.002475\\
50 & 0.1276 & 0.4574 & 0.8126 & 0.4282 & -0.2936652 & \ & 0.1240 & 0.4500 & 0.8236 & 0.4180 & -1.152889\\
100 & 0.0464 & 0.7120 & 0.9818 & 0.6926 & -0.3829872 & \ & 0.0562 & 0.7152 & 0.9822 & 0.6862 & -1.279200\\
150 & 0.0224 & 0.8122 & 0.9986 & 0.7996 & -0.4497613 & \ & 0.0368 & 0.8106 & 0.9984 & 0.7844 & -1.298826\\
200 & 0.0202 & 0.8486 & 0.9998 & 0.8372 & -0.5333925 & \ & 0.0306 & 0.8570 & 0.9998 & 0.8338 & -1.302318\\
500 & 0.0066 & 0.9246 & 1.0000 & 0.9194 & -0.4670475 & \ & 0.0134 & 0.9264 & 1.0000 & 0.9146 & -1.300466\\
\bottomrule
\end{tabular}

}
\begin{minipage}{\textwidth}
    \vspace{.25cm}
    \scriptsize\textit{Notes:} DGP \eqref{eq:adfdgp}, $\rho^\star = 0$, $\vdelta_A^\star := (.4, .3, .2)'$, $\vdelta_B^\star := (.4, .3, .2, 0, 0, 0, -.2, 0, 0, .2)'$, $\vdelta_C^\star := .7$, $\vdelta_D^\star:= (-.4, 0, .7)'$. Model \eqref{eq:adfreg} with $p=\lfloor12(T/100)^{.25}\rfloor$. $J_{\alpha}$ computed with LRV estimate $\widehat{\omega}^2_{\textup{AR}}(k)$, $k$ selected with BIC with $k_{\max}=p$, $\alpha = .1$, and $\sigma_\nu=1$. Probability statements denote inclusion of $y_{t-1}$ ($\widehat\rho_\lambda\neq0$), correct ($\widehat{\mathcal{J}} = \mathcal{J}$) and conservative ($\mathcal{J}\in\widehat{\mathcal{J}}$) estimation of the lag pattern, and correct model selection ($\widehat{\mathcal{M}} = \mathcal{M}$). $\lambda_{0,\,\rho^\star}$ and $\Breve{\lambda}_{0,\,\rho^\star}$ are log scale median activation thresholds of $y_{t-1}$. 5000 replications.
\end{minipage}
\end{table}

\begin{table}[htbp]
\centering
\caption{Model selection outcomes for stationary autoregressions \newline (no adjustment)}
\label{tab:sp_nc_extended_power}
\vspace{.25cm}
\setlength{\tabcolsep}{1.5pt}
\renewcommand{\arraystretch}{1.2}
\resizebox{\textwidth}{!}{
	
\centering
\begin{tabular}{l *{5}{S} c *{5}{S}}
\toprule
\multicolumn{1}{c}{ } & \multicolumn{5}{c}{AL} & \multicolumn{1}{c}{ } & \multicolumn{5}{c}{ALIE} \\
\cmidrule(l{3pt}r{3pt}){2-6} \cmidrule(l{3pt}r{3pt}){8-12}
$T$ & {$\Prob{\widehat{\rho}_\lambda\neq0}$} & {$\Prob{\widehat{\mathcal{J}} = \mathcal{J}}$} & {$\Prob{\mathcal{J}\in\widehat{\mathcal{J}}}$} & {$\Prob{\widehat{\mathcal{M}} = \mathcal{M}}$} & {$\lambda_{0,\,\rho^\star}$} &   & {$\Prob{\widehat{\rho}_\lambda\neq0}$} & {$\Prob{\widehat{\mathcal{J}} = \mathcal{J}}$} & {$\Prob{\mathcal{J}\in\widehat{\mathcal{J}}}$} & {$\Prob{\widehat{\mathcal{M}} = \mathcal{M}}$} & {$\Breve{\lambda}_{0,\,\rho^\star}$}\\
\midrule
\addlinespace[.3cm]
\multicolumn{12}{l}{$\vdelta_A^\star$}\\
25 & 0.4904 & 0.0092 & 0.0836 & 0.0076 & 0.9700149 & \ & 0.4576 & 0.0114 & 0.0896 & 0.0098 & 0.3711118\\
50 & 0.4976 & 0.0578 & 0.1438 & 0.0546 & 1.3583955 & \ & 0.6210 & 0.0624 & 0.1458 & 0.0618 & 2.0740044\\
100 & 0.8010 & 0.1856 & 0.3840 & 0.1822 & 2.2510463 & \ & 0.9348 & 0.2216 & 0.3674 & 0.2208 & 3.5850660\\
150 & 0.9568 & 0.3194 & 0.5678 & 0.3190 & 2.7790056 & \ & 0.9932 & 0.3800 & 0.5416 & 0.3796 & 4.4285766\\
200 & 0.9918 & 0.4182 & 0.6908 & 0.4180 & 3.1042526 & \ & 0.9998 & 0.5024 & 0.6700 & 0.5024 & 4.9966691\\
500 & 1.0000 & 0.6922 & 0.9520 & 0.6922 & 4.1019323 & \ & 1.0000 & 0.7898 & 0.9456 & 0.7898 & 7.0040502\\
\addlinespace[.3cm]
\multicolumn{12}{l}{$\vdelta_B^\star$}\\
25 & 0.4242 &  &  &  & 0.5922924 & \ & 0.4022 &  &  &  & -0.0180465\\
50 & 0.3972 & 0.0096 & 0.0218 & 0.0092 & 1.0974317 & \ & 0.5082 & 0.0086 & 0.0242 & 0.0078 & 1.5175985\\
100 & 0.6702 & 0.0370 & 0.0954 & 0.0360 & 1.9524890 & \ & 0.8594 & 0.0442 & 0.0904 & 0.0440 & 3.0260778\\
150 & 0.8798 & 0.1026 & 0.2292 & 0.1024 & 2.4587803 & \ & 0.9790 & 0.1000 & 0.2072 & 0.1000 & 3.7762423\\
200 & 0.9612 & 0.1712 & 0.3822 & 0.1710 & 2.7742577 & \ & 0.9966 & 0.1836 & 0.3470 & 0.1836 & 4.3447644\\
500 & 1.0000 & 0.4504 & 0.9064 & 0.4504 & 3.7547815 & \ & 1.0000 & 0.4954 & 0.9044 & 0.4954 & 6.3385619\\
\addlinespace[.3cm]
\multicolumn{12}{l}{$\vdelta_C^\star$}\\
25 & 0.4070 & 0.2830 & 0.7712 & 0.0762 & 0.6268666 & \ & 0.4248 & 0.2988 & 0.7792 & 0.1046 & 0.3735229\\
50 & 0.3524 & 0.5728 & 0.9774 & 0.1754 & 1.0650753 & \ & 0.4764 & 0.6152 & 0.9804 & 0.2864 & 1.5557239\\
100 & 0.6112 & 0.6446 & 1.0000 & 0.3820 & 1.9408717 & \ & 0.8158 & 0.7674 & 1.0000 & 0.6276 & 2.9126549\\
150 & 0.8498 & 0.6528 & 1.0000 & 0.5482 & 2.4113903 & \ & 0.9690 & 0.8266 & 1.0000 & 0.8032 & 3.6853470\\
200 & 0.9584 & 0.6858 & 1.0000 & 0.6580 & 2.7116763 & \ & 0.9970 & 0.8634 & 1.0000 & 0.8608 & 4.2259045\\
500 & 1.0000 & 0.8244 & 1.0000 & 0.8244 & 3.6414418 & \ & 1.0000 & 0.9252 & 1.0000 & 0.9252 & 5.8610932\\
\addlinespace[.3cm]
\multicolumn{12}{l}{$\vdelta_D^\star$}\\
25 & 0.4100 & 0.1314 & 0.4302 & 0.0280 & 0.4006397 & \ & 0.3816 & 0.1326 & 0.4280 & 0.0278 & -0.4216340\\
50 & 0.2544 & 0.4644 & 0.8200 & 0.0728 & 0.5631110 & \ & 0.3178 & 0.4600 & 0.8230 & 0.1022 & 0.5482951\\
100 & 0.2612 & 0.7130 & 0.9842 & 0.1356 & 1.2551262 & \ & 0.4158 & 0.7224 & 0.9878 & 0.2666 & 1.6925956\\
150 & 0.3636 & 0.7634 & 0.9978 & 0.2242 & 1.6673796 & \ & 0.5844 & 0.7954 & 0.9994 & 0.4334 & 2.4642687\\
200 & 0.5124 & 0.7832 & 1.0000 & 0.3478 & 1.9816930 & \ & 0.7462 & 0.8370 & 1.0000 & 0.6084 & 3.0512651\\
500 & 0.9602 & 0.7568 & 1.0000 & 0.7180 & 2.8823351 & \ & 0.9984 & 0.9186 & 1.0000 & 0.9170 & 5.1187065\\
\bottomrule
\end{tabular}

}
\begin{minipage}{\textwidth}
    \vspace{.25cm}
    \scriptsize\textit{Notes:} DGP \eqref{eq:adfdgp}, $\rho^\star = -.05$, $\vdelta_A^\star := (.4, .3, .2)'$, $\vdelta_B^\star := (.4, .3, .2, 0, 0, 0, -.2, 0, 0, .2)'$, $\vdelta_C^\star := .7$, $\vdelta_D^\star:= (-.4, 0, .7)'$. Model \eqref{eq:adfreg} with $p=\lfloor12(T/100)^{.25}\rfloor$. $J_{\alpha}$ computed with LRV estimate $\widehat{\omega}^2_{\textup{AR}}(k)$, $k$ selected with BIC with $k_{\max}=p$, $\alpha = .1$, and $\sigma_\nu=1$. Probability statements denote inclusion of $y_{t-1}$ ($\widehat\rho_\lambda\neq0$), correct ($\widehat{\mathcal{J}} = \mathcal{J}$) and conservative ($\mathcal{J}\in\widehat{\mathcal{J}}$) estimation of the lag pattern, and correct model selection ($\widehat{\mathcal{M}} = \mathcal{M}$). $\lambda_{0,\,\rho^\star}$ and $\Breve{\lambda}_{0,\,\rho^\star}$ are log scale median activation thresholds of $y_{t-1}$. 5000 replications.
\end{minipage}
\end{table}

\begin{table}[htbp]
\centering
\caption{Model selection outcomes for non-stationary autoregressions \newline (FD-demeaned data)}
\label{tab:sp_c_extended_size}
\vspace{.25cm}
\setlength{\tabcolsep}{1.5pt}
\renewcommand{\arraystretch}{1.2}
\resizebox{\textwidth}{!}{
    
\centering
\begin{tabular}{l *{5}{S} c *{5}{S}}
\toprule
\multicolumn{1}{c}{ } & \multicolumn{5}{c}{AL} & \multicolumn{1}{c}{ } & \multicolumn{5}{c}{ALIE} \\
\cmidrule(l{3pt}r{3pt}){2-6} \cmidrule(l{3pt}r{3pt}){8-12}
$T$ & {$\Prob{\widehat{\rho}_\lambda\neq0}$} & {$\Prob{\widehat{\mathcal{J}} = \mathcal{J}}$} & {$\Prob{\mathcal{J}\in\widehat{\mathcal{J}}}$} & {$\Prob{\widehat{\mathcal{M}} = \mathcal{M}}$} & {$\lambda_{0,\,\rho^\star}$} &   & {$\Prob{\widehat{\rho}_\lambda\neq0}$} & {$\Prob{\widehat{\mathcal{J}} = \mathcal{J}}$} & {$\Prob{\mathcal{J}\in\widehat{\mathcal{J}}}$} & {$\Prob{\widehat{\mathcal{M}} = \mathcal{M}}$} & {$\Breve{\lambda}_{0,\,\rho^\star}$}\\
\midrule
\addlinespace[.3cm]
\multicolumn{12}{l}{$\vdelta_A^\star$}\\
25 & 0.4186 & 0.0076 & 0.0550 & 0.0042 & 0.6948828 & \ & 0.2778 & 0.0082 & 0.0652 & 0.0058 & -1.3599946\\
50 & 0.2196 & 0.0410 & 0.1014 & 0.0348 & -0.0267899 & \ & 0.1384 & 0.0468 & 0.1088 & 0.0392 & -1.1813331\\
100 & 0.0944 & 0.2070 & 0.3222 & 0.1918 & -0.2426603 & \ & 0.0640 & 0.2046 & 0.3262 & 0.1892 & -1.1374359\\
150 & 0.0506 & 0.3486 & 0.4994 & 0.3338 & -0.3423778 & \ & 0.0394 & 0.3468 & 0.4968 & 0.3304 & -1.1319910\\
200 & 0.0344 & 0.4590 & 0.6174 & 0.4466 & -0.3461226 & \ & 0.0340 & 0.4642 & 0.6218 & 0.4488 & -0.9963676\\
500 & 0.0090 & 0.7962 & 0.9360 & 0.7908 & -0.4592232 & \ & 0.0164 & 0.7818 & 0.9324 & 0.7702 & -0.9954589\\
\addlinespace[.3cm]
\multicolumn{12}{l}{$\vdelta_B^\star$}\\
25 & 0.4710 &  &  &  & 0.7439751 & \ & 0.2960 &  &  &  & -1.0197017\\
50 & 0.2172 & 0.0064 & 0.0146 & 0.0052 & -0.0976818 & \ & 0.1240 & 0.0092 & 0.0252 & 0.0078 & -1.2799472\\
100 & 0.0996 & 0.0410 & 0.0802 & 0.0334 & -0.3155552 & \ & 0.0520 & 0.0382 & 0.0746 & 0.0356 & -1.3733493\\
150 & 0.0708 & 0.0854 & 0.1764 & 0.0760 & -0.3203564 & \ & 0.0388 & 0.0956 & 0.1834 & 0.0888 & -1.3784034\\
200 & 0.0598 & 0.1632 & 0.3006 & 0.1488 & -0.3662121 & \ & 0.0328 & 0.1558 & 0.3014 & 0.1474 & -1.3052866\\
500 & 0.0334 & 0.4754 & 0.8682 & 0.4624 & -0.4272349 & \ & 0.0320 & 0.4780 & 0.8678 & 0.4598 & -1.1252978\\
\addlinespace[.3cm]
\multicolumn{12}{l}{$\vdelta_C^\star$}\\
25 & 0.3566 & 0.3044 & 0.7342 & 0.2596 & 0.3237946 & \ & 0.2830 & 0.3204 & 0.7448 & 0.2820 & -0.6442515\\
50 & 0.1004 & 0.6764 & 0.9790 & 0.6386 & -0.2545404 & \ & 0.0808 & 0.6728 & 0.9840 & 0.6442 & -0.8272121\\
100 & 0.0402 & 0.8152 & 1.0000 & 0.7962 & -0.3841453 & \ & 0.0318 & 0.8188 & 1.0000 & 0.7994 & -0.9425469\\
150 & 0.0206 & 0.8690 & 1.0000 & 0.8584 & -0.4952371 & \ & 0.0252 & 0.8634 & 1.0000 & 0.8486 & -0.8891517\\
200 & 0.0144 & 0.8902 & 1.0000 & 0.8824 & -0.4611488 & \ & 0.0214 & 0.8964 & 1.0000 & 0.8818 & -1.0459619\\
500 & 0.0054 & 0.9264 & 1.0000 & 0.9222 & -0.5351062 & \ & 0.0080 & 0.9258 & 1.0000 & 0.9196 & -0.9980374\\
\addlinespace[.3cm]
\multicolumn{12}{l}{$\vdelta_D^\star$}\\
25 & 0.3546 & 0.1398 & 0.4174 & 0.1190 & 0.1959548 & \ & 0.3228 & 0.1232 & 0.4206 & 0.1036 & -0.6671215\\
50 & 0.1286 & 0.4626 & 0.8060 & 0.4354 & -0.2711460 & \ & 0.1274 & 0.4638 & 0.8210 & 0.4294 & -0.7706157\\
100 & 0.0440 & 0.7122 & 0.9818 & 0.6934 & -0.4350293 & \ & 0.0522 & 0.7142 & 0.9810 & 0.6854 & -0.8952212\\
150 & 0.0232 & 0.8108 & 0.9972 & 0.7986 & -0.4634514 & \ & 0.0410 & 0.8024 & 0.9990 & 0.7778 & -0.8930375\\
200 & 0.0150 & 0.8396 & 0.9996 & 0.8310 & -0.4780848 & \ & 0.0296 & 0.8488 & 0.9998 & 0.8290 & -0.9391260\\
500 & 0.0048 & 0.9216 & 1.0000 & 0.9186 & -0.4871292 & \ & 0.0142 & 0.9180 & 1.0000 & 0.9068 & -0.9432669\\
\bottomrule
\end{tabular}

}
\begin{minipage}{\textwidth}
    \vspace{.25cm}
    \scriptsize\textit{Notes:} DGP \eqref{eq:adfdgp}, $\rho^\star = 0$, $\vdelta_A^\star := (.4, .3, .2)'$, $\vdelta_B^\star := (.4, .3, .2, 0, 0, 0, -.2, 0, 0, .2)'$, $\vdelta_C^\star := .7$, $\vdelta_D^\star:= (-.4, 0, .7)'$. Model \eqref{eq:adfreg} with $p=\lfloor12(T/100)^{.25}\rfloor$. $J_{\alpha}$ computed with LRV estimate $\widehat{\omega}^2_{\textup{AR}}(k)$, $k$ selected with BIC with $k_{\max}=p$, $\alpha = .1$, and $\sigma_\nu=1$. Probability statements denote inclusion of $y_{t-1}$ ($\widehat\rho_\lambda\neq0$), correct ($\widehat{\mathcal{J}} = \mathcal{J}$) and conservative ($\mathcal{J}\in\widehat{\mathcal{J}}$) estimation of the lag pattern, and correct model selection ($\widehat{\mathcal{M}} = \mathcal{M}$). $\lambda_{0,\,\rho^\star}$ and $\Breve{\lambda}_{0,\,\rho^\star}$ are log scale median activation thresholds of $y_{t-1}$. 5000 replications.
\end{minipage}
\end{table}

\begin{table}[htbp]
\centering
\caption{Model selection outcomes for stationary autoregressions \newline (FD-demeaned data)}
\label{tab:sp_c_extended_power}
\vspace{.25cm}
\setlength{\tabcolsep}{1.5pt}
\renewcommand{\arraystretch}{1.2}
\resizebox{\textwidth}{!}{
	
\centering
\begin{tabular}{l *{5}{S} c *{5}{S}}
\toprule
\multicolumn{1}{c}{ } & \multicolumn{5}{c}{AL} & \multicolumn{1}{c}{ } & \multicolumn{5}{c}{ALIE} \\
\cmidrule(l{3pt}r{3pt}){2-6} \cmidrule(l{3pt}r{3pt}){8-12}
$T$ & {$\Prob{\widehat{\rho}_\lambda\neq0}$} & {$\Prob{\widehat{\mathcal{J}} = \mathcal{J}}$} & {$\Prob{\mathcal{J}\in\widehat{\mathcal{J}}}$} & {$\Prob{\widehat{\mathcal{M}} = \mathcal{M}}$} & {$\lambda_{0,\,\rho^\star}$} &   & {$\Prob{\widehat{\rho}_\lambda\neq0}$} & {$\Prob{\widehat{\mathcal{J}} = \mathcal{J}}$} & {$\Prob{\mathcal{J}\in\widehat{\mathcal{J}}}$} & {$\Prob{\widehat{\mathcal{M}} = \mathcal{M}}$} & {$\Breve{\lambda}_{0,\,\rho^\star}$}\\
\midrule
\addlinespace[.3cm]
\multicolumn{12}{l}{$\vdelta_A^\star$}\\
25 & 0.4744 & 0.0134 & 0.0834 & 0.0116 & 0.9598458 & \ & 0.4274 & 0.0114 & 0.0864 & 0.0098 & 0.3210179\\
50 & 0.5012 & 0.0560 & 0.1502 & 0.0524 & 1.3521547 & \ & 0.5980 & 0.0628 & 0.1430 & 0.0610 & 1.8865065\\
100 & 0.7922 & 0.1850 & 0.3896 & 0.1828 & 2.2241236 & \ & 0.9150 & 0.2290 & 0.3776 & 0.2276 & 3.3180499\\
150 & 0.9472 & 0.3048 & 0.5484 & 0.3038 & 2.7416290 & \ & 0.9898 & 0.3744 & 0.5430 & 0.3744 & 4.1360299\\
200 & 0.9870 & 0.3968 & 0.6916 & 0.3968 & 3.0698943 & \ & 0.9994 & 0.4852 & 0.6544 & 0.4852 & 4.6939009\\
500 & 1.0000 & 0.6650 & 0.9458 & 0.6650 & 4.0469473 & \ & 1.0000 & 0.7770 & 0.9394 & 0.7770 & 6.5596329\\
\addlinespace[.3cm]
\multicolumn{12}{l}{$\vdelta_B^\star$}\\
25 & 0.4132 &  &  &  & 0.5737328 & \ & 0.3998 &  &  &  & 0.1032652\\
50 & 0.3856 & 0.0066 & 0.0198 & 0.0064 & 1.0728063 & \ & 0.4718 & 0.0066 & 0.0198 & 0.0066 & 1.3993653\\
100 & 0.6538 & 0.0340 & 0.0912 & 0.0332 & 1.9375112 & \ & 0.8126 & 0.0440 & 0.0906 & 0.0436 & 2.7132478\\
150 & 0.8692 & 0.1022 & 0.2316 & 0.1014 & 2.4159540 & \ & 0.9628 & 0.0986 & 0.2052 & 0.0986 & 3.5152313\\
200 & 0.9488 & 0.1664 & 0.3744 & 0.1662 & 2.7263193 & \ & 0.9958 & 0.1740 & 0.3422 & 0.1740 & 4.0581145\\
500 & 1.0000 & 0.4370 & 0.9012 & 0.4370 & 3.7059966 & \ & 1.0000 & 0.4910 & 0.8978 & 0.4910 & 5.9466890\\
\addlinespace[.3cm]
\multicolumn{12}{l}{$\vdelta_C^\star$}\\
25 & 0.4150 & 0.2838 & 0.7786 & 0.0814 & 0.6646431 & \ & 0.4178 & 0.3230 & 0.7810 & 0.1074 & 0.4341492\\
50 & 0.3448 & 0.5772 & 0.9822 & 0.1760 & 1.0393900 & \ & 0.4302 & 0.6100 & 0.9868 & 0.2400 & 1.3802097\\
100 & 0.5878 & 0.6546 & 1.0000 & 0.3668 & 1.9020125 & \ & 0.7634 & 0.7442 & 0.9998 & 0.5624 & 2.6410276\\
150 & 0.8146 & 0.6572 & 1.0000 & 0.5312 & 2.3497452 & \ & 0.9434 & 0.8060 & 1.0000 & 0.7612 & 3.3702505\\
200 & 0.9392 & 0.6826 & 1.0000 & 0.6392 & 2.6703440 & \ & 0.9908 & 0.8442 & 1.0000 & 0.8374 & 3.8883746\\
500 & 1.0000 & 0.8070 & 1.0000 & 0.8070 & 3.5954843 & \ & 1.0000 & 0.9200 & 1.0000 & 0.9200 & 5.5463017\\
\addlinespace[.3cm]
\multicolumn{12}{l}{$\vdelta_D^\star$}\\
25 & 0.4102 & 0.1336 & 0.4198 & 0.0266 & 0.4482728 & \ & 0.4110 & 0.1332 & 0.4338 & 0.0352 & -0.0322183\\
50 & 0.2506 & 0.4634 & 0.8228 & 0.0710 & 0.5232628 & \ & 0.3160 & 0.4658 & 0.8154 & 0.1020 & 0.6636864\\
100 & 0.2462 & 0.7118 & 0.9868 & 0.1286 & 1.2005122 & \ & 0.3700 & 0.7154 & 0.9852 & 0.2232 & 1.5309406\\
150 & 0.3292 & 0.7734 & 0.9988 & 0.2054 & 1.6135159 & \ & 0.5050 & 0.7896 & 0.9990 & 0.3612 & 2.1948812\\
200 & 0.4332 & 0.7990 & 1.0000 & 0.2964 & 1.8752718 & \ & 0.6530 & 0.8396 & 1.0000 & 0.5264 & 2.7143211\\
500 & 0.9160 & 0.7358 & 1.0000 & 0.6572 & 2.8005359 & \ & 0.9934 & 0.8996 & 1.0000 & 0.8934 & 4.5015884\\
\bottomrule
\end{tabular}

}
\begin{minipage}{\textwidth}
    \vspace{.25cm}
    \scriptsize\textit{Notes:} DGP \eqref{eq:adfdgp}, $\vdelta_A^\star := (.4, .3, .2)'$, $\vdelta_B^\star := (.4, .3, .2, 0, 0, 0, -.2, 0, 0, .2)'$, $\vdelta_C^\star := .7$, $\vdelta_D^\star:= (-.4, 0, .7)'$. $\rho^\star = -.05$. Model \eqref{eq:adfreg} with $p=\lfloor12(T/100)^{.25}\rfloor$. $J_{\alpha}$ computed with LRV estimate $\widehat{\omega}^2_{\textup{AR}}(k)$, $k$ selected with BIC with $k_{\max}=p$, $\alpha = .1$, and $\sigma_\nu=1$. Probability statements denote inclusion of $y_{t-1}$ ($\widehat\rho_\lambda\neq0$), correct ($\widehat{\mathcal{J}} = \mathcal{J}$) and conservative ($\mathcal{J}\in\widehat{\mathcal{J}}$) estimation of the lag pattern, and correct model selection ($\widehat{\mathcal{M}} = \mathcal{M}$). $\lambda_{0,\,\rho^\star}$ and $\Breve{\lambda}_{0,\,\rho^\star}$ are log scale median activation thresholds of $y_{t-1}$. 5000 replications.
\end{minipage}
\end{table}

\begin{table}[htbp]
\centering
\caption{Model selection outcomes for non-stationary autoregressions \newline (FD-detrended data)}
\label{tab:sp_ct_extended_size}
\vspace{.25cm}
\setlength{\tabcolsep}{1.5pt}
\renewcommand{\arraystretch}{1.2}
\resizebox{\textwidth}{!}{
    
\centering
\begin{tabular}{l *{5}{S} c *{5}{S}}
\toprule
\multicolumn{1}{c}{ } & \multicolumn{5}{c}{AL} & \multicolumn{1}{c}{ } & \multicolumn{5}{c}{ALIE} \\
\cmidrule(l{3pt}r{3pt}){2-6} \cmidrule(l{3pt}r{3pt}){8-12}
$T$ & {$\Prob{\widehat{\rho}_\lambda\neq0}$} & {$\Prob{\widehat{\mathcal{J}} = \mathcal{J}}$} & {$\Prob{\mathcal{J}\in\widehat{\mathcal{J}}}$} & {$\Prob{\widehat{\mathcal{M}} = \mathcal{M}}$} & {$\lambda_{0,\,\rho^\star}$} &   & {$\Prob{\widehat{\rho}_\lambda\neq0}$} & {$\Prob{\widehat{\mathcal{J}} = \mathcal{J}}$} & {$\Prob{\mathcal{J}\in\widehat{\mathcal{J}}}$} & {$\Prob{\widehat{\mathcal{M}} = \mathcal{M}}$} & {$\Breve{\lambda}_{0,\,\rho^\star}$}\\
\midrule
\addlinespace[.3cm]
\multicolumn{12}{l}{$\vdelta_A^\star$}\\
25 & 0.5252 & 0.0068 & 0.0488 & 0.0000 & 0.9531510 & \ & 0.3404 & 0.0052 & 0.0574 & 0.0008 & -0.4961798\\
50 & 0.3122 & 0.0384 & 0.0864 & 0.0198 & 0.7999382 & \ & 0.2272 & 0.0332 & 0.0842 & 0.0218 & -0.2116588\\
100 & 0.1622 & 0.1802 & 0.2932 & 0.1478 & 0.7343680 & \ & 0.1530 & 0.1836 & 0.3112 & 0.1418 & 0.0188294\\
150 & 0.1188 & 0.3384 & 0.4862 & 0.2924 & 0.7295023 & \ & 0.1080 & 0.3450 & 0.4858 & 0.2986 & 0.1556094\\
200 & 0.0846 & 0.4504 & 0.6054 & 0.4118 & 0.7396626 & \ & 0.0832 & 0.4538 & 0.6162 & 0.4126 & 0.1663433\\
500 & 0.0374 & 0.7854 & 0.9352 & 0.7634 & 0.7155828 & \ & 0.0420 & 0.7874 & 0.9358 & 0.7608 & 0.2426540\\
\addlinespace[.3cm]
\multicolumn{12}{l}{$\vdelta_B^\star$}\\
25 & 0.4738 &  &  &  & 0.7817504 & \ & 0.3344 &  &  &  & -0.6028623\\
50 & 0.2728 & 0.0048 & 0.0114 & 0.0006 & 0.6870007 & \ & 0.2198 & 0.0046 & 0.0142 & 0.0022 & -0.1262214\\
100 & 0.1620 & 0.0276 & 0.0652 & 0.0152 & 0.7146077 & \ & 0.1306 & 0.0276 & 0.0680 & 0.0212 & -0.0320163\\
150 & 0.1470 & 0.0890 & 0.1652 & 0.0546 & 0.7419194 & \ & 0.1092 & 0.0830 & 0.1728 & 0.0652 & 0.0208022\\
200 & 0.1412 & 0.1582 & 0.2844 & 0.1198 & 0.7261209 & \ & 0.0928 & 0.1546 & 0.2882 & 0.1314 & 0.0027348\\
500 & 0.1164 & 0.4966 & 0.8626 & 0.4512 & 0.7557721 & \ & 0.0840 & 0.4920 & 0.8672 & 0.4522 & 0.1675942\\
\addlinespace[.3cm]
\multicolumn{12}{l}{$\vdelta_C^\star$}\\
25 & 0.4920 & 0.2572 & 0.6844 & 0.1598 & 0.8357863 & \ & 0.3720 & 0.2744 & 0.7130 & 0.2018 & -0.2470600\\
50 & 0.2304 & 0.6366 & 0.9722 & 0.5424 & 0.6634027 & \ & 0.2004 & 0.6494 & 0.9758 & 0.5692 & 0.0734400\\
100 & 0.0992 & 0.8072 & 0.9996 & 0.7538 & 0.6474925 & \ & 0.0996 & 0.8212 & 1.0000 & 0.7614 & 0.2223679\\
150 & 0.0718 & 0.8672 & 1.0000 & 0.8244 & 0.7183333 & \ & 0.0794 & 0.8558 & 1.0000 & 0.8054 & 0.2635713\\
200 & 0.0526 & 0.8822 & 1.0000 & 0.8488 & 0.7147513 & \ & 0.0646 & 0.8834 & 1.0000 & 0.8382 & 0.2686415\\
500 & 0.0246 & 0.9274 & 1.0000 & 0.9098 & 0.7087381 & \ & 0.0350 & 0.9312 & 1.0000 & 0.9036 & 0.2681411\\
\addlinespace[.3cm]
\multicolumn{12}{l}{$\vdelta_D^\star$}\\
25 & 0.5436 & 0.1382 & 0.4008 & 0.0850 & 0.8699954 & \ & 0.4448 & 0.1296 & 0.4132 & 0.0960 & -0.2021179\\
50 & 0.3048 & 0.4732 & 0.8210 & 0.3722 & 0.7391812 & \ & 0.3122 & 0.4632 & 0.8188 & 0.3622 & 0.3659767\\
100 & 0.1442 & 0.7234 & 0.9890 & 0.6490 & 0.7275843 & \ & 0.1542 & 0.7188 & 0.9864 & 0.6324 & 0.4214541\\
150 & 0.0928 & 0.8148 & 0.9972 & 0.7608 & 0.7668522 & \ & 0.1038 & 0.8120 & 0.9970 & 0.7454 & 0.4026418\\
200 & 0.0654 & 0.8586 & 1.0000 & 0.8166 & 0.7170748 & \ & 0.0850 & 0.8376 & 0.9998 & 0.7814 & 0.3410734\\
500 & 0.0266 & 0.9228 & 1.0000 & 0.9034 & 0.7428818 & \ & 0.0348 & 0.9140 & 1.0000 & 0.8872 & 0.3423630\\
\bottomrule
\end{tabular}

}
\begin{minipage}{\textwidth}
    \vspace{.25cm}
    \scriptsize\textit{Notes:} DGP \eqref{eq:adfdgp}, $\rho^\star = 0$, $\vdelta_A^\star := (.4, .3, .2)'$, $\vdelta_B^\star := (.4, .3, .2, 0, 0, 0, -.2, 0, 0, .2)'$, $\vdelta_C^\star := .7$, $\vdelta_D^\star:= (-.4, 0, .7)'$. Model \eqref{eq:adfreg} with $p=\lfloor12(T/100)^{.25}\rfloor$. $J_{\alpha,\,\sigma_\nu}^{\tau^\star}$ computed with LRV estimate $\widehat{\omega}^2_{\textup{AR}}(k)$, $k$ selected with BIC with $k_{\max}=p$, $\alpha = .75$, and $\sigma_\nu=1$. Probability statements denote inclusion of $y_{t-1}$ ($\widehat\rho_\lambda\neq0$), correct ($\widehat{\mathcal{J}} = \mathcal{J}$) and conservative ($\mathcal{J}\in\widehat{\mathcal{J}}$) estimation of the lag pattern, and correct model selection ($\widehat{\mathcal{M}} = \mathcal{M}$). $\lambda_{0,\,\rho^\star}$ and $\Breve{\lambda}_{0,\,\rho^\star}$ are log scale median activation thresholds of $y_{t-1}$. 5000 replications.
\end{minipage}
\end{table}

\begin{table}[htbp]
\centering
\caption{Model selection outcomes for stationary autoregressions \newline (FD-detrended data)}
\label{tab:sp_ct_extended_power}
\vspace{.25cm}
\setlength{\tabcolsep}{1.5pt}
\renewcommand{\arraystretch}{1.2}
\resizebox{\textwidth}{!}{
    
\centering
\begin{tabular}{l *{5}{S} c *{5}{S}}
\toprule
\multicolumn{1}{c}{ } & \multicolumn{5}{c}{AL} & \multicolumn{1}{c}{ } & \multicolumn{5}{c}{ALIE} \\
\cmidrule(l{3pt}r{3pt}){2-6} \cmidrule(l{3pt}r{3pt}){8-12}
$T$ & {$\Prob{\widehat{\rho}_\lambda\neq0}$} & {$\Prob{\widehat{\mathcal{J}} = \mathcal{J}}$} & {$\Prob{\mathcal{J}\in\widehat{\mathcal{J}}}$} & {$\Prob{\widehat{\mathcal{M}} = \mathcal{M}}$} & {$\lambda_{0,\,\rho^\star}$} &   & {$\Prob{\widehat{\rho}_\lambda\neq0}$} & {$\Prob{\widehat{\mathcal{J}} = \mathcal{J}}$} & {$\Prob{\mathcal{J}\in\widehat{\mathcal{J}}}$} & {$\Prob{\widehat{\mathcal{M}} = \mathcal{M}}$} & {$\Breve{\lambda}_{0,\,\rho^\star}$}\\
\midrule
\addlinespace[.3cm]
\multicolumn{12}{l}{$\vdelta_A^\star$}\\
25 & 0.5368 & 0.0080 & 0.0602 & 0.0078 & 1.0845195 & \ & 0.3722 & 0.0072 & 0.0714 & 0.0058 & -0.3313153\\
50 & 0.3864 & 0.0380 & 0.1196 & 0.0314 & 0.9865929 & \ & 0.4418 & 0.0410 & 0.1068 & 0.0384 & 1.0514091\\
100 & 0.5726 & 0.1182 & 0.3250 & 0.1104 & 1.7673488 & \ & 0.7548 & 0.1510 & 0.3112 & 0.1452 & 2.5059634\\
150 & 0.7302 & 0.1956 & 0.4872 & 0.1916 & 2.2878587 & \ & 0.9220 & 0.2580 & 0.4732 & 0.2560 & 3.3047541\\
200 & 0.8200 & 0.2324 & 0.6032 & 0.2300 & 2.6302774 & \ & 0.9796 & 0.3340 & 0.5906 & 0.3334 & 3.8699526\\
500 & 0.9220 & 0.3500 & 0.8972 & 0.3498 & 3.5478674 & \ & 0.9998 & 0.5044 & 0.8802 & 0.5044 & 5.5083469\\
\addlinespace[.3cm]
\multicolumn{12}{l}{$\vdelta_B^\star$}\\
25 & 0.4928 &  &  &  & 0.8838048 & \ & 0.3598 &  &  &  & -0.4834219\\
50 & 0.3504 & 0.0080 & 0.0166 & 0.0074 & 0.9641704 & \ & 0.3996 & 0.0066 & 0.0166 & 0.0054 & 0.9508916\\
100 & 0.5262 & 0.0308 & 0.0804 & 0.0294 & 1.6451563 & \ & 0.6302 & 0.0284 & 0.0754 & 0.0272 & 2.0118373\\
150 & 0.7008 & 0.0864 & 0.1962 & 0.0836 & 2.0880987 & \ & 0.8300 & 0.0830 & 0.1800 & 0.0820 & 2.7644972\\
200 & 0.7926 & 0.1356 & 0.3194 & 0.1328 & 2.3826583 & \ & 0.9420 & 0.1436 & 0.3116 & 0.1434 & 3.2854125\\
500 & 0.9312 & 0.2862 & 0.8292 & 0.2846 & 3.2271960 & \ & 0.9992 & 0.3736 & 0.8316 & 0.3736 & 5.0978499\\
\addlinespace[.3cm]
\multicolumn{12}{l}{$\vdelta_C^\star$}\\
25 & 0.5056 & 0.2662 & 0.7444 & 0.1134 & 0.9489913 & \ & 0.4082 & 0.2730 & 0.7618 & 0.0902 & 0.0386178\\
50 & 0.3808 & 0.5608 & 0.9806 & 0.1876 & 1.1288577 & \ & 0.3488 & 0.5790 & 0.9820 & 0.1722 & 0.9189514\\
100 & 0.4946 & 0.6452 & 1.0000 & 0.3010 & 1.6939232 & \ & 0.5684 & 0.6874 & 1.0000 & 0.3754 & 1.9683561\\
150 & 0.6650 & 0.6328 & 1.0000 & 0.4136 & 2.1226479 & \ & 0.7834 & 0.7354 & 1.0000 & 0.5738 & 2.6440224\\
200 & 0.7772 & 0.6200 & 1.0000 & 0.4844 & 2.3990221 & \ & 0.8984 & 0.7704 & 1.0000 & 0.6892 & 3.1145802\\
500 & 0.9440 & 0.6228 & 1.0000 & 0.6018 & 3.2661534 & \ & 0.9984 & 0.8456 & 1.0000 & 0.8442 & 4.6888955\\
\addlinespace[.3cm]
\multicolumn{12}{l}{$\vdelta_D^\star$}\\
25 & 0.5578 & 0.1306 & 0.4146 & 0.0484 & 0.9779244 & \ & 0.4344 & 0.1382 & 0.4412 & 0.0346 & -0.2015608\\
50 & 0.3586 & 0.4628 & 0.8300 & 0.1108 & 0.9029314 & \ & 0.3796 & 0.4626 & 0.8228 & 0.1202 & 0.8593716\\
100 & 0.2894 & 0.6952 & 0.9844 & 0.1450 & 1.2542202 & \ & 0.3232 & 0.7120 & 0.9864 & 0.1848 & 1.2372046\\
150 & 0.3428 & 0.7706 & 0.9992 & 0.2144 & 1.5715681 & \ & 0.3778 & 0.7860 & 0.9988 & 0.2542 & 1.6803044\\
200 & 0.4118 & 0.7874 & 1.0000 & 0.2686 & 1.7975424 & \ & 0.5112 & 0.8108 & 1.0000 & 0.3776 & 2.1102142\\
500 & 0.7598 & 0.7582 & 1.0000 & 0.5382 & 2.5967784 & \ & 0.9030 & 0.8602 & 1.0000 & 0.7670 & 3.5145534\\
\bottomrule
\end{tabular}

}
\begin{minipage}{\textwidth}
    \vspace{.25cm}
    \scriptsize\textit{Notes:} DGP \eqref{eq:adfdgp}, $\vdelta_A^\star := (.4, .3, .2)'$, $\vdelta_B^\star := (.4, .3, .2, 0, 0, 0, -.2, 0, 0, .2)'$, $\vdelta_C^\star := .7$, $\vdelta_D^\star:= (-.4, 0, .7)'$, $\rho^\star = -.05$. Model \eqref{eq:adfreg} with $p=\lfloor12(T/100)^{.25}\rfloor$. $J_{\alpha,\,\sigma_\nu}^{\tau^\star}$ computed with LRV estimate $\widehat{\omega}^2_{\textup{AR}}(k)$, $k$ selected with BIC with $k_{\max}=p$, $\alpha = .75$, and $\sigma_\nu=1$. Probability statements denote inclusion of $y_{t-1}$ ($\widehat\rho_\lambda\neq0$), correct ($\widehat{\mathcal{J}} = \mathcal{J}$) and conservative ($\mathcal{J}\in\widehat{\mathcal{J}}$) estimation of the lag pattern, and correct model selection ($\widehat{\mathcal{M}} = \mathcal{M}$). $\lambda_{0,\,\rho^\star}$ and $\Breve{\lambda}_{0,\,\rho^\star}$ are log scale median activation thresholds of $y_{t-1}$. 5000 replications.
\end{minipage}
\end{table}

\newpage
\subsection{Proofs} 
\label[appendix]{sec:proofs}

This appendix presents the proofs of the results from the main paper. We proceed chronologically and first consider \Cref{lem:crates}, \Cref{prop:plugandplim}, and \Cref{thm:aronelpnull}. Subsequently, we introduce the Lemmas \ref{lem:onesquaresummability}, \ref{lem:rhostarstar}, and \ref{lem:knifeCLT}, which establish asymptotic properties of the OLS estimator in a potentially underspecified stationary ADF(0) model. These lemmas are used to prove \Cref{thm:aronelpalt}. We then consider \Cref{prop:lambdanaught} and use some of the previous statements to show the remaining results in \Cref{sec:alieirw} of the main text, namely \Cref{lem:ZeroShrinkage}, \Cref{cor:knot1asymptotics}, and \Cref{thm:permactive}. We conclude with the proofs to \Cref{thm:nmmlmtwo} and \Cref{cor:qlimnmm}. 

We frequently use the below symbols and abbreviations. For the remainder of the notation, we refer to the main paper.\\ 

\noindent\textbf{Symbols and abbreviations}\\
\renewcommand*{\arraystretch}{1}
\begin{longtable}
{r@{\hspace{.5cm}}l@{\hspace{.5cm}}l}
$\lvert S \rvert$ &:& cardinality of $S$\\
i.i.d. &:& independent, identically distributed\\
r.v. &:& random variable\\
$\mathcal{M}$ &:& data-generating model\\
$\widehat{\mathcal{M}}$ &:& selected model\\
$\vbeta^\star$ &:& coefficient vector of the data generating model\\
$\widehat{\vbeta}_\lambda$ &:& adaptive Lasso estimator of $\vbeta^\star$\\
$\widehat{\vbeta}$ &:& least squares estimator of $\vbeta^\star$\\
$\mathbb{R}^+$ &$:=$& $(0,\infty)$\\

$f(T)=\Theta(g(T))$ &:$\Leftrightarrow$& $\displaystyle 0<\liminf_{T\rightarrow\infty}\left\lvert\frac{f(T)}{g(T)}\right\rvert\leq\limsup_{T\rightarrow\infty}\left\lvert\frac{f(T)}{g(T)}\right\rvert<\infty$;\\ &&$f,g:\ \mathbb{N}\to\mathbb{R}$\\

$f(T) = O(g(T))$ &:$\Leftrightarrow$& $\displaystyle\limsup_{T\to\infty}\left\lvert\frac{f(T)}{g(T)}\right\rvert<\infty;\quad f,g:\ \mathbb{N}\to\mathbb{R}$\\

$f(T) = o(g(T))$ &:$\Leftrightarrow$& $\displaystyle\lim_{T\to\infty}\left\lvert\frac{f(T)}{g(T)}\right\rvert = 0; \quad f,g:\ \mathbb{N}\to\mathbb{R}$\\

$X_T = o_p(g(T))$ &:$\Leftrightarrow$& $\displaystyle\lim_{T\to\infty}\Prob{\left\lvert\frac{X_T}{g(T)}\right\rvert\geq\epsilon} = 0;\quad \epsilon>0, \quad g:\ \mathbb{N}\to\mathbb{R}$\\

$X_T = O_p(g(T))$ &:$\Leftrightarrow$& $\displaystyle \forall\epsilon>0\ \reflectbox{E}\ \eta_\epsilon,N_\epsilon>0:\ \Prob{\left\lvert\frac{X_T}{g(T)}\right\rvert\geq\eta_\epsilon} < \epsilon,\quad T\geq N_\epsilon$;\\ && $g:\mathbb{N}\to\mathbb{R}$\\

\end{longtable}
\renewcommand*{\arraystretch}{1}

\newpage

\begin{proof}[\bfseries Proof of \Cref{lem:crates}]
\noindent
\begin{enumerate}
    \item It is a well-known result that $\widehat{\rho} = o_p(1)$ and $\widehat{\rho} = O_p(T^{-1})$ for $\rho^\star=0$, see \textcite{Hamilton1994}. By Theorem 3.1. (a) in \textcite{Phillips1986}, $\widehat{\zeta} = O_p(1)$ such that empirical quantiles $\widehat{\zeta}_{\alpha}$, $\alpha\in(0,1)$ are bounded and therefore $J_\alpha= O_p(1)$.

    Note that
    \begin{align*}
        \Breve{w}_1^{-1} =&\, 1/J_\alpha \cdot \lvert\widehat{\rho}\rvert\\ =&\,O_p(1) \cdot O_p(T^{-1})\\ 
        =&\,O_p(T^{-1})
    \end{align*}
    and, by application of the CMT, we have $\Breve{w}_1^{-1} = o_p(1)$. Thus $\Prob{\Breve{w}_1 > b} \rightarrow 1$ for $b\in\mathbb{R}^+$. We conclude that the modified weight $\Breve{w}_1$ diverges towards infinity in probability whereas $\Breve{w}_1^{-1}$ converges to zero in probability as $T\rightarrow\infty$.

    \item For stationary models, $\rho^\star\in(-2,0)$ and $(\widehat{\rho} - \rho^\star) = O_p(T^{-1/2})$. Following Proposition 17.1 (b), (e), and the preceding exposition in \textcite{Hamilton1994}, $\widehat{\zeta} \xrightarrow[]{p} 0$ at rate $O_p(T^{-1})$. Thus, $\widehat{\zeta}_{\alpha}= O_p(T^{-1})$ for empirical quantiles $\widehat{\zeta}_{\alpha}$, $\alpha\in(0,1)$ and so $J_{\alpha}\xrightarrow[]{p}0$ at rate $O_p(T^{-1})$.
    
    For the stochastic order, we hence have
    \begin{align*}
        \Breve{w}_1 = \biggr\lvert\frac{J_\alpha}{\widehat{\rho}}\biggr\rvert =  \biggr\lvert\frac{J_\alpha}{\rho^\star} \cdot  \frac{\rho^\star}{\widehat{\rho}}\biggr\rvert=& \, \Bigg\lvert\ \frac{J_\alpha}{\rho^\star} + \frac{J_\alpha}{\rho^\star} \cdot \left(\frac{\rho^\star}{\widehat{\rho}} - 1\right) \Bigg\rvert \\
        =&\, O_p(T^{-1}) + O_p(T^{-1}) \cdot o_p(1) \\ 
        =&\, O_p(T^{-1}) + o_p(T^{-1})
        \\
        =&\, O_p(T^{-1}),
    \end{align*}
    and it follows that $\Breve{w}_1 = o_p(1)$ by application of the CMT.\qedhere    
\end{enumerate}
\end{proof}
\vspace{1em}

\begin{proof}[\bfseries Proof of \Cref{prop:plugandplim}]
We verify the conditions required for ALIE to be asymptotically equivalent to the AL estimator of \textcite{kock2016consistent}. For brevity, we confine the exposition to the relevant deviations from the proofs on the oracle properties of AL and consistent tuning using BIC in Theorems 1 and 2 of \textcite{kock2016consistent} that arise for ALIE.  Part (a) considers the asymptotic equivalence of AL and ALIE for $\rho^\star = 0$ and part (b) for $\rho^\star\in(-2,0)$, highlighting deviating results due to the modified penalty weight $\Breve{w}_1$, respectively. Part (c) considers the consistent tuning of ALIE.
\begin{enumerate}
    \item[(a)]
    For $\rho^\star=0$ we may rewrite the Lasso loss function \eqref{eq:adaptive lassooptim} by substituting $\dot{\rho} = u_1/T$ and $\dot{\delta}_j = \delta_{j,\,p} + u_{2,\,j} / \sqrt{T}$,
    \begin{align}
        \begin{split}
        \Psi_{T,\,p}(u_1,\vu_2) :=& \sum_{t=1}^T \left(\Delta y_t - \frac{u_1}{T} y_{t-1} - \sum_{j=1}^p \left( \delta_{j,\,p} + \frac{u_{2,\,j}}{\sqrt{T}} \right) \Delta y_{t-j} \right)^2 \\
        &+ 2\lambda_\mathcal{O}\left( \Breve{w}_1^{\gamma_1} \left\lvert \frac{u_1}{T} \right\rvert +  \sum_{j=1}^p w_{2,\,j}^{\gamma_2} \left\lvert \delta_{j,\,p} + \frac{u_{2,\,j}}{\sqrt{T}} \right\rvert\right),
        \end{split} \label{eq:lassooptrew}
    \end{align}
    with constants $u_1 \in \mathbb{R}$ and $\vu_2 \in \mathbb{R}^p$. Note how only the part $\lambda_\mathcal{O} \Breve{w}_1^{\gamma_1} \left\lvert u_1/T \right\rvert$ of the $\ell_1$ penalty term is affected by our penalty weight $\Breve{w}_1$. Therefore, the investigation of the effects of $\Breve{w}_1$ onto the difference of the empirical and the oracle loss functions $V_{T,\,p}(u_1, \vu_2) := \Psi_{T,\,p}(u_1,\vu_2) - \Psi_{T,\,p}(0,\vzero)$ is confined to this term.

    Under \Cref{assum:sparsity} (sparsity) we may replace $\delta_{j,\,p}$ with $\delta_j^\star$ so that $\dot{\delta}_j = \delta_j^\star + u_{2,\,j} / \sqrt{T}$. Then, $\Psi_{T,\,p}\equiv\Psi_T$ $V_{T,\,p}\equiv V_T$, yielding expressions similar as in \textcite{kock2016consistent}. Under \Cref{assum:lperrors} (approximate sparsity), we have $\lim_{T\to\infty}V_{T,\,p} = \lim_{T\to\infty}V_T$ since $\lim_{T\to\infty}\delta_{j,\,p}=\delta_j^\star$ and $(\widehat{\rho},\widehat{\vdelta})\xrightarrow[]{p}(0,\vdelta^\star)$, provided the lag order $p$ satisfies $1/p+p^3/T\to0$ as $T\to\infty$, cf. \textcite{ChangPark2002}. 
    
    Establishing the oracle property for ALIE under sparsity or approximate sparsity considers the limits of \eqref{eq:lassooptrew} depending on the deviation $u_1$, following the same logic as in equation (4) in the proof of Theorem 1 in \textcite{kock2016consistent}. Specifically, the $\Breve{w}_1$-dependent part of $V_{T,\,p}$ satisfies
    \begin{align*}
        \lambda_\mathcal{O} \Breve{w}_1^{\gamma_1} \left\lvert \frac{u_1}{T} \right\rvert
        =& \, \frac{\lambda_\mathcal{O}}{T} \lvert u_1 \rvert T^{\gamma_1} T^{-\gamma_1} \Breve{w}_1^{\gamma_1}\\ 
        =& \, \frac{\lambda_\mathcal{O}}{T^{1-\gamma_1}} \lvert u_1 \rvert (T \Breve{w}_1^{-1})^{-\gamma_1}\\
        =& \, \frac{\lambda_\mathcal{O}}{T^{1-\gamma_1}} \lvert u_1 \rvert O_p(1)\\
    \overset{p}{\to} &\, 
        \begin{cases}
            \infty &\text{if } u_1 \neq 0\\
            0      &\text{if } u_1 = 0\\
        \end{cases},
    \end{align*}
    where $\lambda_\mathcal{O}/T^{1-\gamma_1}\rightarrow\infty$, $\gamma_1 > 1/2$, and $\lambda_\mathcal{O}/\sqrt{T}\rightarrow 0$ by assumption. Note that $\Breve{w}_1^{-1} = O_p(T^{-1})$ by part \ref{item:lem:crates:H0} of \cref{lem:crates}, implying $\Breve{w}_1 T^{-1} = O_p(1)$. 
    
    Both $w_1$ and $\Breve{w}_1$ hence give identical limits for the above expression so that the implementation of $\Breve{w}_1$ does not affect the optimal choice of $u_1$ asymptotically, i.e., $u_1 \to 0$ if $\rho^\star = 0$ as $T \to \infty$. Furthermore, since the $w_{2,\,j}$, $j \in \{1, \dots, p\}$ are unaffected by $\Breve{w}_1$, the $u_{2,\,j}$ are identical for AL and ALIE. Therefore, the results on consistency and asymptotic normality of the AL solution carry over to ALIE.

    \item[(b)] Under the alternative $\rho^\star\in(-2,0)$, we again make use of the rewritten Lasso optimisation problem but define ${\dot\rho = \rho_p + u_1/\sqrt{T}}$ in accordance with $\sqrt{T}$-consistency. Analogously to part (a), the only $\Breve{w}_1$-dependent part of the loss function difference $V_{T,\,p}(u_1,\vu_2)$ is $ \lambda_\mathcal{O} \Breve{w}_1^{\gamma_1} \left( \left\lvert \rho_p + u_1/\sqrt{T} \right\rvert - \lvert \rho_p \rvert \right)$. We have
    \begin{align*}
       \lambda_\mathcal{O} \Breve{w}_1^{\gamma_1} \left( \left\lvert \rho_p + \frac{u_1}{\sqrt{T}} \right\rvert - \lvert \rho^\star \rvert \right) 
       &= \frac{\lambda_\mathcal{O}}{\sqrt{T}} \, \Breve{w}_1^{\gamma_1} \, \sqrt{T} \left( \left\lvert \rho_p + \frac{u_1}{\sqrt{T}} \right\rvert - \lvert \rho_p \rvert \right) \\
        &= o(1) \, o_p(1) \, \sqrt{T} \left( \left\lvert \rho_p + \frac{u_1}{\sqrt{T}} \right\rvert - \lvert \rho_p \rvert \right) \\
        & \overset{p}{\to} 0,
    \end{align*}
    noting that $\sqrt{T} \left( \left\lvert \rho_p + u_1/\sqrt{T} \right\rvert - \lvert \rho_p \rvert \right) \overset{p}{\to} u_1 \sgn(\rho^\star)$, $\Breve w_1 = o_p(1)$ by part~\ref{item:lem:crates:H1} of \cref{lem:crates} and since  $\lambda_\mathcal{O}/\sqrt{T}\rightarrow 0$ by assumption of the proposition.
    
    \item[(c)] As discussed in parts (a) and (b), consistency of ALIE is ensured for the same rate conditions on $\lambda_\mathcal{O}$ as for AL, which are stated in this proposition. Hence,
    \begin{align*}
        \lim_{T\to\infty}\widehat\vbeta_{\lambda_\mathcal{O}}^{\text{AL}} = \lim_{T\to\infty}\widehat\vbeta_{\lambda_\mathcal{O}}^{\text{ALIE}} = \lim_{T\to\infty}\widehat\vbeta_{\mathcal{M}}^\text{OLS} = \vbeta^\star,
    \end{align*}
    where $\widehat\vbeta_{\mathcal{M}}^\text{OLS}$ denotes the OLS estimator of the true model $\mathcal{M}$. By Theorem 2 in \textcite{kock2016consistent}, we have
    \begin{align*}
        \lim_{T\to\infty}\widehat\vbeta_{\lambda_\mathcal{O}}^{\text{AL}} = \argmin\lim_{T\to\infty} \textup{BIC}(\vbeta).
    \end{align*}
    Combining both results, we conclude that $\widehat\vbeta_{\lambda_\mathcal{O}}^{\text{ALIE}}$ minimises $\textup{BIC}(\vbeta)$ in the limit, just as $\widehat\vbeta_{\lambda_\mathcal{O}}^{\text{AL}}$. Conversely, we have $\lim_{T\to\infty}\lambda_{\text{BIC}}\in\lambda_\mathcal{O}$ for ALIE, too.\qedhere
\end{enumerate}
\end{proof}
\vspace*{1em}

\begin{proof}[\bfseries Proof of \Cref{thm:aronelpnull}]
We state $\Prob{\widehat\rho_\lambda = 0}$ in terms of the corresponding FOC to $\min_{\dot\rho}\Psi(\dot\rho)$,
	\begin{align*}
 	  \Prob{\widehat\rho_\lambda = 0} =&\,\prob{\frac{\widehat\rho^2}{J_\alpha} \sum_{t=1}^T y_{t-1}^2 \leq \lambda}\\
    =& \, \Prob{\frac{(T\widehat\rho)^2}{J_\alpha}\frac{1}{T^2}\sum_{t=1}^T y_{t-1}^2\leq\lambda}.
	\end{align*}
	 By \textcite{Phillips1986, PhillipsPerron1988}, we obtain
    \begin{align*}
        T\widehat\rho \ \xrightarrow[]{d}& \ \frac{W(1)^2-1}{2 \int_0^1 W(r)^2\mathrm{d}r} + \frac{\sigma^2 \phi(1)^2 - \iota_0}{2\sigma^2 \phi(1)^2 \int_0^1 W(r)^2\mathrm{d}r},\\
        \frac{1}{T^2}\sum_{t=1}^T y_{t-1}^2 \ \xrightarrow[]{d}& \ \sigma^2\phi(1)^2\int_0^1 W(r)^2\mathrm{d}r, \\
        J_\alpha \ \xrightarrow[]{d}& \ J,
        \end{align*}
        where $\iota_0 = \E(u_t^2)$ and $J$ is the weak limit of $J_\alpha$.
	By application of the CMT, we conclude that
	\begin{align}
        \begin{split}
            \Breve{\mathcal{H}}_T :=\frac{(T\widehat\rho)^2}{J_\alpha} \frac{1}{T^2}\sum_{t=1}^T y_{t-1}^2 \ \xrightarrow[]{d} \ \Breve{\mathcal{H}} =&\, \frac{1}{J} \frac{\left(\sigma^2\phi(1)^2W(1)^2 -\iota_0\right)^2}{4\sigma^2\phi(1)^2\int_0^1W(r)^2\mathrm{d}r}\\
            =&\, \frac{1}{J} \frac{\left(\omega^2 W(1)^2 -\iota_0\right)^2}{4\omega^2\int_0^1W(r)^2\mathrm{d}r},
        \end{split}\label{eq:ar1lpnulllimit}
	\end{align}
    with $\omega^2 = \sigma^2\phi(1)^2$ the LRV of $u_t$ and $\iota_0$ its variance.\footnote{Note that $\Breve{\mathcal{H}} = \mathcal{H}/J$ with $\mathcal{H}$ the corresponding limiting r.v. for AL.} Noting that $\Breve{\mathcal{H}}$ is an absolutely continuous r.v. that is bounded on $\mathbb{R}^+$ and has CDF $F_{\Breve{\mathcal{H}}}$ obtains the individual statements of the lemma:
	\begin{enumerate}
		\item $\lambda\rightarrow0$: $\lim_{T\rightarrow\infty}\Prob{\widehat\rho_\lambda=0} = F_{\Breve{\mathcal{H}}}(0) = 0$ by continuity of $\mathcal{\Breve{H}}$.
		\item $\lambda\rightarrow c\in(0,\infty)$: $\lim_{T\rightarrow\infty}\Prob{\widehat\rho_\lambda=0} = F_{\Breve{\mathcal{H}}}(c)$ by \eqref{eq:ar1lpnulllimit}.
		\item $\lambda\rightarrow\infty$: $\lim_{T\rightarrow\infty}\Prob{\widehat\rho_\lambda=0} = \lim_{\lambda\rightarrow\infty}F_{\Breve{\mathcal{H}}}(\lambda) = 1$ since $\Breve{\mathcal{H}}= O_p(1)$.\qedhere
	\end{enumerate}
\end{proof}
\vspace*{1em}

\begin{restatable}[]{lem}{onesquaresummability}
    \label{lem:onesquaresummability}
    Let $\Delta y_{t} = \rho^\star y_{t-1} + u_t$ with $\rho^\star\in(-2,0)$ and $u_t$ as in \Cref{assum:lperrors}. Then $y_t = B(L)\varepsilon_t$ with coefficients $b_s$ such that $\sum_{s=0}^\infty s b_s^2<\infty$.
\end{restatable}
\begin{proof}[\bfseries Proof]
    Note that $y_t = \varrho y_{t-1} + u_t$ where $\varrho=\rho^\star+1$ and $\lvert\varrho\vert<1$ by assumption of the lemma. By \Cref{assum:lperrors}, we have
	\begin{align*}
		y_t = \sum_{i=0}^\infty \varrho^i u_{t-i} =&\, \sum_{i=0}^\infty \sum_{j=0}^\infty \varrho^i\phi_j\varepsilon_{t-i-j}\\
			 =&\, \sum_{s=0}^\infty \sum_{j\leq s} \varrho^{s-j}\phi_j\varepsilon_{t-s}\\
			 =&\, \sum_{s=0}^\infty b_s \varepsilon_{t-s},
	\end{align*}
	where $s=i+j$ and $b_s:=\sum_{j\leq s}^\infty \varrho^{s-j}\phi_j$. We next show that $\sum_{s=0}^\infty sb_s^2<\infty$. Note that 
	\begin{align}
		\sum_{s=0}^\infty sb_s^2 \leq \left(\sum_{s=0}^\infty s\lvert b_s\rvert\right)\left(\sum_{s=0}^\infty s\lvert b_s\rvert\right)<\infty, \label{eq:pscondition}
	\end{align}	
	provided that $\sum_{s=0}^\infty s\lvert b_s\rvert\ < \infty$. We next prove \eqref{eq:pscondition} by showing one-summability of $b_s$:
	\begin{align*}
		\sum_{s=0}^\infty s\lvert b_s\rvert =&\, \sum_{s=0}^\infty s \left\lvert \sum_{j\leq s} \varrho^{s-j}\phi_j \right\rvert\\
            \leq&\, \sum_{s=0}^\infty \sum_{j \leq s} \left\lvert s \varrho^{s-j} \phi_j \right\rvert\\
		=&\, \sum_{i=0}^\infty\sum_{j=0}^\infty(i+j)\lvert\varrho\rvert^{i}\left\lvert\phi_j\right\rvert\\
		=&\, \sum_{i=0}^\infty\sum_{j=0}^\infty j\left\lvert\phi_j\right\rvert \lvert\varrho\rvert^i + \sum_{i=0}^\infty\sum_{j=0}^\infty i \lvert\varrho\rvert^i \left\lvert\phi_j\right\rvert\\
		=&\, \sum_{i=0}^\infty \lvert\varrho\rvert^i\sum_{j=0}^\infty j\left\lvert\phi_j\right\rvert + \sum_{j=0}^\infty i \lvert\varrho\rvert^i\sum_{i=0}^\infty\left\lvert\phi_j\right\rvert\\
		<&\,\infty,
	\end{align*}
	where the last inequality follows from one-summability of $\phi_j$ and convergence of the geometric and arithmetico-geometric sums over $i$, since $\lvert\varrho\rvert<1$.\qedhere
 \end{proof}   
\vspace*{1em}

\begin{restatable}[]{lem}{rhostarstar}
\label{lem:rhostarstar}
Let $\Delta y_t = \rho^\star y_{t-1} + u_t$ with errors $u_t$ satisfying \Cref{assum:lperrors} and be $\widehat{\rho}$ the OLS estimator of $\rho^\star$. If $\rho^\star\in(-2,0)$, then \begin{enumerate*} \item $\widehat\rho \xrightarrow{p}\rho^{\star\star}\in(-2,0)$ and \item $\sqrt{T}(\widehat\rho-\rho^{\star\star})\xrightarrow{d}\mathcal{G}$,
	where $\mathcal{G}$ is a zero-mean Gaussian r.v. 
\end{enumerate*}
\end{restatable}
\begin{proof}[\bfseries Proof]
\noindent
\begin{enumerate}
	\item\label{item:lem:rhostarstar:P1} Under \Cref{assum:lperrors} and for $\rho^\star\in(-2,0)$, \Cref{lem:onesquaresummability} states that $y_t$ has the MA($\infty$) representation
	\begin{align}
		y_t = B(L) \varepsilon_t = \sum_{s=0}^\infty b_s \varepsilon_{t-s}, \quad \text{with} \quad \sum_{s=0}^\infty s b_s^2 < \infty.
        \label{eq:sb2sum}
	\end{align}
	By standard results (cf. \cite{Hamilton1994}), $\widehat{\rho}$ the OLS estimator of $\rho^\star$ in $\Delta y_t = \rho^\star y_{t-1} + u_t$ satisfies
        \begin{align*}
             \widehat{\rho} \xrightarrow{p}&\, \rho^{\star \star} = \frac{\sum_{s=0}^\infty b_s b_{s+1}}{\sum_{s=0}^\infty b_s^2} - 1.
             \intertext{By the summability condition in \eqref{eq:sb2sum},}
             \sum_{s=0}^\infty b_s b_{s+1} \leq&\, \left(\sum_{s=0}^\infty b_s^2\right)^{1/2} \left(\sum_{s=0}^\infty b_{s+1}^2\right)^{1/2}\\ 
             <&\, \left(\sum_{s=0}^\infty b_s^2\right)^{1/2} \left(\sum_{s=0}^\infty b_s^2\right)^{1/2}\\ 
             =&\, \sum_{s=0}^\infty b_s^2<\infty.
        \end{align*}
        Therefore, $$ \frac{\sum_{s=0}^\infty b_s b_{s+1}}{\sum_{s=0}^\infty b_s^2}\in(-1,1),$$ and we conclude that $\rho^{\star\star}\in(-2, 0)$ if $\rho^\star \in (-2, 0)$.
	\item\label{item:lem:rhostarstar:P2} The result follows from Theorem 3.4 in \textcite{PhillipsSolo1992}.\qedhere
\end{enumerate}
\end{proof}
\vspace*{1em}

\begin{restatable}[Gaussian knife edge]{lem}{knifeCLT}
	\label{lem:knifeCLT}
	Under the conditions of \Cref{lem:rhostarstar}, it holds that
	\begin{equation*}
		\Xi_T := \sqrt{T} \left( -\rho^{\star\star} - \sqrt{\frac{{\rho^{\star\star}}^2\E\left(y_{t-1}^2\right)}{\frac{1}{T} \sum_{t=1}^{T} y_{t-1}^2}} \right) \xrightarrow{d} \Xi \sim N\left(0, \sigma^2_{\scriptscriptstyle\Xi}\right),
	\end{equation*}
    with $\sigma^2_{\scriptscriptstyle\Xi}$ as given in the proof of this lemma.
\end{restatable}
\begin{proof}[\bfseries Proof]
Since $\rho^{\star\star}\in(-2,0)$ by part \ref{item:lem:rhostarstar:P1} of \Cref{lem:rhostarstar}, we may write
	\begin{align*}
		\Xi_T =&\, \sqrt{T}\left(-\rho^{\star\star} - \frac{\lvert\rho^{\star\star}\rvert\sqrt{\E\left(y_{t-1}^2\right)}}{\sqrt{\frac{1}{T}\sum_{t=1}^Ty_{t-1}^2}} \right)\\
	   		  =&\, \lvert\rho^{\star\star}\rvert\sqrt{\E\left(y_{t-1}^2\right)}\sqrt{T} \left(\frac{1}{\sqrt{\E\left(y_{t-1}^2\right)}} - \frac{1}{\sqrt{\frac{1}{T}\sum_{t=1}^Ty_{t-1}^2}}\right) \\
	   		  =&\, \lvert\rho^{\star\star}\rvert\sqrt{\E\left(y_{t-1}^2\right)}\sqrt{T} \left( g\left(\E\left(y_{t-1}^2\right)\right) - g\left(\frac{1}{T}\sum_{t=1}^Ty_{t-1}^2\right)\right),
	\end{align*}
	where $g(z) := z^{-1/2}$. By \Cref{assum:lperrors}, the process $y_t$ satisfies the conditions of Theorem 3.8 of \textcite{PhillipsSolo1992} so that their CLT applies to $\frac{1}{T}\sum_{t=1}^Ty_t^2$, i.e.,
	\begin{align*}
		\varpi_T:=\sqrt{T}\left(\E\left(y_{t-1}^2\right) -\frac{1}{T}\sum_{t=1}^Ty_{t-1}^2\right) \xrightarrow{d} \varpi := N\left(0, \sigma^2_\varpi \right). 
	\end{align*}
	Furthermore, $g'(z) = -\frac{1}{2}z^{-3/2}$ is continuous on $\mathbb{R}^+$ and $g'\left(\E\left(y_{t-1}^2\right)\right)<0$ since $\E\left(y_{t-1}^2\right)\in(0,\infty)$. By the delta method, we thus establish that
	\begin{align*}
		\sqrt{T} \left( g\left(\E\left(y_{t-1}^2\right)\right) - g\left(\frac{1}{T}\sum_{t=1}^Ty_{t-1}^2\right)\right) \xrightarrow{d} N\left(0,g'\left(\E\left(y_{t-1}^2\right)\right)^2\sigma^2_{\varpi}\right).
	\end{align*}
	Consequently, $\Xi_T \xrightarrow{d} \Xi \sim N\left(0, \sigma^2_{\scriptscriptstyle\Xi}\right)$ with variance
	\begin{align*}
	 \ \ \sigma^2_{\scriptscriptstyle\Xi} = \left(\frac{\lvert\rho^{\star\star}\rvert\sigma_\varpi}{2\E\left(y_{t-1}^2\right)}\right)^2,	
	\end{align*}
	which completes the proof.\qedhere
\end{proof}
\vspace*{1em}

\begin{proof}[\bfseries Proof of \Cref{thm:aronelpalt}]
Following the proof to Theorem 4 in \textcite{kock2016consistent} we state $\Prob{\widehat\rho_\lambda = 0}$ in terms of the corresponding FOC to $\min_{\dot\rho}\Psi(\dot\rho)$,
\begin{align*}
	\frac{\widehat{\rho}^2}{J_\alpha} \sum_{t=1}^T y_{t-1}^2 \leq \lambda.
\end{align*}
Then,
\begin{align*}
\Prob{\widehat\rho_\lambda = 0} =&\,\prob{\frac{\widehat\rho^2}{J_\alpha} \sum_{t=1}^T y_{t-1}^2 \leq \lambda}\\
	=&\, \prob{\sqrt{T}\lvert\widehat\rho\rvert \leq \sqrt{\frac{J_\alpha\lambda}{\frac{1}{T}\sum_{t=1}^T y_{t-1}^2}}}\\
	=&\,\prob{\sqrt{T}(\lvert\widehat{\rho}\rvert-\lvert\rho^{\star\star}\rvert) + \sqrt{T}\lvert\rho^{\star\star}\rvert \leq \sqrt{\frac{J_\alpha\lambda}{\frac{1}{T}\sum_{t=1}^T y_{t-1}^2}}}.
\intertext{Since $p^{\star\star}\in(-2,0)$ by part \ref{item:lem:rhostarstar:P1} of \Cref{lem:rhostarstar},}
\Prob{\widehat\rho_\lambda = 0} =&\, \prob{\sqrt{T}(\lvert\widehat{\rho}\rvert-\lvert\rho^{\star\star}\rvert) - \sqrt{T} \rho^{\star\star} - \sqrt{\frac{J_\alpha\lambda}{\frac{1}{T}\sum_{t=1}^T y_{t-1}^2}} \leq 0}.
\intertext{Let $Z_T := \sqrt{T}(\lvert\widehat{\rho}\rvert-\lvert\rho^{\star\star}\rvert)$, so that}
	\Prob{\widehat\rho_\lambda = 0} =&\, \Prob{Z_T - \sqrt{T} \rho^{\star\star} - \sqrt{\frac{J_\alpha\lambda}{\frac{1}{T}\sum_{t=1}^T y_{t-1}^2}} \leq 0}\\
	=&\,\prob{Z_T + \sqrt{T}\left(-\rho^{\star\star} - \sqrt{\frac{J_\alpha\lambda/T}{\frac{1}{T}\sum_{t=1}^T y_{t-1}^2}}\right) \leq 0}\\
	=&\, \prob{U_T \leq 0},
\intertext{where}
	U_T :=&\, Z_T + \sqrt{T}\left(-\rho^{\star\star} - \sqrt{\frac{J_\alpha\lambda/T}{\frac{1}{T}\sum_{t=1}^T y_{t-1}^2}}\right).
\end{align*}
Note that $\sqrt{T}(\widehat{\rho} - \rho^{\star\star})\xrightarrow{d}N\left(0,\sigma^2_{\rho^\star,\,B(s)}\right)$ by \Cref{lem:rhostarstar}, result \ref{item:lem:rhostarstar:P2}. Also $\rho^{\star\star}\in(-2,0)$ by result \ref{item:lem:rhostarstar:P1} of \Cref{lem:rhostarstar}, so that
	\begin{align*}
	   Z_T = \sqrt{T}\left(\left\lvert\widehat{\rho}\right\rvert-\left\lvert\rho^{\star\star}\right\rvert\right) \xrightarrow{d} Z \sim N\left(0,\sigma^2_Z\right),	
	\end{align*}
	by the delta method. Also, note that by Theorem 3.8 in \textcite{PhillipsSolo1992},
	\begin{align*}
	\frac{1}{T}\sum_{t=1}^Ty_{t-1}^2\xrightarrow{p} \E\left(y_{t-1}^2\right)>0,	
	\end{align*}
	and $J_{\alpha}\geq0$ by definition, i.e.,
\begin{align*}
	\sqrt{\frac{J_\alpha\lambda/T}{\frac{1}{T} \sum_{t=1}^Ty_{t-1}^2}}\geq0,
\end{align*}
for any $\lambda\in\mathbb{R}^+$.

It remains to examine $\lim_{T\rightarrow\infty} \Prob{\widehat\rho_\lambda = 0} = \lim_{T\to\infty}\prob{U_T\leq0}$ given sequences $\lambda=\Theta(T^{1+\kappa})$ for segments of $\kappa\in\mathbb{R}$. Since $J_\alpha= O_p(T^{-1})$ by part \ref{item:lem:crates:H1} of \Cref{lem:crates},
\begin{align*}
	J_\alpha\lambda/T=&\,O_p\left(T^{-1}\right)\Theta\left(T^{1+\kappa}\right)\Theta\left(T^{-1}\right)\\
 =&\,O_p\left(T^{\kappa-1}\right),
\end{align*}
and so
\begin{align}
\sqrt{
	\frac{J_\alpha\lambda/T}{\frac{1}{T} \sum_{t=1}^Ty_{t-1}^2}} =&\, \sqrt{\frac{O_p\left(T^{\kappa-1}\right)}{\E\left(y_{t-1}^2\right) + o_p(1)}} = O_p\left(T^{\frac{\kappa-1}{2}}\right).\label{eq:Jterm}
\end{align}
Therefore, $\lim_{T\rightarrow\infty}\Prob{U_T\leq0}$ depends on the limiting behaviour of the non-negative expression in \eqref{eq:Jterm}, with the stochastic order being governed by $\kappa$. We next address the individual statements of the theorem.

\begin{enumerate}
    \item $\kappa<1$: Since $J_\alpha\xrightarrow{p}0$, \eqref{eq:Jterm} is $o_p(1)$ so that $U_T \rightarrow \infty$ and $\lim_{T\rightarrow\infty}\Prob{\widehat\rho_\lambda = 0} = 0$, which proves result 1.
    
    \item $\kappa=1$: $\lambda=\Theta(T^2)$ yields $J_\alpha\lambda/T\xrightarrow{d}\mathcal{C}$. We examine the limiting behaviour of $\Prob{U_T\leq0\vert\mathcal{C}=c} = \Prob{\widehat\rho_\lambda=0\vert\mathcal{C}=c}$ for $c\in(0,\infty)$. 
    
    \begin{itemize}
        \item[(a)] $c\in\left(0,\,{\rho^{\star\star}}^2\E\left(y_{t-1}^2\right)\right)$: 
            \begin{align*}
               U_T = Z_T + \sqrt{T}\left(-\rho^{\star\star} - \sqrt{\frac{c}{\frac{1}{T} \sum_{t=1}^T y_{t-1}^2}}\right) \to \infty
            \end{align*}
            Therefore, $\lim_{T\rightarrow\infty}\Prob{\widehat{\rho}_\lambda=0\vert\mathcal{C}=c} = 0$.
        \item[(b)] $c = {\rho^{\star\star}}^2\E\left(y_{t-1}^2\right)$:
            \begin{align*}
                \Xi_T = \sqrt{T}\left(-\rho^{\star\star} - \sqrt{\frac{c}{\frac{1}{T} \sum_{t=1}^T y_{t-1}^2}}\right) \xrightarrow{d} \Xi\sim N(0,\sigma^2_{\scriptscriptstyle\Xi})
            \end{align*}
        by \Cref{lem:knifeCLT}. Using Slutsky's theorem, we find that 
        \begin{align*}
	        U_T = Z_T + \Xi_T \xrightarrow{d} U := Z + \Xi,
        \end{align*}
 		 and so $U$ is the sum of two zero-mean Gaussians. Denoting $F_U$ the CDF of $U$, we conclude that
        \begin{align*}
        	\lim_{T\rightarrow\infty}\Prob{\widehat{\rho}_\lambda=0\vert\mathcal{C}=c} = F_U(0) = .5,
        \end{align*}
		which proves (b).
        \item[(c)] $c\in\left({\rho^{\star\star}}^2\E\left(y_{t-1}^2\right),\,\infty\right)$:        \begin{align*}
             	U_T = Z_T + \sqrt{T}\left(-\rho^{\star\star} -\sqrt{\frac{c}{\frac{1}{T} \sum_{t=1}^T y_{t-1}^2}}\right) \to -\infty
            \end{align*}
       		 Hence, $\lim_{T\rightarrow\infty}\Prob{\widehat{\rho}_\lambda=0\vert\mathcal{C}=c} = 1$, which proves (c).
    \end{itemize}
    \item $\kappa>1$: \eqref{eq:Jterm} diverges such that $U_T\rightarrow -\infty$ and therefore $\lim_{T\rightarrow\infty}\Prob{\widehat\rho_\lambda = 0} = 1$, which yields result 3 and completes the proof.\qedhere
\end{enumerate}
\end{proof}
\vspace{1em}

\begin{proof}[\bfseries Proof of \Cref{prop:lambdanaught}]
    Note that the FOC for $\dot\rho_\lambda = 0$ yields
		\begin{align*}
    	\begin{split}
        	\lambda_{0,\,\rho^\star} 
        =&\, \left(\frac{1}{w_1}\right)^{\gamma_1} \left\lvert \sum_t y_{t-1} \left(\rho^\star y_{t-1} + \sum_{j=1}^{p} (\delta_j^\star - \dot{\delta}_j) \Delta y_{t-j} + \varepsilon_{p,\,t} \right) \right\rvert \\
        =:&\, \left(\frac{1}{w_1}\right)^{\gamma_1} \left\lvert \sum_t \xi_{t,\,\rho^\star} \right\rvert
	    \end{split}
	\end{align*}
	and let $\xi_{T,\,\rho^\star} := \frac{1}{T}\sum_t \xi_{t,\,\rho^\star}$.
	\begin{enumerate}
		\item If $\rho^\star=0$, then $w_1^{-1}$ and $\Breve w_1^{-1}$ are $O_p(T^{-1})$ by \Cref{lem:crates}, part \ref{item:lem:crates:H0}. Also, $\xi_{T,\,\rho^\star}= O_p(1)$. We thus have
		\begin{align*}
			\lambda_{0,\,\rho^\star} =&\, \left(\frac{1}{w_1}\right)^{\gamma_1} \biggr\lvert T \xi_{T,\rho^\star}\biggr\rvert\\
			=&\, \left(T\frac{1}{w_1}\right)^{\gamma_1} T^{1-\gamma_1} \biggr\lvert\xi_{T,\,\rho^\star}\biggr\rvert\\
			=&\,O_p(1)\Theta(T^{1-\gamma_1})O_p(1)\\
			=&\,O_p(T^{1-\gamma_1}).
		\end{align*}
		Using the same argument we conclude that $\Breve\lambda_{0,\,\rho^\star} = O_p(T^{1-\gamma_1})$, too.
		\item If $\rho^\star\in(-2,0)$, then $\xi_{T,\,\rho^\star} = O_p(1)$. For AL, we obtain
		\begin{align}
			\lambda_{0,\,\rho^\star}=&\, \left(\frac{1}{w_1}\right)^{\gamma_1} \lvert T\xi_{T,\,\rho^\star}\rvert\notag\\
			=&\, T\lvert\widehat{\rho}\rvert^{\gamma_1}\lvert\xi_{T,\,\rho^\star}\rvert\notag\\
			=&\, \Theta(T)\Theta(1)\Theta(1)\notag,
		\end{align}
		and so the leading term will be $\Theta(T)$, regardless of $\gamma_1$. For ALIE we find by part \ref{item:lem:crates:H1} of \Cref{lem:crates} and the above argument that
		\begin{align*}
			\Breve\lambda_{0,\,\rho^\star}=&\, \left(\frac{1}{\Breve w_1}\right)^{\gamma_1} \lvert T\xi_{T,\,\rho^\star}\rvert\notag\\
			=&\, \left(\frac{\lvert\widehat\rho\rvert}{T\cdot J_\alpha}\right)^{\gamma_1} T^{1+\gamma_1} \lvert\xi_{T,\,\rho^\star}\rvert\\
			=&\, \Theta(1)\Theta\left(T^{1+\gamma_1}\right)\Theta(1)\\
			=&\, \Theta(T^{1+\gamma_1}).
		\end{align*}
		\item From the FOC for $\dot\delta_j = 0$ we obtain		
		\begin{align*}
			\lambda_{0,\,\delta_j^\star\,}=&\, \left(\frac{1}{w_{2,\,j}}\right)^{\gamma_2}\biggl\lvert \sum_t \Delta y_{t-j}\biggl((\rho^\star-\dot\rho)y_{t-1} + \delta_j^\star\Delta y_{t-j} + \sum_{i:\, i\neq j}(\delta_i^\star-\dot\delta_i)\Delta y_{t-i} + \varepsilon_{p,\,t}\biggr) \biggr\rvert.\\
			\lambda_{0,\,\delta_j^\star\,}=&\, \left(\frac{1}{w_{2,\,j}}\right)^{\gamma_2}\biggl\lvert\sum_t\xi_{t,\,\delta_j^\star}\biggr\rvert.
		\end{align*}
		Let $\xi_{T,\,\delta_j^\star} := \frac{1}{T}\sum_t\xi_{t,\,\delta_j^\star}$ and note that $\xi_{T,\,\delta_j^\star}= O_p(T^{-1/2})$ for $\rho^\star\in(-2,0]$ and irrespective of $\delta_i^\star,\ i\neq j$. If $\delta_j^\star = 0$, then
		\begin{align*}
			\lambda_{0,\,\delta_j^\star\,} =&\ \left\lvert\widehat\delta_j\right\rvert^{\gamma_2} \left\lvert T\xi_{T,\,\delta_j^\star}\right\rvert\\
			=&\, \left\lvert\sqrt{T}\widehat\delta_j\right\rvert^{\gamma_2} T^{\frac{1-\gamma_2}{2}} \left\lvert\sqrt{T}\xi_{T,\,\delta_j^\star}\right\rvert\\
			=&\,O_p(1)\Theta\left(T^{\frac{1-\gamma_2}{2}}\right)O_p(1)\\
			=&\,O_p\left(T^{\frac{1-\gamma_2}{2}}\right).
		\end{align*}
	
		If $\delta_j^\star\neq0$, then $\xi_{T,\,\delta_j^\star}= O_p(1)$ for $\rho^\star\in(-2,0]$ and irrespective of the $\delta_i^\star,\ i\neq j$. Hence,
		\begin{align*}
			\lambda_{0,\,\delta_j^\star}=&\, \left\lvert\widehat\delta_j\right\rvert^{\gamma_2} \left\lvert T\xi_{T,\delta_j^\star}\right\rvert\\
			=&\, \Theta(1)\Theta(T)\Theta(1)\\
			=&\, \Theta(T).\qedhere
		\end{align*}
	\end{enumerate}
\end{proof}

\begin{proof}[\bfseries Proof of \Cref{lem:ZeroShrinkage}]
	Investigating the same (just-binding or non-binding) FOC derived from the KKT conditions as presented in the proof of \Cref{prop:lambdanaught}, we prove that the FOC for $y_{t-1}$ can asymptotically never be non-binding by deriving the limits of both sides. Rewrite the FOC as
	\begin{align*}
		\lambda \Breve{w}_1^{\gamma_1} \,\geq&\, \left\lvert \sum_t y_{t-1}  \widehat{\varepsilon}_{p,\,t,\,\lambda} \right\rvert \label{eq:rhohat0_foc},\\
		\frac{\lambda}{\Breve{\lambda}_{0,\,\rho^\star \in (-2,0)}} \,\frac{\Breve{w}_1^{\gamma_1} \, \Breve{\lambda}_{0,\,\rho^\star \in (-2,0)}}{T} \geq&\, \left\lvert \frac{1}{T} \sum_t y_{t-1} \widehat{\varepsilon}_{p,\,t,\,\lambda} \right\rvert \, .
		\intertext{Since $\lambda = o\left(\Breve{\lambda}_{0,\,\rho^\star \in (-2,0)}\right)$ by assumption, evaluating the limiting behaviour of the left-hand side using \Cref{lem:crates} and \Cref{prop:lambdanaught} yields}
		o(1) \frac{O_p(T^{-\gamma_1}) \, \Theta(T^{1+\gamma_1})}{T} \geq&\, \left\lvert \frac{1}{T} \sum_t y_{t-1} \widehat{\varepsilon}_{p,\,t,\,\lambda} \right\rvert \\
		o_p(1) \,\geq&\, \left\lvert \frac{1}{T} \sum_t y_{t-1} \widehat{\varepsilon}_{p,\,t,\,\lambda} \right\rvert \, .
	\end{align*}
	Since the left-hand side has a zero limit and the right-hand side has the lower bound zero, the FOC must be asymptotically always binding for $\lambda = o(\Breve{\lambda}_{0,\,\rho^\star \in (-2,0)})$ at
	\begin{equation*}
		\lim_{T \to \infty}  \frac{1}{T} \sum_t y_{t-1} \widehat{\varepsilon}_{p,\,t,\,\lambda} = 0 \, .\qedhere
	\end{equation*}
\end{proof}

\begin{proof}[\bfseries Proof of \Cref{cor:knot1asymptotics}]
    \noindent
    \begin{enumerate}
        \item By result \ref{item:prop:lambdanaught:rho<0} of \Cref{prop:lambdanaught},
        \begin{equation*}
            \frac{\Breve\lambda_{0,\, \rho^\star\in(-2,0)}}{T^{1+\gamma_1}} \overset{p}{\to} c_{\rho^\star} \in (0, \infty).
        \end{equation*}
        By result \ref{item:prop:lambdanaught:delta=0} of \Cref{prop:lambdanaught},
        \begin{equation*}
            \frac{\lambda_{0,\,\delta_j^\star \neq 0}}{T}=\Theta(1)\quad\textup{and}\quad\frac{\lambda_{0,\,\delta_j^\star = 0}}{T^{1-\gamma_2}} = O_p(1), 
        \end{equation*}    
        and because $\gamma_1 > 0$ and $\gamma_2 > 0$,
        \begin{equation*}
        \frac{\lambda_{0,\,\delta_j^\star}}{T^{1+\gamma_1}} \overset{p}{\to} 0.
        \end{equation*}
        By definition, $\lambda^{(1)} = \max(\Breve{\lambda}_{0,\,\rho^\star}, \lambda_{0,\,\delta_1^\star}, \dots, \lambda_{0,\,\delta_p^\star})$, see \textcite{Efronetal2004}. By linearity and continuity of the $\max$ function and since $\sgn(T) = +1$,
        \begin{equation*}
            \max\left( \Breve{\lambda}_{0,\,\rho^\star}, \lambda_{0,\,\delta_1^\star}, \dots, \lambda_{0,\,\delta_p^\star} \right) = T^{1 + \gamma_1} \, 
            \max\left(\frac{\Breve{\lambda}_{0,\, \rho^\star}}{T^{1 + \gamma_1}}, \frac{\lambda_{0,\, \delta_1^\star}}{T^{1 + \gamma_1}}, \dots, \frac{\lambda_{0,\,\delta_p^\star}}{T^{1 + \gamma_1}} \right),
        \end{equation*}
        Application of the CMT yields
        \begin{align*}
            \textup{plim}\Bigg( & \max\left( \frac{\Breve{\lambda}_{0,\,\rho^\star}}{T^{1+\gamma_1}}, \frac{\lambda_{0,\,\delta_1^\star}}{T^{1+\gamma_1}}, \dots, \frac{\lambda_{0,\,\delta_p^\star}}{T^{1+\gamma_1}} \right) \Bigg)\\
            =& \max\left( \textup{plim} \left( \frac{\Breve{\lambda}_{0,\,\rho^\star}}{T^{1+\gamma_1}}\right), \textup{plim}\left(\frac{\lambda_{0,\,\delta_1^\star}}{T^{1+\gamma_1}}\right), \dots, \textup{plim}\left(\frac{\lambda_{0,\,\delta_p^\star}}{T^{1+\gamma_1}}\right) \right)\\
            =& \max \left(c_{\rho^\star}, 0, \dots, 0 \right)\\
            =& c_{\rho^\star}.
        \end{align*}
        We conclude that $c_{\rho^\star}$ is the unique maximum among \textit{all} possible activation knots, obtaining the desired result.
        
        \item By result \ref{item:cor:knot1asymptotics:P1} of this proposition, $y_{t-1}$ is activated first as $T \to \infty$. By continuity of $\widehat{\rho}_\lambda$ in $\lambda$ (cf. \cite{Efronetal2004}), $\lim_{T\to\infty}\widehat{\rho}_{\lambda^{(2)}}$ is the only non-zero coefficient at $\lambda^{(2)}$: we must have $\lambda^{(2)} = \lambda_{0,\,\delta_j^\star}$ for some $j$ which consequently cannot be a \textit{deactivation} knot for $y_{t-1}$. From part \ref{item:prop:lambdanaught:delta=0} of \Cref{prop:lambdanaught} we infer that $\lambda^{(2)} = \lambda_{0,\, \delta_j^\star} = O_p(T)$, and so $\lambda^{(2)} = o_p(T^{1+\gamma_1})$. By \Cref{lem:ZeroShrinkage}, we conclude that
        \begin{equation}
            \lim_{T \to \infty} \frac{1}{T} \sum_{t} y_{t-1} \widehat{\varepsilon}_{t,\,p,\,\lambda^{(2)}} = 0.\label{eq:ortholambda2}
        \end{equation}
        Since $y_{t-1}$ is the only regressor with a non-zero coefficient, the estimated model corresponds to the ADF(0) model considered in \Cref{thm:aronelpalt} and \Cref{lem:onesquaresummability}. Hence, the orthogonality property \eqref{eq:ortholambda2} implies asymptotic equivalence of $\widehat{\rho}_{\lambda^{(2)}}$ to the OLS estimator $\widehat{\rho}$ in $\Delta y_{t-1} = \rho y_{t-1} + u_t$. The result follows by \Cref{lem:rhostarstar}, part \ref{item:lem:rhostarstar:P1}.
    \end{enumerate}
\end{proof}

\begin{proof}[\bfseries Proof of \Cref{thm:permactive}]
	We prove the result using partitions of $[0, \lambda^{(1)}]$:
	\begin{enumerate}
		\item For $\lambda\in(\lambda^{(2)},\lambda^{(1)}]$, the result follows immediately from \Cref{cor:knot1asymptotics} in conjunction with the linearity of $\dot\rho(\lambda)$ between $\lambda^{(2)}$ and $\lambda^{(1)}$.
		\item For $\lambda\in[0,\lambda^{(2)}]$, inactivity does asymptotically require $y_{t-1}$ to be removed from the active set since it has already been activated. We need to distinguish between two cases related to a hypothetical deactivation knot $\lambda^{-}$ and its closest reactivation knot $\lambda^{+} \leq \lambda^{-}$:
		\begin{enumerate}
			\item\label{itm:permactive:2case1} $\lambda^{+} = \lambda^{-}$
			\item\label{itm:permactive:2case2} $\lambda^{+} < \lambda^{-}$
		\end{enumerate}
		Following the considerations on the LARS algorithm in \textcite{Tibshirani2013lasso}, who refers to $\widehat{\mathcal{M}}_\lambda$ as the \enquote{equicorrelation set}, case~(a) does not cause $y_{t-1}$ to be dropped from the active set, because the FOC remains binding despite $\dot{\rho}_{\lambda^{-}} = 0$,
		$$ \lambda^{+} \Breve{w}_1^{\gamma_1} = \lambda^{-} \Breve{w}_1^{\gamma_1} = \left\lvert \sum_t y_{t-1} \widehat{\varepsilon}_{p,\,t,\,\lambda^{-}} \right\rvert \, .\footnote{\textcite{Tibshirani2013lasso} denotes such $\lambda^{-}$ as $\lambda^\mathrm{cross}$ because it emulates a sign change without deactivation, similar as for a LAR solution path.} $$
		Case~(b) does lead to a removal of $y_{t-1}$ from the active set. Since $y_{t-1}$ is due to be removed at $\lambda^{-}$ and re-enters the equicorrelation set not before $\lambda^{+}$, it does \emph{not} belong to the equicorrelation set for any $\lambda' \in (\lambda^{+}, \lambda^{-})$. According to the defining equation (4) on page 1461 in \textcite{Tibshirani2013lasso}, 
	\begin{equation*}
	   \left\lvert \sum_t y_{t-1} \widehat{\varepsilon}_{p,\,t,\,\lambda^{'}} \right\rvert \neq \lambda' \Breve{w}_1^{\gamma_1}  .
	\end{equation*}
	That is why the FOC needs to hold with strict inequality for $y_{t-1}$ to be inactive. However, since every $\lambda' \in [0, \lambda^{(2)}]$ is $o(\Breve{\lambda}_{0,\,\rho^\star\in(-2,0)})$, \Cref{lem:ZeroShrinkage} proves the FOC to be asymptotically always binding at zero:
		\begin{align*}
			\lim_{T\to\infty} \Prob{y_{t-1}\not\in \widehat{\mathcal{M}}_{\lambda'}} =&\, \lim_{T\to\infty}\Prob{\lambda'\Breve{w}_1^{\gamma_1} > \left\lvert\sum_t y_{t-1}\widehat{\varepsilon}_{p,\,t,\,\lambda^{'}}\right\rvert}\\
			\leq&\, 1-\lim_{T\to\infty}\Prob{\lambda'\Breve{w}_1^{\gamma_1}\sgn(\dot\rho_{\lambda'}) = \sum_t y_{t-1} \widehat{\varepsilon}_{p,\,t,\,\lambda^{'}}}\\
			=&\, 0.
		\end{align*}
		Therefore,
		$$\lim_{T\to\infty}\Prob{y_{t-1}\in\widehat{\mathcal{M}}_\lambda} = 1 \quad \forall \quad \lambda\in[0,\lambda^{(2)}].$$
	\end{enumerate}
\end{proof}

\begin{proof}[\bfseries Proof of \Cref{thm:nmmlmtwo}]
    \noindent
    \begin{enumerate}
        \item
         For result \ref{item:thm:nmmlmtwo:P1:a}, we examine the activation probabilities of $\widehat\mu_{w}$ and $\widehat\mu_{\Breve w}$ for the tuning parameter $\lambda_\textup{IC}\sim F_{\lambda_\textup{IC}}$.
        \begin{align}
         	\begin{split}
          	\Prob{\widehat\mu_{w}\neq0} =&\,\Prob{\lambda_0>\lambda_\textup{IC}}\\
         	=&\,1-\int_{0}^\infty F_{\lambda_0}(\lambda_\textup{IC})\,\mathrm{d}F_{\lambda_\textup{IC}},
         	\end{split} \label{eq:actprobeq}\\
         	\notag\\
         	\begin{split}
          	\Prob{\widehat\mu_{\Breve w}\neq0} =&\,\Prob{\Breve\lambda_0>\lambda_\textup{IC}}\\
         	=&\,1-\int_{0}^\infty F_{\Breve\lambda_0}(\lambda_\textup{IC})\,\mathrm{d}F_{\lambda_\textup{IC}}.
         	\end{split}\notag
        \end{align}
     
 		To see why $\lim_{q\to\infty}\E(\Breve{\lambda}_0)=0$ is a sufficient condition, note that
 		\begin{equation*}
            \lim_{q\to\infty}\Prob{\Breve\lambda_0 > c} \leq \lim_{q\to\infty}\frac{E(\Breve\lambda_0)}{c} = 0,
 		\end{equation*}
 		for $c\in(0,\infty)$ by Markov's inequality, implying $\lim_{q\to\infty} F_{\Breve{\lambda}_0}=\varrho_0$, where $\varrho_0$ is the Dirac measure at 0. Therefore,
            \begin{equation*}
               \lim_{q\to\infty}\Prob{\Breve{\lambda}_0>\lambda_\textup{IC}} = 0 \quad \forall \  \lambda_\textup{IC} > 0,
            \end{equation*} 
            	which proves result~\ref{item:thm:nmmlmtwo:P1:b}.
        \item 
        Since $F_{\lambda_0}$ and $F_{\lambda_\textup{IC}}$ have common support $[0,\infty)$, $\lambda_\textup{IC}$ can be represented by the injection $\lambda_\textup{IC}=F_{\lambda_0}^{-1}(1-a)$ with $a\in[0,1]$. We then have by \eqref{eq:actprobeq},
         \begin{align}
              &\Prob{\widehat\mu_{w}\neq0} - \Prob{\widehat\mu_{\Breve{w}}\neq0}\notag\\
              &=\,\textup{Pr}(\lambda_0 > \lambda_\textup{IC}) - \textup{Pr}(\Breve\lambda_0 > \lambda_\textup{IC})\notag\\
              &= \int_{a=1}^{a=0} F_{\Breve{\lambda}_0}\left(F^{-1}_{\lambda_0}(1-a)\right)\,\mathrm{d} F_{\lambda_\textup{IC}}\left(F^{-1}_{\lambda_0}(1-a)\right)\notag\\
              &\,\quad- \int_{a=1}^{a=0} F_{\lambda_0}\left(F^{-1}_{\lambda_0}(1-a)\right)\,\mathrm{d} F_{\lambda_\textup{IC}}\left(F^{-1}_{\lambda_0}(1-a)\right)\notag\\
              &= \int_{a=1}^{a=0}\left[F_{\Breve{\lambda}_0}\left(F^{-1}_{\lambda_0}(1-a)\right) - (1-a)\right] \,\mathrm{d} F_{\lambda_\textup{IC}}\left(F^{-1}_{\lambda_0}(1-a)\right).\label{eq:lambdaprobs}
         \end{align}
         Now consider
         \begin{equation*}
          	A_q := \left\{a\in(0,1): F^{-1}_{\Breve{\lambda}_0}(1-a) < F^{-1}_{\lambda_0}(1-a)\right\},
        \end{equation*}
        for some $q\in(0,\infty)$. For any $a\in A_q$,
        \begin{equation}
            F_{\Breve{\lambda}_0}\left(F^{-1}_{\lambda_0}(1-a)\right)\geq F_{\lambda_0}\left(F^{-1}_{\lambda_0}(1-a)\right) = 1-a.\label{eq:blambdaineq}
        \end{equation}
        Using \eqref{eq:blambdaineq} we conclude from \eqref{eq:lambdaprobs} that
         \begin{equation*}
             0\leq\int_{a\in A_q} \left[ F_{\Breve{\lambda}_0}\left(F^{-1}_{\lambda_0}(1-a)\right) - (1-a)\right]\,\mathrm{d} F_{\lambda_\textup{IC}}\left(F^{-1}_{\lambda_0}(1-a)\right).
         \end{equation*}
         Hence, $\Prob{\widehat\mu_{\Breve{w}}\neq0}\leq \Prob{\widehat\mu_w\neq0}=a$ for $a\in A_q$. \qedhere
    \end{enumerate}
\end{proof}
\vspace*{1em}

	\begin{proof}[\bfseries Proof of \Cref{cor:qlimnmm}]
    \noindent
    \begin{enumerate}
        \item 
		By independence and Gaussianity, $\Breve{\lambda}_0$ obtains as the product of transformed half-Gaussian r.v.s,
		\begin{align*}
			\breve\lambda_0 =&\, \left\lvert\epsilon\cdot \prod_{j=1}^q \nu_j\right\rvert^\gamma \lvert\epsilon\rvert = \lvert\epsilon\rvert^\gamma \lvert\epsilon\rvert\cdot\prod_{j=1}^q\left\lvert \nu_j\right\rvert^\gamma,
			\intertext{so that}
			\E(\breve\lambda_0)=& \, \E\left(\lvert\epsilon\rvert^\gamma \lvert\epsilon\rvert\cdot\prod_{j=1}^q\left\lvert \nu_j\right\rvert^\gamma\right)\\ 
			=&\,\E(\lvert\epsilon\rvert^\gamma \lvert\epsilon\rvert)\cdot\prod_{j=1}^q \E(\lvert \nu_j\rvert^\gamma).
		\end{align*}
		If $\prod_{j=1}^q\E(\lvert\nu_j\rvert^\gamma) = o(1)$, we have
		\begin{equation}
			\lim_{q\to\infty}\E(\breve\lambda_0) = \lim_{q\to\infty} \E(\lvert\epsilon\rvert^\gamma \lvert\epsilon\rvert)\cdot\prod_{j=1}^q \E(\lvert \nu_j\rvert^\gamma) = 0,
		\end{equation} 
		and hence $\lim_{q\rightarrow\infty} \Prob{\widehat\mu_{\Breve w}\neq0} = 0$, by result \ref{item:thm:nmmlmtwo:P1:b} of \Cref{thm:nmmlmtwo}.
        \item
        For $\gamma=1$ we obtain
        \begin{align*}
			\E(\breve\lambda_0)=&\, \sigma_\epsilon^2\prod_{j=1}^q \sqrt{\frac{2}{\pi}}\sigma_{\nu,\,j} = \sigma_\epsilon^2\left(\frac{2}{\pi}\right)^{q/2} \prod_{j=1}^q \sigma_{\nu,\,j}.
		\end{align*}
		Hence, if $\prod_{j=1}^q \sigma_{\nu,\,j} = o\left( (\pi/2)^{q/2}\right)$, then 
		\begin{equation}
			\lim_{q\to\infty}\E(\breve\lambda_0) = \lim_{q\to\infty} \sigma_\epsilon^2\left(\frac{2}{\pi}\right)^{q/2} \prod_{j=1}^q \sigma_{\nu,\,j} = 0,
		\end{equation}
		so that $\lim_{q\rightarrow\infty} \Prob{\widehat\mu_{\Breve w}\neq0} = 0$, by result \ref{item:thm:nmmlmtwo:P1:b} of \Cref{thm:nmmlmtwo}.\qedhere
    \end{enumerate}
\end{proof}

%
\newpage
\printbibliography
\end{document}